\documentclass[a4paper,11pt]{article}
\usepackage{amsmath,amssymb}
\usepackage{comment}
\usepackage{bm}
\usepackage{mathtools}
\usepackage[dvips]{graphicx}
\usepackage{here}
\usepackage[mathscr]{euscript}
\newcommand{\llangle}{\bigl\langle\!\bigl\langle}
\newcommand{\rrangle}{\bigr\rangle\!\bigr\rangle}

\begin{document}

\begin{titlepage}

\rightline{\tt UT-Komaba/16-9}
\rightline{\tt YITP-16-130}
\begin{center}
\vskip 2.5cm
{\Large \bf {Fermion scattering amplitudes from gauge-invariant actions \\
for open superstring field theory}\\}
\vskip 1.0cm
{\large {Hiroshi Kunitomo,${}^1$ Yuji Okawa,${}^2$ Hiroki Sukeno${}^2$ and Tomoyuki Takezaki${}^2$}}
\vskip 1.0cm
{${}^1$\it{Center for Gravitational Physics, Yukawa Institute for Theoretical Physics,\\ 
Kyoto University, Kyoto 606-8502, Japan}}\\
{${}^2$\it {Institute of Physics, The University of Tokyo, Komaba, Meguro-ku, Tokyo 153-8902, Japan}}\\
\vskip 1.0cm
{kunitomo@yukawa.kyoto-u.ac.jp, okawa@hep1.c.u-tokyo.ac.jp, \\sukeno@hep1.c.u-tokyo.ac.jp, takezaki@hep1.c.u-tokyo.ac.jp}

\vskip 2.0cm

{\bf Abstract}
\end{center}

\noindent

We calculate on-shell scattering amplitudes involving fermions
at the tree level in open superstring field theory.
We confirm that four-point and five-point amplitudes
in the world-sheet path integral
with the standard prescription using picture-changing operators
are reproduced.
For the four-point amplitudes,
we find that the quartic interaction required by gauge invariance
adjusts the different assignment of picture-changing operators
in the $s$-channel and in the $t$-channel
of Feynman diagrams with two cubic vertices.
For the five-point amplitudes,
the correct amplitudes are reproduced
in a more intricate way via the quartic and quintic interactions.
Our calculations can be interpreted
as those for a complete action
with a constraint on the Ramond sector
or as those for the covariant formulation
developed by Sen
with spurious free fields.

\end{titlepage}

\tableofcontents

\newpage

\baselineskip=17pt

\section{Introduction}

In formulating superstring field theory, construction of an action including the Ramond sector
has been a major obstacle for about thirty years.
One alternative approach was to impose a constraint
on the equation of motion after it was derived from an action
just as in the case of type IIB supergravity.
This approach was applied to open superstring field theory
including the Ramond sector in~\cite{Michishita:2004by},
and a set of Feynman rules was proposed.
While four-point amplitudes at the tree level
were correctly reproduced based on these Feynman rules,
it was reported that five-point amplitudes at the tree level
were not correctly reproduced~\cite{Michishita:Riken}.
This issue was later resolved by a refined set of Feynman rules proposed in~\cite{Kunitomo:2014qla}.\footnote{
This approach with a constraint to be imposed on the equation of motion
was also applied to heterotic string field theory including the Ramond sector~\cite{Kunitomo:2015hda}.
}

Very recently, the situation drastically changed.
A gauge-invariant action for open superstring field theory
including both the Neveu-Schwarz (NS) sector and the Ramond sector
was constructed in~\cite{Kunitomo:2015usa}.
The string field in the Ramond sector is in a restricted subspace
of the Hilbert space of the boundary conformal field theory (CFT),
and this restriction allows a covariant kinetic term
for the string field in the Ramond sector.
Interactions are written in a closed form,
and it was shown that the action is invariant under gauge transformations
which are compatible with the restriction.

At almost the same time,
another approach to constructing covariant actions including the Ramond sector
was proposed by Sen~\cite{Sen:2015uaa, Sen:2015hha, Sen:2014dqa, Sen:2014pia}.
Extra free fields are introduced in this approach, and with these free fields
covariant kinetic terms are realized.
While the discussion in~\cite{Sen:2015uaa, Sen:2015hha} is in the context of heterotic string field theory
and type II superstring field theory,
the idea can be applied to open superstring field theory as well,
and for the open superstring
the interactions constructed in~\cite{Kunitomo:2015usa}
can be combined with the kinetic terms in~\cite{Sen:2015uaa, Sen:2015hha}
to obtain a gauge-invariant action in a covariant form.

The interaction constructed in~\cite{Kunitomo:2015usa}
is based on the Wess-Zumino-Witten-like (WZW-like) form
for the string field in the NS sector
and does not exhibit an $A_\infty$ structure
which plays a crucial role in the Batalin-Vilkovisky quantization.
The construction of multi-string products
with an $A_\infty$ structure for the open string
or an $L_\infty$ structure for the closed string
developed by Erler, Konopka and Sachs~\cite{Erler:2013xta, Erler:2014eba, Erler:2015lya}
has recently adapted to be compatible
with the restriction for the string field in the Ramond sector,
and a gauge-invariant action with a cyclic $A_\infty$ structure
was constructed in~\cite{Erler:2016ybs, Konopka:2016grr}.
By extending the method developed for the NS sector in~\cite{Erler:2015uoa,Erler:2015rra,Erler:2015uba},
it was also shown in~\cite{Erler:2016ybs} that the action with a cyclic $A_\infty$ structure~\cite{Erler:2016ybs, Konopka:2016grr}
is equivalent to the one constructed in~\cite{Kunitomo:2015usa}
via partial gauge fixing and field redefinition.

We now have gauge-invariant actions for open superstring field theory
including both the NS sector and the Ramond sector,
and in this paper we calculate scattering amplitudes
which involve spacetime fermions.
We use the action constructed in~\cite{Kunitomo:2015usa}
and calculate four-point and five-point amplitudes at the tree level.
Unlike the case with an action supplemented
by a constraint we mentioned before,
it is straightforward to derive Feynman rules from the action.
There is no ambiguity in the resulting Feynman rules,
and we find that four-point and five-point amplitudes
calculated from the world-sheet theory of the open superstring
are correctly reproduced.
Our calculations can also be viewed as those from the action
based on the approach by Sen~\cite{Sen:2015uaa}.
Our calculations further serve as indirect confirmation
that the correct amplitudes are reproduced from the action
with a cyclic $A_\infty$ structure~\cite{Erler:2016ybs,Konopka:2016grr}
because it is related to the action in~\cite{Kunitomo:2015usa} by field redefinition.\footnote{
See~\cite{Konopka:2015tta} for an alternative discussion
that correct scattering amplitudes are reproduced from theories
constructed by the approach developed in~\cite{Erler:2013xta,Erler:2016ybs,Konopka:2016grr}.
}
By the way, an explicit construction of multi-string products
by carrying out the program in~\cite{Erler:2013xta,Erler:2016ybs,Konopka:2016grr} is exceedingly complicated
at higher orders,
and direct calculations of four-point and five-point amplitudes
would be a formidable task.
This is an advantage of using the WZW-based formulation~\cite{Kunitomo:2015usa, Berkovits:1995ab}
where the interaction is beautifully written in a closed form.

It is generally believed in the framework of string field theory
that the extension from a free theory to an interacting theory is unique
up to field redefinition if the gauge invariance in the free theory
is nonlinearly extended and the interacting theory is invariant
under the nonlinearly extended gauge transformation. 
Therefore, it is not surprising that the correct scattering amplitudes
are reproduced from the action in~\cite{Kunitomo:2015usa},
although it is still important to confirm this explicitly.
Our motivation for the calculations is not only
to confirm this
but also to see how the correct results are reproduced.
In particular, scattering amplitudes in the world-sheet theory
are based on the integration over the supermoduli space
of super-Riemann surfaces, and it is important to understand
how this integration over the supermoduli space is implemented
in superstring field theory based on spacetime gauge invariance.

This point was partially elucidated in the calculation
of four-point amplitudes in the NS sector
for the Berkovits formulation~\cite{Berkovits:1999bs, Iimori:2013kha} of open superstring field theory.
In~\cite{Iimori:2013kha} the four-point amplitudes were calculated
in such a way that the relation to the supermoduli space
can be understood more manifestly
than the previous calculation in~\cite{Berkovits:1999bs}.
While the bosonic direction of the supermoduli space
was covered by contributions from Feynman diagrams
with two cubic vertices and one propagator,
the assignment of picture-changing operators
turned out to be different in the $s$-channel and in the $t$-channel.
Since the insertion of a picture-changing operator corresponds
to the integration of a fermionic modulus of the supermoduli space,
this mismatch can be interpreted that the supermoduli space
of disks with four punctures is not correctly covered
by contributions generated from the cubic interaction
of the Berkovits formulation.
It was found, however, that
the different behavior in the $s$-channel and in the $t$-channel
is adjusted by the contribution from Feynman diagrams
with a quartic vertex.\footnote{
In other words, the quartic interaction precisely implements
the {\it vertical integration}~\cite{Sen:2015hia}.
}
Therefore, the quartic interaction, which is required 
for gauge invariance in the Berkovits formulation, is also necessary
for the covering the supermoduli space.

In this paper we extend this insight to include the Ramond sector.
We again find that the contributions to four-point amplitudes from Feynman diagrams
with two cubic vertices and one propagator cover the bosonic direction
of the supermoduli space, but the assignment of picture-changing operators
is different in the $s$-channel and in the $t$-channel,
and this different behavior is precisely adjusted
by the contribution from Feynman diagrams with a quartic vertex.
One source of picture-changing operators is a cubic vertex
with three string fields in the NS sector as in~\cite{Berkovits:1995ab},
but picture-changing operators are also provided from propagators
in the Ramond sector,
and the adjustment is more elaborate.
This is one of the main results of this paper
and presented in section~\ref{four-point-amplitudes}.
We further calculate a class of five-point amplitudes in section~\ref{five-point-amplitudes}
and present calculations for other five-point amplitudes in appendix~\ref{appendix}.
In all cases we find agreement with amplitudes in the world-sheet theory.

On one hand, use of the large Hilbert space
of the superconformal ghost sector has turned out to be magically useful
in constructing gauge-invariant actions~\cite{Kunitomo:2015usa,Berkovits:1995ab}.
On the other hand, use of the large Hilbert space
obscures the relation to the supermoduli space,
and it would be fair to say that we do not yet understand
why the use of the large Hilbert space is so successful.
For further developments of superstring field theory,
it would be important to understand the relation
between the large Hilbert space and the supermoduli space,
and we hope that the analysis in this paper will provide
a clue to decode the magic of the large Hilbert space.

\section{Propagators and interactions}
\setcounter{equation}{0}

In this section
we derive propagators and present interaction terms
which are necessary for the calculations of four-point and five-point amplitudes.

\subsection{The propagator in the NS sector}

We denote the string field in the NS sector by $\phi$.
It is a Grassmann-even state in the large Hilbert space
and carries ghost number $0$ and picture number $0$.
The kinetic term is given by
\begin{equation}
 S_{NS}^{(0)}\ =\ -\frac{1}{2} \bigl\langle\, \phi,Q\eta\phi\,\bigr\rangle \,,
\label{NS-kinetic}
\end{equation}
where $\bigl\langle\, \, A, B \, \,\bigr\rangle$ is the BPZ inner product
in the large Hilbert space for a pair of states $A$ and $B$, 
$Q$ is the BRST operator,
and $\eta$ is the zero mode of the superconformal ghost $\eta (z)$.
The kinetic term is invariant under the following gauge transformations:
\begin{equation}
\delta \phi = Q \Lambda +\eta \Omega \,,
\end{equation}
where $\Lambda$ and $\Omega$ are the gauge parameters.

We introduce a source $J$ and add a source term to the kinetic term as
\begin{equation}
 S_{NS}^{(0)} [J] \ =\ -\frac{1}{2}\bigl\langle\, \phi,Q\eta\phi\,\bigr\rangle + \bigl\langle\, \phi,J\,\bigr\rangle\,.
\end{equation}
The equation of motion is
\begin{equation}
 Q\eta\phi\ =\ J\, .
\end{equation}
We choose the following gauge-fixing conditions:
\begin{equation}
 b_0\phi\ =0 \,, \qquad \xi_0\phi\ =\ 0\,,
\end{equation}
where $b_0$ is the zero mode of the conformal ghost $b (z)$
and $\xi_0$ is the zero mode of the superconformal ghost $\xi (z)$.
The solution to the equation of motion
under these gauge-fixing conditions is
\begin{equation}
 \phi\ =\ \frac{\xi_0\, b_0}{L_0}J\, .
\end{equation}
Evaluating $S_{NS}^{(0)} [J]$ for this solution, we find
\begin{equation}
S_{NS}^{(0)} [J] = \frac{1}{2} \bigl\langle\, J,\frac{\xi_0b_0}{L_0}J\,\bigr\rangle \,.
\end{equation}
The propagator for $\phi$ is thus given by
\begin{equation}
 \overbracket[0.5pt]{\!\!\!\! \big|\,\phi\,\bigr\rangle\bigl\langle\,\phi\,\big|\!\!\!\!}\ \,=\
\frac{\xi_0 b_0}{L_0}\,.
\end{equation}
The condition $\xi_0 \phi = 0$ for gauge fixing
can be generalized by replacing $\xi_0$
with a more general line integral of $\xi (z)$.
See~\cite{Iimori:2013kha,Kroyter:2012ni} for detailed discussions on the gauge fixing
in the NS sector and the calculation of the propagator.

We will compare on-shell scattering amplitudes calculated from open superstring field theory
with the corresponding amplitudes in the world-sheet theory
where on-shell vertex operators in the small Hilbert space are used.
Let us discuss the relation between the open superstring field $\phi$ in the large Hilbert space
and the on-shell vertex operator $\Phi$ in the small Hilbert space.
Since
\begin{equation}
\{ \, \xi_0, \eta \, \} = 1 \,,
\end{equation}
the state $\phi$ in the large Hilbert space can be decomposed as follows:
\begin{equation}
\phi = \xi_0 \eta \phi +\eta \xi_0 \phi \,.
\end{equation}
Using the gauge transformation generated by $\eta$,
we can bring $\phi$ to the form
\begin{equation}
\phi = \xi_0 \Phi \,,
\label{phi-to-Phi}
\end{equation}
where $\Phi$ carrying ghost number $1$ and picture number $-1$
is in the small Hilbert space and is given by
\begin{equation}
\Phi = \eta \phi \,.
\label{Phi-to-phi}
\end{equation}
In fact, the state $\phi$ in the form~\eqref{phi-to-Phi} satisfies
the gauge-fixing condition $\xi_0 \phi = 0$.
Under this gauge-fixing condition,
the equation of motion for $\phi$ in the free theory reduces as follows:
\begin{equation}
Q \eta \phi = Q \eta \xi_0 \Phi = Q \, \{ \, \eta, \xi_0 \, \} \, \Phi = Q \Phi = 0 \,.
\end{equation}
We thus identify $\Phi$ annihilated by $Q$
with the on-shell vertex operator of $-1$ picture in the small Hilbert space.
The state $\phi$ in the large Hilbert space and $\Phi$ in the small Hilbert space
are related as~\eqref{phi-to-Phi} and~\eqref{Phi-to-phi}.

We will use the BPZ inner product in the small Hilbert space
in addition to the BPZ inner product in the large Hilbert space.
We denote the BPZ inner product in the small Hilbert space
by $\llangle \, A, B \, \rrangle$ defined for a pair of states $A$ and $B$ in the small Hilbert space.
It is related to the BPZ inner product $\langle \, A, B \, \rangle$ in the large Hilbert space as
\begin{equation}
\llangle\, A, B \,\rrangle = \bigl\langle\, \xi_0 A, B \,\bigr\rangle \,.
\label{small-BPZ-definition}
\end{equation}

\subsection{The propagator in the Ramond sector}

As we mentioned in the introduction,
there are two approaches to constructing a covariant kinetic term
in the Ramond sector.
In~\cite{Kunitomo:2015usa}, a restricted subspace of the Hilbert space was used.
In the approach by Sen~\cite{Sen:2015uaa}, an additional free string field is introduced.
We discuss gauge fixing for both cases
and derive the propagator in this subsection.

\subsubsection{The restricted space}

Let us first consider the construction in~\cite{Kunitomo:2015usa}.
The restriction of the Hilbert space in the Ramond sector
is described in the following way.
We say that a state $A$ of picture $-1/2$ is in the {\it restricted space}
when it satisfies
\begin{equation}
\mathscr{X} \mathscr{Y} A = A \,,
\end{equation}
with
\begin{equation}
 \mathscr{X}\ =\ -\,\delta(\beta_0)\,G_0 + \delta'(\beta_0)\,b_0 \,, \qquad
 \mathscr{Y}\ =\ -\,c_0\,\delta'(\gamma_0)\,,
\end{equation}
where $\beta_0$ is the zero mode of the superconformal ghost $\beta (z)$, $\gamma_0$ is the zero mode of the superconformal ghost $\gamma (z)$, and $G_0$ is the zero mode of the supercurrent. 
Since
\begin{equation}
\mathscr{X} \mathscr{Y} \mathscr{X} = \mathscr{X} \,,
\end{equation}
the operator $\mathscr{X} \mathscr{Y}$ is a projector.
The operator $\mathscr{X}$ commutes with the BRST operator,
\begin{equation}
[ \, Q, \mathscr{X} \, ] = 0 \,,
\end{equation}
and we can show that $Q A$ is in the restricted space
when $A$ is in the restricted space.
See~\cite{Kunitomo:2015usa, Erler:2016ybs} for more details on the restricted space.

We denote the string field in the Ramond sector by $\Psi$.
It is a Grassmann-odd state in the small Hilbert space
and carries ghost number $1$ and picture number $-1/2$.
Furthermore, the string field $\Psi$ is in the restricted space:
\begin{equation}
\mathscr{X} \mathscr{Y} \Psi = \Psi \,.
\end{equation}
The kinetic term for $\Psi$ is given by
\begin{equation}
 S_{R}^{(0)}\ =\ -\frac{1}{2}\llangle\, \Psi, \mathscr{Y} Q\Psi\,\rrangle \,,
\label{Ramond-kinetic}
\end{equation}
and it is invariant under the following gauge transformation:
\begin{equation}
\delta \Psi = Q \lambda \,,
\end{equation}
where $\lambda$ is the gauge parameter in the restricted space.

We introduce a source $J$ for $\Psi$ and add a source term to the kinetic term as
\begin{equation}
 S_{R}^{(0)} [J] \ =\ -\frac{1}{2}\llangle\, \Psi, \mathscr{Y} Q\Psi\,\rrangle + \llangle\, \Psi,J\,\rrangle\,. 
\end{equation}
The equation of motion is
\begin{equation}
\mathscr{Y} Q \Psi = J \,.
\end{equation}
We multiply the equation of motion by $\mathscr{X}$ to obtain
\begin{equation}
\mathscr{X} \mathscr{Y} Q \Psi = \mathscr{X} J \,.
\end{equation}
Since $Q \Psi$ is in the restricted space, we have
$\mathscr{X} \mathscr{Y} Q \Psi = Q \Psi \,$.
The equation of motion is then
\begin{equation}
Q \Psi = \mathscr{X} J \,.
\end{equation}
We choose the following gauge-fixing condition:
\begin{equation}
 b_0\Psi\ =\ 0\,.
\label{Ramond-gauge-fixing}
\end{equation}
The solution to the equation of motion under this gauge-fixing condition is
\begin{equation}
 \Psi\ =\ \frac{b_0 \mathscr{X}}{L_0}\,J\, .
\end{equation}
Evaluating $S_{R}^{(0)} [J]$ for this solution, we find
\begin{equation}
S_{R}^{(0)} [J] = \frac{1}{2}\llangle\, J,\frac{b_0\mathscr{X}}{L_0}J\,\rrangle \,.
\end{equation}
The propagator for $\Psi$ in the small Hilbert space is given by
\begin{equation}
 \big|\,\,  \overbracket[0.5pt]{ \!\! \Psi \, \,\rrangle \llangle\, \,  \Psi \!\!}  \,\, \big| = \,  \frac{b_0 \mathscr{X}}{L_0} \,.
\end{equation}

Since the interaction terms are written as the BPZ inner product in the large Hilbert space, 
it is convenient to work in the large Hilbert space for the Ramond sector as well.
Let us therefore derive the propagator in the large Hilbert space.
We introduce a source $J$ in the large Hilbert space and add a source term
to the kinetic term as
\begin{align}
 S_{R}^{(0)} [J] \ =&\ -\frac{1}{2}\llangle\, \Psi, \mathscr{Y} Q\Psi\,\rrangle + \bigl\langle\, \Psi,J\,\bigr\rangle\, 
\nonumber\\
=&\ -\frac{1}{2}\bigl\langle\,\xi_0\Psi, \mathscr{Y} Q\Psi\,\bigr\rangle + \bigl\langle\, \Psi,J\,\bigr\rangle\,
\nonumber\\
=&\ \frac{1}{2}\,\bigl\langle\,\Psi, \xi_0 \mathscr{Y} Q\Psi\,\bigr\rangle + \bigl\langle\, \Psi,J\,\bigr\rangle\,,
\end{align}
where we used the fact that $\xi_0$ is BPZ even.
The equation of motion for $\Psi$ is then
\begin{equation}
 \xi_0 \mathscr{Y} Q\Psi\ =\ -\,J\, .
\end{equation}
The solution to the equation of motion under the gauge-fixing condition~\eqref{Ramond-gauge-fixing} is
\begin{equation}
 \Psi\ =\ -\frac{b_0 \mathscr{X} \eta}{L_0}\,J\, .
\end{equation}
Evaluating $S_{R}^{(0)} [J]$ for this solution, we find
\begin{equation}
S_{R}^{(0)} [J] = \frac{1}{2}\bigl\langle\, J,\frac{b_0 \mathscr{X}\eta}{L_0}J\,\bigr\rangle \,.
\end{equation}
The propagator for $\Psi$ in the large Hilbert space is given by
\begin{equation}
 \overbracket[0.5pt]{\!\!\!\!\! \big|\, \Psi\,\bigr\rangle\bigl\langle\,\Psi \,\big|\!\!\!\!\!}\ \ \, =\
-\,\frac{b_0 \mathscr{X} \eta}{L_0}\, .
\label{Ramond-propagator}
\end{equation}

\subsubsection{The approach by Sen}

Let us next consider the approach by Sen~\cite{Sen:2015uaa}.
In this approach, two string fields $\Psi$ and $\tilde{\Psi}$
in the Ramond sector are used.
The string field $\Psi$ is a Grassmann-odd state in the small Hilbert space
and carries ghost number $1$ and picture number $-1/2$.
The string field $\tilde{\Psi}$ is a Grassmann-odd state in the small Hilbert space
and carries ghost number $1$ and picture number $-3/2$.
While $\Psi$ appears in the interaction terms,
$\tilde{\Psi}$ appears only in the kinetic terms and is a free field.
Unlike the construction in~\cite{Kunitomo:2015usa}, these string fields $\Psi$ and $\tilde{\Psi}$
are not in the restricted space.

The kinetic terms for $\Psi$ and $\tilde{\Psi}$ are given by
\begin{equation}
S_{R}^{(0)} =\frac{1}{2} \llangle\,  \tilde{\Psi}, \, Q X_0 \tilde{\Psi} \,\rrangle
- \llangle\, \tilde{\Psi}, \, Q \Psi \, \rrangle \,,
\label{Sen-kinetic}
\end{equation}
where $X_0$ is the zero mode of the picture-changing operator.
It is BPZ even and commutes with the BRST operator:
\begin{equation}
[ \, Q, X_0 \, ] = 0 \,.
\end{equation}
The action $S_{R}^{(0)}$ is invariant under the gauge transformations:
\begin{equation}
\delta \Psi = Q \lambda \,, \qquad
\delta \tilde{\Psi} = Q \tilde{\lambda} \,,
\end{equation}
where $\lambda$ and $\tilde{\lambda}$ are the gauge parameters.
When we include interactions, the gauge transformation with the parameter $\lambda$
is nonlinearly extended, while the gauge transformation with the parameter $\tilde{\lambda}$ remains intact.

We introduce $J$ for $\Psi$ and $\tilde{J}$ for $\tilde{\Psi}$ as sources
in the small Hilbert space
and add source terms to $S_{R}^{(0)}$ as follows:
\begin{equation}
S_{R}^{(0)} [ \, J, \tilde{J} \, ] =\ \frac{1}{2} \llangle\, \tilde{\Psi}, \, Q X_0 \tilde{\Psi} \,\rrangle
- \llangle\, \tilde{\Psi}, \, Q \Psi \,\rrangle
+ \llangle\, \Psi, J \,\rrangle
+ \llangle\, \tilde{\Psi}, \tilde{J} \,\rrangle \,.
\end{equation}
The equations of motion are
\begin{equation}
\begin{split}
&Q\Psi-Q X_0 \tilde{\Psi}=\tilde{J} \,, \\
& Q\tilde{\Psi}=J \,.
\end{split}
\end{equation}
We choose the following gauge-fixing conditions:
\begin{equation}
b_0 \Psi = 0 \,, \qquad b_0 \tilde{\Psi} = 0 \,.
\end{equation}
The string fields $\Psi$ and $\tilde{\Psi}$
satisfying the equations of motion
under these gauge-fixing conditions are
\begin{equation}
\Psi = \frac{b_0}{L_0}(X_0J+\tilde{J}) \,, \quad
\tilde{\Psi} = \frac{b_0}{L_0}J \,.
\end{equation}
Evaluating $S_{R}^{(0)} [J]$ for these string fields, we find
\begin{equation}
S_{R}^{(0)} [ \, J, \tilde{J} \, ] = \frac{1}{2}\llangle\, J, \frac{b_0X_0}{L_0}J\,\rrangle + \llangle\, J, \frac{b_0}{L_0}\tilde{J}\,\rrangle \,.
\end{equation}
The propagators in the small Hilbert space are thus given by
\begin{align}
& \big|\,\,  \overbracket[0.5pt]{\!\! \Psi \,\rrangle \llangle\,  \Psi \!\!}  \,\, \big| = \,   \frac{b_0 X_0}{L_0} \,,
\label{Psi-Psi-propagator} \\
& \big|\,\,  \overbracket[0.5pt]{\!\! \Psi \,\rrangle \llangle \, \tilde{ \Psi} \!\!}\,  \, \big| = \,   \frac{b_0}{L_0} \,.
\label{Psi-tilde-Psi-propagator}
\end{align}
Since $\tilde{\Psi}$ does not appear in the interaction terms,
we never use~\eqref{Psi-tilde-Psi-propagator}
and we will only use~\eqref{Psi-Psi-propagator}.
As the string field $\phi$ for the NS sector is in the large Hilbert space,
again it will be convenient to work in the large Hilbert space
for the Ramond sector as well.
The propagator in the large Hilbert space can be derived as before,
and we find
\begin{equation}
 \overbracket[0.5pt]{\!\!\!\!\! \big|\, \Psi\,\bigr\rangle\bigl\langle\,\Psi\,\big|\!\!\!\!\!}\ \ =\
-\,\frac{b_0 X_0 \eta}{L_0}\,.
\label{Sen-propagator}
\end{equation}

\subsection{Interactions}\label{interaction}

A gauge-invariant action for open superstring field theory was constructed in~\cite{Kunitomo:2015usa}.
The action $S$ for the interacting theory consists of $S_{NS}$ and $S_{R}$:
\begin{equation}
S = S_{NS} +S_{R} \,,
\end{equation}
where $S_{NS}$ does not contain string fields in the Ramond sector.
We expand $S_{NS}$ and $S_{R}$ as follows:
\begin{align}
S_{NS} & = S_{NS}^{(0)}+g S_{NS}^{(1)}+g^2S_{NS}^{(2)} + \mathcal{O}(g^3) \,, \\
S_{R} & = S_{R}^{(0)}+gS_{R}^{(1)}+g^2S_{R}^{(2)}+g^3 S_{R}^{(3)}+\mathcal{O}(g^4) \,,
\end{align}
where $g$ is the open string coupling constant.
For the NS sector, we use $S_{NS}^{(0)}$ in~\eqref{NS-kinetic}.
For the Ramond sector, we use $S_{R}^{(0)}$ in~\eqref{Ramond-kinetic}
or $S_{R}^{(0)}$ in~\eqref{Sen-kinetic}.
The interaction terms to be used in the calculations
of four-point and five-point amplitudes are given by
\begin{align}\label{interactions}
 S_{NS}^{(1)}\ 
=&\ -\frac{1}{6}\, \bigl\langle\,\phi,\, 
Q [ \, \phi,\, \eta \phi \, ] \,\bigr\rangle \,, \\
S_{NS}^{(2)}\ =&\ {}-\frac{1}{24}\,\bigl\langle\,\phi,
Q [ \, \phi,\, [ \, \phi, \, \eta \phi \, ] \, ]
\,\bigr\rangle\,,\\
S_R^{(1)}\ 
=&\ {}-\bigl\langle\,\phi, \, \Psi^2 \, \bigr\rangle\,,\\
S_R^{(2)}\ =&\
-\frac{1}{2}\,\bigl\langle\,\phi,\,
\{ \, \Psi,\, \Xi\, \{ \, \eta \phi,\, \Psi \, \} \, \} \,\bigr\rangle\,,
\label{S_R^(2)} 
\\
S_R^{(3)}\ =&\ 
-\frac{1}{3}\,\bigl\langle\,\phi,\,
\{ \, \Psi,\, \Xi\, \{ \, \eta \phi,\, \Xi \{ \, \eta \phi,\, \Psi \, \} \, \} \, \} \,\bigr\rangle \nonumber\\
&\
-\frac{1}{6}\,\bigl\langle\,\phi,\,
\{ \, \Psi,\, \Xi\, \{ \, [ \, \eta \phi,\, \phi \, ],\, \Psi \, \} \, \} \,\bigr\rangle \nonumber \\
&\ -\frac{1}{3}\,\bigl\langle\,\phi,\,
(\, \Xi\, \{ \eta \phi\,, \Psi \,  \} \, )^2 \,\bigr\rangle\,,
\end{align}
where the products of string fields are defined by the star product~\cite{Witten:1985cc}.
The operator $\Xi$ is Grassmann odd and carries ghost number $-1$ and picture number $1$.
In the context of the approach by Sen~\cite{Sen:2015uaa},
we choose
\begin{equation}
\Xi = \xi_0 \,.
\end{equation}
With this choice, the operator $\Xi$ is BPZ even and satisfies
\begin{equation}
\{ \, \eta, \Xi \, \} = 1 \,.
\end{equation}
In the context of the approach in~\cite{Kunitomo:2015usa},
we choose the following operator constructed in~\cite{Erler:2016ybs}:
\begin{equation}
\Xi\ =\ \xi_0 +  (\Theta(\beta_0)\,\eta\xi_0 - \xi_0)\,\mathcal{P}_{-3/2}
+ (\xi_0\eta\,\Theta(\beta_0) - \xi_0)\,\mathcal{P}_{-1/2}\,,
\label{def-Xi}
\end{equation}
where $\mathcal{P}_n$ projects onto states at picture $n$.
With this choice, it was shown in~\cite{Erler:2016ybs} that
the operator $\Xi$ is BPZ even and satisfies
\begin{equation}
\{ \, \eta, \Xi \, \} = 1 \,.
\end{equation}
In either cases, using the algebraic property $\{ \eta , \Xi \} = 1$, we can relate the BPZ inner product in the small Hilbert space and the large Hilbert space as
\begin{equation}\label{large-small-BPZ}
\llangle\, A, B \,\rrangle = \bigl\langle\, \Xi A, B \,\bigr\rangle \,.
\end{equation}
Let us now consider the operator $X$ defined by
\begin{equation}\label{x-q-xi}
X = \{ \, Q, \Xi \, \} \,.
\end{equation}
In the context of the approach by Sen~\cite{Sen:2015uaa}, we have
\begin{equation}
X = X_0 \,,
\end{equation}
and with this property we can show the gauge invariance of the action.
See~\cite{Sen:2015uaa} for details.
In the context of the approach in~\cite{Kunitomo:2015usa},
it was shown in~\cite{Erler:2016ybs} that
$X$ is identical to $\mathscr{X}$
when it acts on a state $A$ in the small Hilbert space at picture $-3/2$,
\begin{equation}
X A = \mathscr{X} A \,,
\end{equation}
and with this property we can show the gauge invariance of the action.
Note also that
the propagator in the large Hilbert space for $\Psi$
takes the common form
for both cases~\eqref{Ramond-propagator} and~\eqref{Sen-propagator}
using this operator $X$:
\begin{equation}
 \overbracket[0.5pt]{\!\!\!\!\! \big|\,\Psi\,\bigr\rangle\bigl\langle\,\,\Psi \, \big|\!\!\!\!\!}\ \ =\
-\,\frac{b_0 X \eta}{L_0}\,.
\end{equation}

\subsection{Summary}

To summarize, we will use the propagators
\begin{align}
 \overbracket[0.5pt]{\!\!\!\!\! \big|\,\phi\,\bigr\rangle\bigl\langle\,\phi\,\big|\!\!\!\!\!}\ \ \, =&\
\frac{\xi_0 b_0}{L_0}\,, \\
 \overbracket[0.5pt]{\!\!\!\!\! \big|\,\Psi\,\bigr\rangle\bigl\langle\,\Psi\,\big|\!\!\!\!\!}\ \ \, =&\
-\,\frac{b_0 X \eta}{L_0}\,,
\end{align}
and the interactions in~\eqref{interactions}.
The operator $\Xi$ in the interactions is defined either by~\eqref{def-Xi}
or by $\Xi = \xi_0$,
and in either case it is related to the operator $X$ in the propagator
as $X = \{ \, Q, \Xi \, \} \,$.
When we calculate four-point and five-point amplitudes,
we only use the following algebraic relations regarding $\Xi$ and $X$:
\begin{align}
\bigl\langle\, \Xi A, B \, \bigr\rangle = (-1)^A \bigl\langle\,  A, \Xi B  \,\bigr\rangle \,, \qquad
\{ \, Q, \Xi \, \} = X \,, \qquad
\{ \, \eta, \Xi \, \} = 1 \,,  \label{eta-xi} \\
\bigl\langle\,  X A, B \,\bigr\rangle = \bigl\langle\, A, X B \,\bigr\rangle \,, \qquad
[ \, Q, X \, ] = 0 \,, \qquad
[ \, \eta, X \, ] = 0 \,,
\end{align}
where $(-1)^A = 1$ when $A$ is a Grassmann-even state
$(-1)^A = -1$ when $A$ is a Grassmann-odd state.
We can thus interpret our following calculations
either in the context of the approach in~\cite{Kunitomo:2015usa}
or in the context of the approach by Sen~\cite{Sen:2015uaa}. \par

\section{Four-point amplitudes}\label{four-point-amplitudes}
\setcounter{equation}{0}

In this section we calculate on-shell four-point amplitudes
which involve spacetime fermions described by the Ramond sector
in open superstring field theory at the tree level.
We show that amplitudes in the world-sheet theory are correctly reproduced,
and we in particular elucidate the role of quartic interactions
in the context of covering the supermoduli space of super-Riemann surfaces.
Four-point amplitudes of four spacetime bosons were calculated
in a similar way in~\cite{Iimori:2013kha},
and our calculations generalize those in~\cite{Iimori:2013kha} to incorporate the Ramond sector.

\subsection{A brief review on the bosonic string}
\label{bosonic-string-field-theory-review} 

Let us begin with a brief review on the calculation of four-point amplitudes
at the tree level in open bosonic string field theory
following subsections~4.1 and~4.2 of~\cite{Iimori:2013kha}.
The four-point amplitude~$\mathcal{A}^{{\rm WS}}$ in the world-sheet theory is given
by the following correlation function on the upper half-plane (UHP):\footnote{
For the open superstring, we distinguish correlation functions
in the large Hilbert space $\bigl\langle\, \, \ldots \, \,\bigr\rangle_{{\rm UHP}}$
from correlation functions in the small Hilbert space as $\llangle\, \, \ldots \, \,\rrangle_{{\rm UHP}}$.
For the open bosonic string, the superconformal ghost sector is absent
and there is no such distinction, but we will denote correlation functions
as $\llangle\, \, \ldots \, \,\rrangle_{{\rm UHP}}$ because they are more analogous
to those in the small Hilbert space for the open superstring.
}
\begin{equation}
\mathcal{A}^{{\rm WS}}=\int _{-\infty}^{\infty} dt \, \llangle\, \Psi_A(0) \, V_B(t)\,\Psi_C(1)\, \Psi_D(\infty) \,\rrangle_{{\rm UHP}} +(C \leftrightarrow D) \,,
\end{equation}
where $A$, $B$, $C$, and $D$ label four external states.
We use three unintegrated vertex operators $\Psi_A(0)$, $\Psi_C(1)$, and $\Psi_D(\infty)$,\footnote{
A proper treatment of vertex operators inserted at the point $t = \infty$
requires another coordinate patch, and we implicitly assume such a treatment
when we write, for example, $\Psi_D(\infty)$.
}
and the vertex operator $V_B(t)$ is integrated over the modulus $t$
of disks with four punctures on the boundary.
The amplitude~$\mathcal{A}^{{\rm WS}}$ can be decomposed
with respect to the cyclic ordering of external states as follows:
\begin{equation}
\mathcal{A}^{{\rm WS}}=\mathcal{A}^{{\rm WS}}_{ABCD}+\mathcal{A}^{{\rm WS}}_{ ACBD}+\mathcal{A}^{{\rm WS}}_{ABDC}+\mathcal{A}^{{\rm WS}}_{ ACDB}+\mathcal{A}^{{\rm WS}}_{ADBC }+\mathcal{A}^{{\rm WS}}_{ADCB }\,\,.
\end{equation}
The term with the ordering $[A, \, B, \, C, \, D \,]$ , for example, is given by
\begin{equation}
\mathcal{A}^{{\rm WS}}_{ABCD }=\int _{0}^{1} dt \, \llangle\, \Psi_A(0) \, V_B(t)\,\Psi_C(1)\, \Psi_D(\infty)\,\rrangle_{{\rm UHP}}.
\end{equation}
This is the amplitude we want to reproduce in open bosonic string field theory.

The action of open bosonic string field theory is given by
\begin{equation}
S = -\frac{1}{2} \, \llangle\,\Psi , \, Q\Psi \,\rrangle
-\frac{g}{3} \, \llangle\, \Psi ,\, \Psi * \Psi \,\rrangle \,,
\end{equation}
where $\Psi$ is the open bosonic string field of ghost number $1$,
$\llangle\, \, A, B \, \rrangle$ is the BPZ inner product
for a pair of states $A$ and $B$,\footnote{
As in the case of correlation functions,
we will denote the BPZ inner product of states $A$ and $B$
as $\llangle\, A, \, B \,\rrangle$ for the open bosonic string
because it is more analogous
to that in the small Hilbert space for the open superstring.
}
and $g$ is the open string coupling constant.
We choose the Siegel-gauge condition,
\begin{equation}
b_0 \Psi = 0 \,,
\end{equation}
and the propagator is given by
\begin{equation}
\big|\,\,  \overbracket[0.5pt]{ \!\!  \Psi \,\rrangle \llangle\,  \Psi \!\!} \, \, \big| = \, \frac{b_0}{L_0} \,.
\end{equation}
Let us treat $\Psi_A$ and $\Psi_B$ as incoming states
and $\Psi_C$ and $\Psi_D$ as outgoing states.
There are two contributions to the four-point amplitude
with the cyclic ordering $[A, \, B, \, C, \, D \,]$.
One of them is the contribution $\mathcal{A}_{ABCD}^{(s)}$ in the $s$-channel given by
\begin{equation}
\mathcal{A}_{ABCD}^{(s)}=\llangle\, \Psi_A *   \Psi_B ,\, \frac{b_0}{L_0}\, (\,\Psi_C  *  \Psi_D \,) \,\rrangle \,,
\end{equation}
and the other one is the contribution $\mathcal{A}_{ABCD}^{(t)}$ in the $t$-channel: 
\begin{equation}
\mathcal{A}_{ABCD}^{(t)}=\llangle\, \Psi_B *  \Psi_C, \,   \frac{b_0}{L_0}\, ( \, \Psi_D  * \Psi_A \, )\,\rrangle \,.
\end{equation}
It is known that they cover the following regions of the moduli integral \cite{Giddings:1986iy}:
\begin{equation}\label{transformation}
\begin{split}
\mathcal{A}_{ABCD}^{(s)}=\llangle\, \Psi_A * \Psi_B , \,  \frac{b_0}{L_0}\,(\,\Psi_C * \Psi_D \,) \,\rrangle
& =\int_{0}^{1/2}dt \, \llangle\, \Psi_A(0) \, V_B(t)\, \Psi_C(1)\, \Psi_D(\infty) \,\rrangle_{{\rm UHP}} \,, \\
\mathcal{A}_{ABCD}^{(t)}=\llangle\, \Psi_B * \Psi_C , \, \frac{b_0}{L_0}\, ( \, \Psi_D * \Psi_A \, )\,\rrangle
& =\int_{1/2}^{1}dt \, \llangle\, \Psi_A(0) \, V_B(t)\, \Psi_C(1)\, \Psi_D(\infty) \,\rrangle_{{\rm UHP}} \,.
\end{split}
\end{equation}
Therefore, the sum of the contributions in the $s$-channel and in the $t$-channel
precisely covers the whole region of the moduli integral,
and the amplitude $\mathcal{A}^{{\rm WS}}_{ABCD }$ is reproduced correctly:
\begin{equation}
\mathcal{A}^{{\rm WS}}_{ABCD} = \mathcal{A}_{ABCD}^{(s)} +\mathcal{A}_{ABCD}^{(t)} \,.
\end{equation}
Note in particular that we do not need quartic interactions in open bosonic string field theory.

\subsection{The world-sheet theory in the superstring}

Let us move on to the world-sheet theory of the open superstring in the Ramond-Neveu-Schwarz (RNS) formalism.
We consider on-shell disk amplitudes
involving the Ramond sector.
As in the open bosonic string,
we can use one integrated vertex operator and three unintegrated vertex operators for four-point amplitudes.
The number of integrated vertex operators in general corresponds
to the number of even moduli of disks with punctures on the boundary.
The number of even moduli of disks with $n_B$ NS punctures and $n_F$ Ramond punctures is given by
\begin{equation}
\#({\rm ~even~moduli~})=n_B + n_F -3.
\end{equation}
In the open superstring, the supermoduli space of super-Riemann surfaces contain fermionic directions.
The number of odd moduli of disks is
\begin{equation}
\#({\rm ~odd~moduli~})=n_B + \frac{n_F}{2} -2. 
\end{equation}
The integration of odd moduli
with associated ghost insertions
yields
insertions of picture-changing operators.\footnote{
For the supermoduli space of super-Riemann surfaces at higher genus,
we cannot integrate over odd moduli independently of the integration over even moduli in general,
and this simple prescription using picture-changing operators does not work \cite{Donagi:2013dua}.
One possible prescription is to perform the integration over odd moduli in local patches
and combine them using ``vertical integration'' \cite{Sen:2015hia}.
As we will see, the calculation of on-shell scattering amplitudes
in open superstring field theory based on the large Hilbert space
is closely related to this prescription. 
}
Vertex operators in the $-1$ picture for the NS sector
and vertex operators in the $-1/2$ picture for the Ramond sector
correspond to unintegrated vertex operators with respect to odd moduli.
We use these vertex operators and insert one picture-changing operator for each odd modulus.
The total picture number for disk amplitudes is then $-2$,
which coincides with the picture number required by the anomaly of the picture number current.

\subsubsection*{Four-fermion amplitude}
Let us first consider disk amplitudes with four external fermions. 
In this case there are no odd moduli so that we do not need to insert picture-changing operators.
The on-shell disk amplitude of external states
with the cyclic ordering $[A,B,C,D]$ is written as
\begin{equation}
\mathcal{A}_{FFFF}^{{\rm WS}}=\int_{0}^{1} dt \, \llangle\, \Psi_A(0)\,V_{B}^{(-1/2)}(t) \, \Psi_C(1)\, \Psi_D (\infty)  \,\rrangle_{{\rm UHP}} \,,
\end{equation}
where $V^{(-1/2)}(t)$ is the integrated Ramond vertex operator in the $-1/2$ picture. 
We will compare this with the amplitude calculated from open superstring field theory.
It is convenient to transform the amplitude to the form
\begin{equation} \label{WS-FFFF}
\mathcal{A}_{FFFF}^{{\rm WS}}=
\llangle\,\Psi_A*\Psi_B\, , \frac{b_0}{L_0}\,( \Psi_C * \Psi_D ) \,\rrangle +
\llangle\,\Psi_B*\Psi_C\,, \frac{b_0}{L_0}\,(\Psi_D * \Psi_A )\,\rrangle\,,
\end{equation}
where we used~\eqref{transformation}, which depends only on properties
with respect to bosonic conformal transformations.

\subsubsection*{Fermion-boson-fermion-boson amplitude}

Next consider on-shell disk amplitudes with two external bosons and two external fermions.
We need one picture-changing operator in this case.
There are two inequivalent cyclic orderings,
and the amplitude~$\mathcal{A}_{FBFB}^{{\rm WS}}$ for the fermion-boson-fermion-boson ordering is written as
\begin{equation}
\mathcal{A}_{FBFB}^{{\rm WS}}
= \int_{0}^{1} dt \, \llangle\, \Psi_A(0)\, X_0 \cdot V_{B}^{(-1)}(t) \, \Psi_C(1) \, \Phi_D(\infty) \,\rrangle_{ \rm UHP} \,,
\end{equation}
where $V_B^{(-1)}(t) $ is the integrated NS vertex operator in the $-1$ picture 
and we used the zero mode $X_0$ of the picture-changing operator.
Note that the action of $X_0$ on an on-shell vertex operator $\Phi (0)$
is the same as the colliding limit of the local picture-changing operator:
\begin{equation}
X_0 \cdot \Phi (0)
= \oint_C \frac{dz}{2\pi i } \frac{X(z)}{z} \Phi(0)
= \lim_{\epsilon \to 0} X(\epsilon) \Phi(0) \,,
\end{equation}
where the contour $C$ encircles the origin counterclockwise.
As in the four-fermion amplitude, we transform this amplitude using~\eqref{transformation} as follows:
\begin{equation} \label{WS-FBFB}
\mathcal{A}_{FBFB}^{{\rm WS}}= 
\llangle\, \Psi_A * X_0 \Phi_B \, , \, \frac{b_0}{L_0}\, ( \Psi_C * \Phi_D) \,\rrangle +
\llangle\, X_0\Phi_B * \Psi_C \, , \, \frac{b_0}{L_0}\, (\Phi_D * \Psi_A) \,\rrangle\,.
\end{equation}
Note that the operator $X_0$ acts on the same state $\Phi_B$ in both channels.

\subsubsection*{Fermion-fermion-boson-boson amplitude}
For the amplitude $\mathcal{A}_{FFBB}^{{\rm WS}}$
with the fermion-fermion-boson-boson ordering,
we choose $X_0$ to act on the state $\Phi_C$ in the NS sector.
We then have
\begin{equation}
\mathcal{A}_{FFBB}^{{\rm WS}} \label{WS-FFBB}
= \llangle\, \Psi_A * \Psi_B \, , \, \frac{b_0}{L_0} \, (X_0 \Phi_C * \Phi_D) \,\rrangle
+ \llangle\, \Psi_B * X_0 \Phi_C \, , \, \frac{b_0}{L_0} \, ( \Phi_D * \Psi_A) \,\rrangle \,.
\end{equation}
\par

\subsubsection*{Summary and remarks on the notation}

Let us summarize the four-point amplitudes presented in this section.
To simplify notations, we omit the star symbol and the comma for the BPZ inner product in what follows
and we use parentheses to avoid ambiguities:
\begin{align}
\mathcal{A}_{FFFF}^{{\rm WS}}=&
\llangle\,\Psi_A \, \Psi_B\, \frac{b_0}{L_0}\,( \Psi_C \,  \Psi_D ) \,\rrangle +
\llangle\,\Psi_B \, \Psi_C\, \frac{b_0}{L_0}\,(\Psi_D \,  \Psi_A )\,\rrangle ,
\label{A_FFFF^WS} \\ 
\mathcal{A}_{FBFB}^{{\rm WS}}=&
\llangle\, \Psi_A \,  X_0 \Phi_B  \, \frac{b_0}{L_0}\, ( \Psi_C \, \Phi_D) \,\rrangle +
\llangle\, X_0\Phi_B \, \Psi_C  \, \frac{b_0}{L_0}\, (\Phi_D \,   \Psi_A) \,\rrangle , \\
\mathcal{A}_{FFBB}^{{\rm WS}}=&
 \llangle\, \Psi_A \,  \Psi_B  \, \frac{b_0}{L_0} \, (X_0 \Phi_C \,  \Phi_D) \,\rrangle
+ \llangle\, \Psi_B \,  X_0 \Phi_C  \, \frac{b_0}{L_0} \, ( \Phi_D \,  \Psi_A) \,\rrangle \, .
\label{A_FFBB^WS} 
\end{align}

\subsection{String field theory in the superstring }

In this subsection we calculate four-point amplitudes including external fermions in open superstring field theory. 
In open bosonic string field theory, as we reviewed in subsection~\ref{bosonic-string-field-theory-review},
the four-point amplitudes in the world-sheet theory are reproduced
by Feynman diagrams with two cubic vertices and one propagator.
The moduli space of disks with four punctures on the boundary
is precisely covered by the sum of the contributions in the $s$-channel and in the $t$-channel
of these diagrams.
In contrast, four-point amplitudes of external bosons in the Berkovits formulation
of open superstring field theory were calculated in~\cite{Berkovits:1999bs, Iimori:2013kha},
and the contribution from Feynman diagrams with a quartic vertex
was necessary to reproduce the amplitudes in the world-sheet theory.
In~\cite{Iimori:2013kha} the role of the quartic interaction
in the covering of the supermoduli space of super-Riemann surfaces was elucidated,
and it was found that the contributions from Feynman diagrams with a quartic vertex
adjusts the different assignment of picture-changing operators on the external states
between the $s$-channel and the $t$-channel
of Feynman diagrams with two cubic vertices and one propagator.
Our analysis in this subsection generalizes that of~\cite{Iimori:2013kha}
to incorporate fermions.

\subsubsection{Four-fermion amplitudes}

\begin{figure}[htb]
  \begin{center}
    \begin{tabular}{c}
    
      \begin{minipage}{0.33\hsize}
        \begin{center}
          \includegraphics[clip, width=3.5cm]{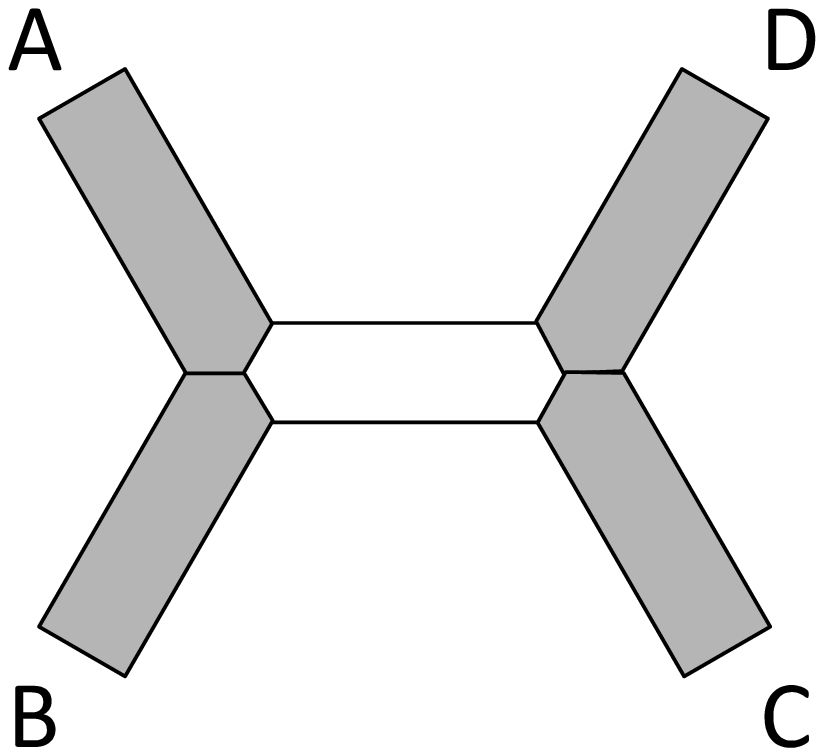}
          \\ $s$-channel
        \end{center}
      \end{minipage}
      \begin{minipage}{0.33\hsize}
        \begin{center}
          \includegraphics[clip, width=3.2cm]{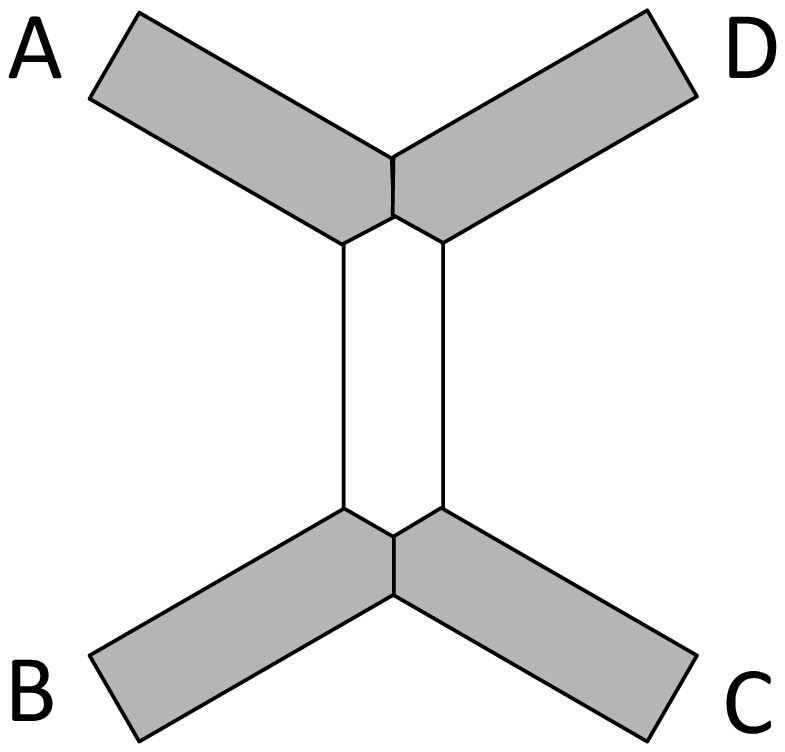}
          \\ $t$-channel
        \end{center}
      \end{minipage}
      
    \end{tabular}
    \caption{Feynman diagrams for $\mathcal{A}_{FFFF}$.}
    \label{FFFF}
  \end{center}
\end{figure}

Let us first calculate scattering amplitudes with four external fermions denoted by $\mathcal{A}_{FFFF}$.
Since there are no four-fermion vertices in \eqref{S_R^(2)},
we only need to calculate Feynman diagrams with two cubic vertices and one propagator.
The $s$-channel diagram and the $t$-channel diagram are depicted in figure~\ref{FFFF}.
In our Feynman diagrams, fermion legs and propagators are represented in gray,
and boson legs and propagators in white.

The $s$-channel contribution~$\mathcal{A}_{FFFF}^{(s)}$
consists of two cubic vertices from~$S_{R}^{(1)}$
and one NS propagator:
\begin{align}
 \mathcal{A}_{FFFF}^{(s)}\ =&\ \big(-\bigl\langle\,\Psi_A\, \Psi_B\, \ 
 \overbracket[0.5pt]{\!\! \phi \,\bigr\rangle\big)\big(-\bigl\langle\,\phi\!\!}\  \,
\Psi_C\,\Psi_D\,\bigr\rangle\big)
\nonumber\\
=&\ \bigl\langle\,\Psi_A\,\Psi_B\,\frac{\xi_0 b_0}{L_0}\,(\Psi_C\,\Psi_D)\,\bigr\rangle
\nonumber\\
=&\ \llangle\,\Psi_A\,\Psi_B\,\frac{b_0}{L_0}\,(\Psi_C\,\Psi_D)\,\rrangle\,  . 
\end{align}
In the last step, we translated the BPZ inner product in the large Hilbert space
into the one in the small Hilbert space by the relation~\eqref{small-BPZ-definition}.
Similarly, the $t$-channel contribution~$\mathcal{A}_{FFFF}^{(t)}$ is calculated as
\begin{align}
 \mathcal{A}_{FFFF}^{(t)}\ 
=&\ \bigl\langle\,\Psi_B\,\Psi_C\,\frac{\xi_0 b_0}{L_0}\,(\Psi_D\,\Psi_A)\,\bigr\rangle
= \llangle\,\Psi_B\,\Psi_C\,\frac{b_0}{L_0}\,(\Psi_D\,\Psi_A)\,\rrangle\,.
\end{align}
The four-fermion amplitude~$\mathcal{A}_{FFFF}$ is therefore given by 
\begin{align}
 \mathcal{A}_{FFFF}\ =&\ 
\mathcal{A}_{FFFF}^{(s)} + \mathcal{A}_{FFFF}^{(t)}
\nonumber\\
=&\
\llangle\,\Psi_A\,\Psi_B\,\frac{b_0}{L_0}\,(\Psi_C\,\Psi_D)\,\rrangle +
\llangle\,\Psi_B\,\Psi_C\,\frac{b_0}{L_0}\,(\Psi_D\,\Psi_A)\,\rrangle\, .
\end{align}
This coincides with the amplitude~\eqref{A_FFFF^WS} in the world-sheet theory.

\subsubsection{Fermion-fermion-boson-boson amplitudes}

\begin{figure}[htb]
  \begin{center}
    \begin{tabular}{c}
    
      \begin{minipage}{0.33\hsize}
        \begin{center}
          \includegraphics[clip, width=3.5cm]{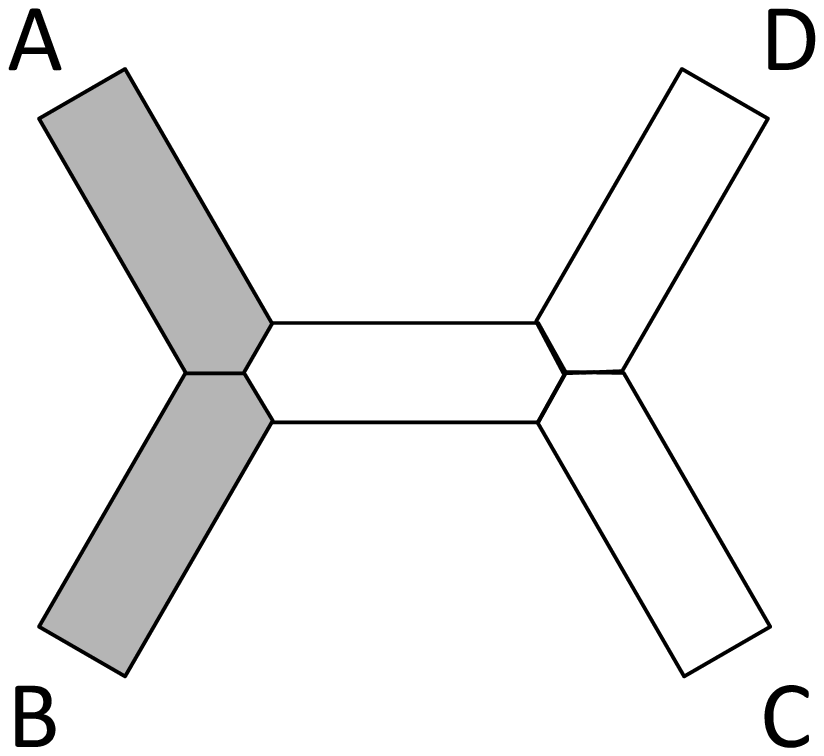}
          \\$s$-channel
        \end{center}
      \end{minipage}
      \begin{minipage}{0.33\hsize}
        \begin{center}
          \includegraphics[clip, width=3.2cm]{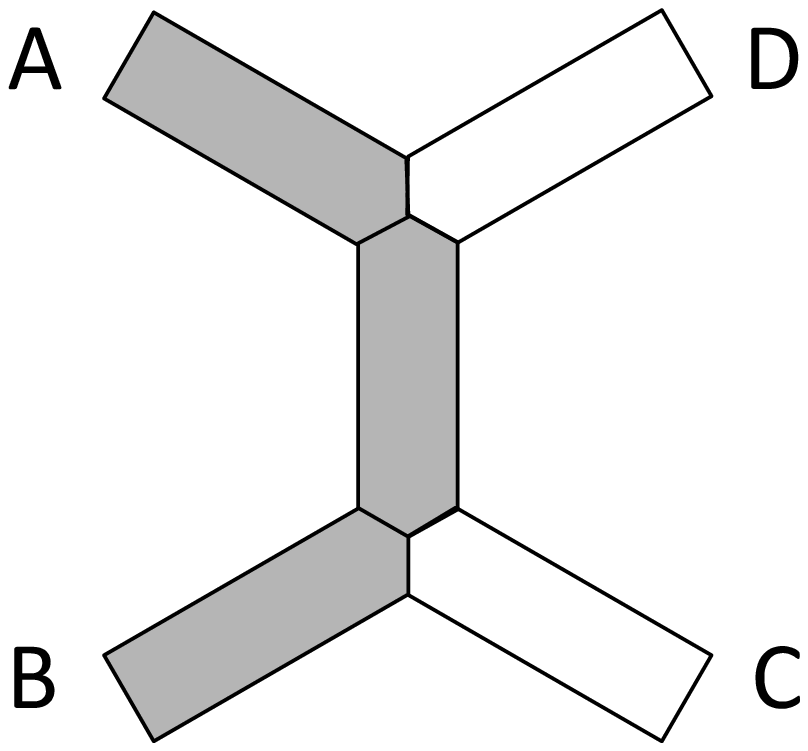}
           \\$t$-channel
        \end{center}
      \end{minipage}
      
    \end{tabular}
    \caption{Feynman diagrams with two cubic vertices and one propagator for $\mathcal{A}_{FFBB}$.}
    \label{FFBB}
  \end{center}
\end{figure}

Next consider the four-point amplitude of two bosons and two fermions
with the fermion-fermion-boson-boson ordering denoted by $\mathcal{A}_{FFBB}$.
Feynman diagrams with two cubic vertices and one propagator for $\mathcal{A}_{FFBB}$
are depicted in figure~\ref{FFBB}.

The diagram in the $s$-channel consists of one cubic vertex from $S_{NS}^{(1)}$,
one cubic vertex from $S_R^{(1)}$, and one NS propagator.
Note that the cubic interaction $S_{NS}^{(1)}$ has the following property:
\begin{equation}
 \delta\bigl\langle\,\phi\,(Q\phi\,\eta\phi + \eta\phi\,Q\phi)\,\bigr\rangle\ =\ 3\, \bigl\langle\,\delta\phi\,(Q\phi\,\eta\phi + \eta\phi\,Q\phi)\,\bigr\rangle\,.
 \end{equation} 
The $s$-channel contribution~$\mathcal{A}_{FFBB}^{(s)}$ is thus calculated as
\begin{align}
\mathcal{A}_{FFBB}^{(s)}\ =&\ 
\frac{1}{2}\, \big(-\bigl\langle\,\Psi_A\,\Psi_B\,
\overbracket[0.5pt]{\! \phi \,\bigr\rangle\big) \big(-\bigl\langle\, \phi \!\!}\,\,\,
(Q\phi_C\,\eta\phi_D + \eta\phi_C\,Q\phi_D)\,\bigr\rangle\big)
\nonumber\\
=&\
\frac{1}{2}\,\bigl\langle\,\Psi_A\,\Psi_B\,\frac{\xi_0 b_0}{L_0}\,
\left(Q\phi_C\,\eta\phi_D + \eta\phi_C\,Q\phi_D\right)\,\bigr\rangle.
\end{align}
Following the relation~\eqref{Phi-to-phi} between the open superstring field in the large Hilbert space
and the on-shell vertex operator in the small Hilbert space,
we replace $\eta \phi_C$ with $\Phi_C$ and $\eta \phi_D$ with $\Phi_D$.
For $Q \phi_C$ we express it in terms of $\Phi_C$ as follows:
\begin{equation}
Q \phi_C = Q \xi_0 \Phi_C = \{\, Q , \xi_0 \,\} \Phi_C = X_0 \Phi_C \,,
\end{equation}
where we used the on-shell condition $Q \Phi_C = 0$.
This is the origin of the zero mode $X_0$ of the picture-changing operator in the Berkovits formulation
based on the large Hilbert space.
Similarly, we have
\begin{equation}
Q \phi_D = X_0 \Phi_D.
\end{equation}
Since all the string fields are in the small Hilbert space,
we can represent $\mathcal{A}_{FFBB}^{(s)}$
in terms of the BPZ inner product in the small Hilbert space
using the relation~\eqref{small-BPZ-definition} as follows:
\begin{align}
\mathcal{A}_{FFBB}^{(s)}
=\frac{1}{2} \llangle\, \Psi_A \, \Psi_B \,\frac{b_0}{L_0}\, (X_0 \Phi_C \, \Phi_D ) \,\rrangle
+ \frac{1}{2} \llangle\, \Psi_A \, \Psi_B \, \frac{b_0 }{L_0 } \, ( \Phi_C \, X_0 \Phi_D) \,\rrangle.
\end{align}

The diagram in the $t$-channel consists of two cubic vertices from $S_R^{(1)}$
and one Ramond propagator.
The $t$-channel contribution~$\mathcal{A}_{FFBB}^{(t)}$ is calculated as
\begin{align} \label{FFBB-t}
\mathcal{A}_{FFBB}^{(t)}\ 
= \bigl\langle\,\Psi_B\,\phi_C  \ \,
 \overbracket[0.5pt]{\!\! \Psi \,\bigr\rangle\bigl\langle\,\Psi\!\!}\ \,
\phi_D\,\Psi_A\,\bigr\rangle 
={}-\bigl\langle\,\Psi_B\,\phi_C\,\frac{b_0 X \eta}{L_0}\,(\phi_D\,\Psi_A)\,\bigr\rangle.
\end{align}
Using the derivation property of $\eta$ and the condition that $\Psi_A$ is in the small Hilbert space
$\eta \Psi_A =0$, we obtain
\begin{equation}
\mathcal{A}_{FFBB}^{(t)}= {}-\bigl\langle\, \Psi_B \, \phi_C \, \frac{b_0 X}{L_0} \, (\eta \phi_D \, \Psi_A) \,\bigr\rangle. 
\end{equation}
We replace $\phi_C$ with $\xi_0 \Phi_C$ and $\eta \phi_D$ with $\Phi_D$,
and then we express $\mathcal{A}_{FFBB}^{(t)}$ in terms of the BPZ inner product
in the small Hilbert space as follows:
\begin{align}
\mathcal{A}_{FFBB}^{(t)}=
\llangle\, \Psi_B \,\Phi_C \, \frac{b_0 X}{L_0}\,( \Phi_D\,\Psi_A) \,\rrangle.
\end{align}

The total contribution from Feynman diagrams with two cubic vertices and one propagator
is thus given by
\begin{align}
&\mathcal{A}_{FFBB}^{(s)} +\mathcal{A}_{FFBB}^{(t)} \nonumber \\
&=  \frac{1}{2} \llangle\, \Psi_A \, \Psi_B \,\frac{b_0}{L_0}\, (X_0 \Phi_C \, \Phi_D ) \,\rrangle
+ \frac{1}{2} \llangle\, \Psi_A \, \Psi_B \, \frac{b_0 }{L_0 } \, ( \Phi_C \, X_0 \Phi_D) \,\rrangle
+\llangle\, \Psi_B \,\Phi_C \, \frac{b_0 X}{L_0}\,( \Phi_D\,\Psi_A) \,\rrangle. 
\label{A_FFBB^(s)+A_FFBB^(t)} 
\end{align} 
We find that the assignment of picture-changing operators is different in the $s$-channel and in the $t$-channel
as opposed to the amplitude~\eqref{A_FFBB^WS} in the world-sheet theory,
where the operator $X_0$ acts on the same state $\Phi_C$ in both channels.
The situation is analogous to the contribution to four-point amplitudes
in the NS sector from Feynman diagrams with two cubic vertices and one propagator~\cite{Iimori:2013kha}.
A new feature in our case is the appearance of the operator $X$ from the Ramond propagator
in addition to $X_0$ acting on an external state.

Following the method in~\cite{Iimori:2013kha}, let us try to move the picture-changing operators
so that they act on the same state $\Phi_C$ in both channels.
For the second term on the right-hand side of~\eqref{A_FFBB^(s)+A_FFBB^(t)},
we want to move $X_0$ from $\Phi_D$ to $\Phi_C$.
In this case we uplift the expression in the small Hilbert space to the large Hilbert space
by replacing the target state $\Phi_C$ with $\xi_0 \Phi_C$ with a possible sign when necessary,
and write $X_0$ as $\{ Q, \xi_0 \}$:
\begin{equation}
\llangle\, \Psi_A \, \Psi_B \, \frac{b_0 }{L_0 } \, ( \Phi_C \, X_0 \Phi_D) \,\rrangle
= {}-\langle\, \Psi_A \, \Psi_B \, \frac{b_0 }{L_0 } \, ( \, \xi_0 \Phi_C \, \{ Q, \xi_0 \} \Phi_D)  \,\rangle \,.
\end{equation}
We then move $Q$ to obtain
\begin{equation}
\begin{split}
& \llangle\, \Psi_A \, \Psi_B \, \frac{b_0 }{L_0 } \, ( \Phi_C \, X_0 \Phi_D) \,\rrangle
= {}-\langle\, \Psi_A \, \Psi_B \, \frac{b_0 }{L_0 } \, ( \, \xi_0 \Phi_C \, \{ Q, \xi_0 \} \Phi_D ) \,\rangle \\
& = \langle\, \Psi_A \, \Psi_B \, \frac{b_0 }{L_0 } \, ( \, X_0 \Phi_C \, \xi_0 \Phi_D \, ) \,\rangle
{}-\langle\, \Psi_A \, \Psi_B  \, \xi_0 \Phi_C \, \xi_0 \Phi_D  \,\rangle \\
& = \llangle\, \Psi_A \, \Psi_B \, \frac{b_0 }{L_0 } \, ( \, X_0 \Phi_C \, \Phi_D \, ) \,\rrangle
{}-\langle\, \Psi_A \, \Psi_B \, \xi_0 \Phi_C \, \xi_0 \Phi_D  \,\rangle \,.
\end{split}
\label{D-to-C}
\end{equation}
We find that $X_0$ has moved from $\Phi_D$ to $\Phi_C$,
but an extra term $\langle\, \Psi_A \, \Psi_B \, \xi_0 \Phi_C \, \xi_0 \Phi_D \,\rangle$
without the propagator~$b_0 / L_0$ has appeared.

For the third term on the right-hand side of~\eqref{A_FFBB^(s)+A_FFBB^(t)},
we want to transmute $X$ to $X_0$ acting on $\Phi_C$.
As before, we uplift the expression in the small Hilbert space to the large Hilbert space
by replacing the target state $\Phi_C$ with $\xi_0 \Phi_C$ with a possible sign when necessary,
and this time we write $X$ as~$\{ Q, \Xi \}$:
\begin{equation}
\llangle\, \Psi_B \,\Phi_C \, \frac{b_0 X}{L_0}\,( \Phi_D\,\Psi_A) \,\rrangle
= {}-\langle\, \Psi_B \, \xi_0 \Phi_C  \, \frac{b_0 \{ Q, \Xi \}}{L_0}\,( \Phi_D\,\Psi_A) \, \rangle \,. 
\end{equation}
We then move $Q$ to obtain
\begin{equation}
\begin{split}
& \llangle\, \Psi_B \,\Phi_C \, \frac{b_0 X}{L_0}\,( \Phi_D\,\Psi_A) \,\rrangle
= {}-\langle\, \Psi_B \, \xi_0 \Phi_C \, \frac{b_0 \{ Q, \Xi \}}{L_0}\,( \Phi_D\,\Psi_A) \, \rangle \\
& = {}-\langle\, \Psi_B \, X_0 \Phi_C \, \frac{b_0 \Xi}{L_0}\,( \Phi_D\,\Psi_A) \, \rangle
{}-\langle\, \Psi_B \, \xi_0 \Phi_C \, \Xi \, ( \Phi_D\,\Psi_A) \, \rangle \\
& = \llangle\, \Psi_B \, X_0 \Phi_C \, \frac{b_0 \Xi}{L_0}\,( \Phi_D\,\Psi_A) \, \rrangle
{}-\langle\, \Psi_B \, \xi_0 \Phi_C \, \Xi \, ( \Phi_D\,\Psi_A) \, \rangle \,.
\end{split}
\label{X-to-X_0}
\end{equation}
We find that $X$ has transmuted to $X_0$ acting on $\Phi_C$,
but an extra term $\langle\, \Psi_B \, \xi_0 \Phi_C \, \Xi \, ( \Phi_D\,\Psi_A) \, \rangle$
without the propagator~$b_0 / L_0$ has appeared.
Recall that $\Xi$ can be either the operator written in~\eqref{def-Xi}
in the context of the approach in~\cite{Kunitomo:2015usa}
or the zero mode $\xi_0$ of $\xi (z)$ in the context of the approach by Sen~\cite{Sen:2015uaa},
as explained in subsection~\ref{interaction}.

Now the total contribution $\mathcal{A}_{FFBB}^{(s)}+\mathcal{A}_{FFBB}^{(t)}$ from Feynman diagrams
with two cubic vertices and one propagator is expressed as
\begin{align}
\mathcal{A}_{FFBB}^{(s)}+\mathcal{A}_{FFBB}^{(t)}
& = \llangle\, \Psi_A \, \Psi_B \, \frac{b_0}{L_0} \,(X_0\Phi_C\,\Phi_D)\,\rrangle 
+\llangle\, \Psi_B\,X_0\Phi_C\, \frac{b_0}{L_0} \,(\Phi_C\,\Psi_A)\,\rrangle \nonumber \\
& \quad~ {}-\frac{1}{2}\,\bigl\langle\,\Psi_A\,\Psi_B\, \xi_0 \Phi_C \, \xi_0 \Phi_D  \,\bigr\rangle\
-\bigl\langle\,\Psi_B\, \xi_0 \Phi_C \,\Xi\, ( \Phi_D\,\Psi_A)\,\bigr\rangle\ .
\end{align}
Compared with the amplitude~$\mathcal{A}_{FFBB}^{{\rm WS}}$ in the world-sheet theory,
the difference $\Delta \mathcal{A}_{FFBB}$ is given by
\begin{equation}
\begin{split}
\Delta \mathcal{A}_{FFBB}
& \equiv \mathcal{A}_{FFBB}^{(s)}+\mathcal{A}_{FFBB}^{(t)}-\mathcal{A}_{FFBB}^{{\rm WS}} \\
& = {}-\frac{1}{2}\,\bigl\langle\,\Psi_A\,\Psi_B\, \xi_0 \Phi_C \, \xi_0 \Phi_D \,\bigr\rangle\
-\bigl\langle\,\Psi_B\, \xi_0 \Phi_C \,\Xi\, ( \Phi_D\,\Psi_A)\,\bigr\rangle\ .
\end{split}
\end{equation}
This consists of terms without the propagator $b_0 / L_0$.
In other words, the difference $\Delta \mathcal{A}_{FFBB}$ is localized
at the boundary of the moduli space
between the region covered by the $s$-channel contribution
and the region covered by the $t$-channel contribution.

\begin{figure}[t]
  \begin{center}
              \includegraphics[clip, width=2.5cm]{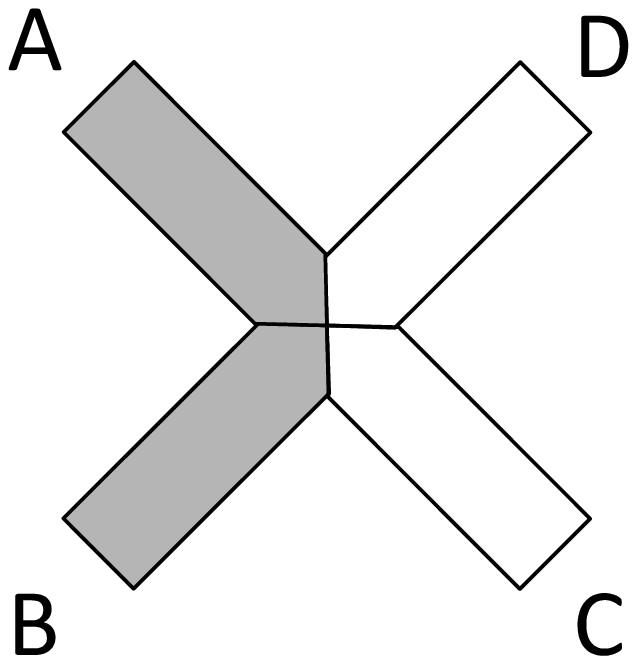}
    \caption{Feynman diagram with a quartic vertex for $\mathcal{A}_{FFBB}$.}
    \label{FFBB4}
  \end{center}
\end{figure}

So far we have only considered Feynman diagrams with two cubic vertices and one propagator.
There is an additional contribution to~$\mathcal{A}_{FFBB}$ from the Feynman diagram
with a quartic vertex depicted in figure~\ref{FFBB4}.
The contribution from this diagram is denoted by~$\mathcal{A}_{FFBB}^{(4)}$
and is calculated as
\begin{align}
\mathcal{A}_{FFBB}^{(4)}\ =&\
-\frac{1}{2}\,\bigl\langle\,\phi_D\,\Psi_A\,\Xi\,(\Psi_B\,\eta\phi_C)\,\bigr\rangle
+\frac{1}{2}\,\bigl\langle\,\Psi_B\,\phi_C\,\Xi\,(\eta\phi_D\,\Psi_A)\,\bigr\rangle.
\end{align}
Let us compare this with the difference $\Delta \mathcal{A}_{FFBB}$.
We replace $\phi_C$ with $\xi_0 \Phi_C$, $\phi_D$ with $\xi_0 \Phi_D$,
and $\eta \phi_D$ with $\Phi_D$ to obtain
\begin{equation}
\mathcal{A}_{FFBB}^{(4)}
= {}-\frac{1}{2}\,\bigl\langle\, \xi_0 \Phi_D \,\Psi_A\, \Xi \, (\Psi_B\,\eta \xi_0 \Phi_C)\,\bigr\rangle
+\frac{1}{2}\,\bigl\langle\,\Psi_B\, \xi_0 \Phi_C \,\Xi\,( \Phi_D\,\Psi_A)\,\bigr\rangle \,.
\end{equation}
We move $\eta$ using $ \{ \eta , \xi_0 \} =1$ and $ \{ \eta , \Xi \} =1$ to find
\begin{equation}
\mathcal{A}_{FFBB}^{(4)}\
= \frac{1}{2}\,\bigl\langle\,\Psi_A\,\Psi_B\, \xi_0 \Phi_C \, \xi_0 \Phi_D \,\bigr\rangle
+\bigl\langle\,\Psi_B\, \xi_0 \Phi_C \,\Xi\, ( \Phi_D\,\Psi_A)\,\bigr\rangle \,.
\label{FFBB-np}
\end{equation}
This precisely cancels the difference~$\Delta \mathcal{A}_{FFBB}$,
so the amplitude~$\mathcal{A}_{FFBB}$ in open superstring field theory,
\begin{equation}
\mathcal{A}_{FFBB} = \mathcal{A}_{FFBB}^{(s)} + \mathcal{A}_{FFBB}^{(t)} + \mathcal{A}_{FFBB}^{(4)} \,,
\end{equation}
coincides with the amplitude~$\mathcal{A}_{FFBB}^{{\rm WS}}$ in the world-sheet theory:
\begin{equation}
\mathcal{A}_{FFBB} = \mathcal{A}_{FFBB}^{{\rm WS}} \,.
\end{equation}

\begin{figure}[H]
  \begin{center}
    \begin{tabular}{c}    
      \begin{minipage}{0.40\hsize}
        \begin{center}
          \includegraphics[clip, width=6cm]{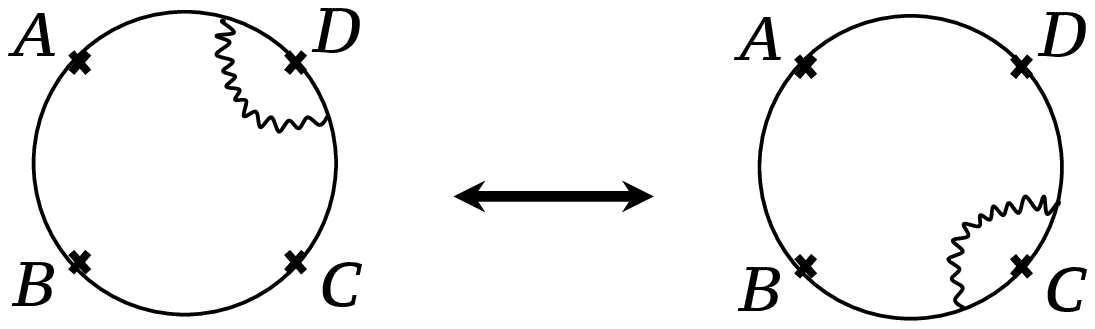}
          \caption{The first term on the right-hand side of~(\ref{FFBB-np}) contains $\xi_0$ acting on $\Phi_D$
          and $\xi_0$ acting on $\Phi_C$. This term exchanges $X_0$ acting on $\Phi_D$
          and $X_0$ acting on $\Phi_C$ as seen in~\eqref{D-to-C}.} 
          \label{FFBB-xi-1}
        \end{center}
      \end{minipage}
      \hspace{0.1\hsize}
      \begin{minipage}{0.40\hsize}
        \begin{center}
          \includegraphics[clip, width=6cm]{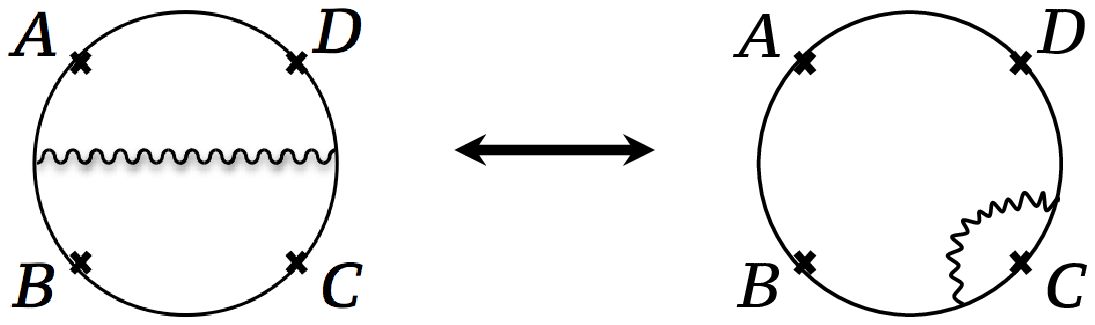}
          \caption{The second term on the right-hand side of~(\ref{FFBB-np}) contains $\Xi$
          acting on an intermediate state in the Ramond sector
          and $\xi_0$ acting on $\Phi_C$.
          This term exchanges $X$ in the Ramond propagator and $X_0$ acting on $\Phi_C$ 
          as seen in~\eqref{X-to-X_0}.} 
          \label{FFBB-xi-2}
        \end{center}
      \end{minipage}    
    \end{tabular}
  \end{center}
\end{figure}

We have seen that the contribution~$\mathcal{A}_{FFBB}^{(4)}$ adjusted
the different assignment of picture-changing operators in the $s$-channel and in the $t$-channel.
The role of each term of $\mathcal{A}_{FFBB}^{(4)}$ in~\eqref{FFBB-np}
is illustrated in figures~\ref{FFBB-xi-1} and~\ref{FFBB-xi-2}.
Compared with the analysis of four-point amplitudes of the NS sector in~\cite{Iimori:2013kha},
the new aspect we found for the analysis including the Ramond sector
is that the quartic interaction adjusts
not only insertions of $X_0$ on external states
but also insertions of $X$ in the Ramond propagator.
We will see more examples of this in the next case.

\subsubsection{Fermion-boson-fermion-boson amplitudes}
\begin{figure}[tb]
  \begin{center}
    \begin{tabular}{c}
    
      \begin{minipage}{0.33\hsize}
        \begin{center}
          \includegraphics[clip, width=3.5cm]{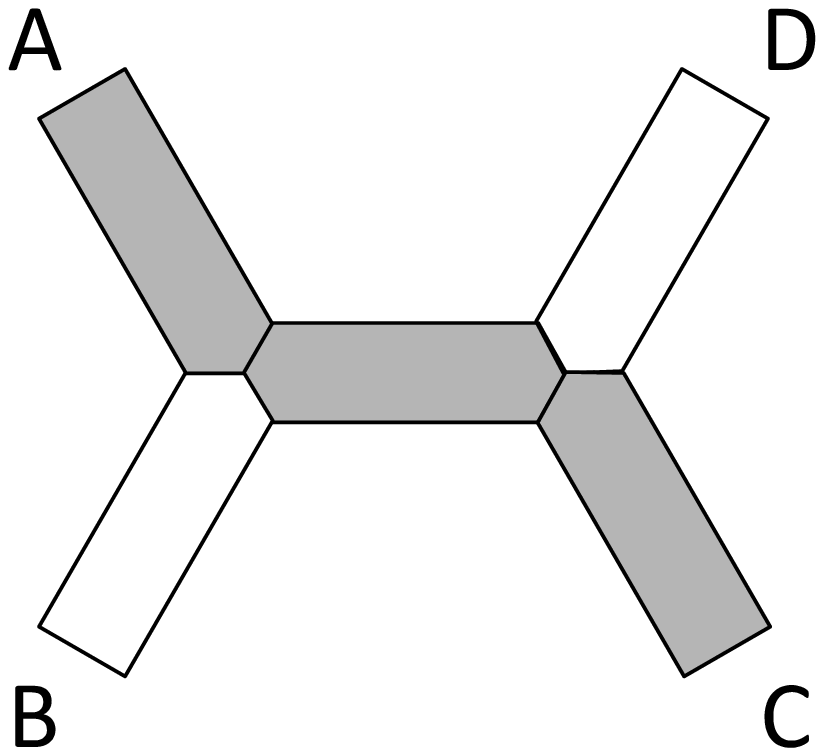}
          \\ $s$-channel        
        \end{center}
      \end{minipage}
      \begin{minipage}{0.33\hsize}
        \begin{center}
          \includegraphics[clip, width=3.2cm]{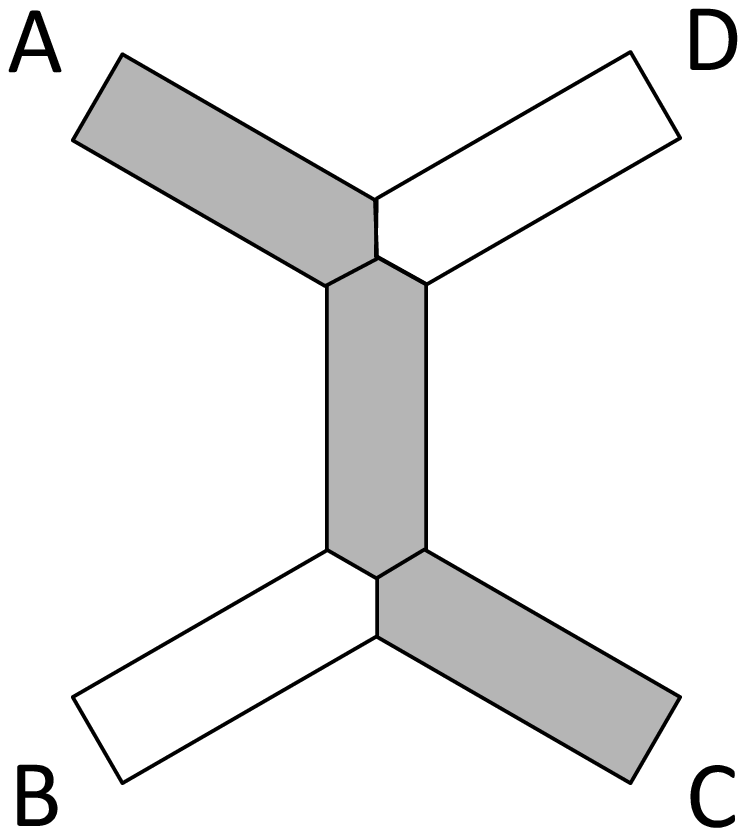}
          \\ $t$-channel          
        \end{center}
      \end{minipage}
      
    \end{tabular}
    \caption{Feynman diagrams with two cubic vertices and one propagator for $\mathcal{A}_{FBFB}$.}
    \label{FBFB}
  \end{center}
\end{figure}

Let us finally consider the four-point amplitude of two bosons and two fermions
with the fermion-boson-fermion-boson ordering denoted by $\mathcal{A}_{FBFB}$.
Feynman diagrams with two cubic vertices and one propagator for $\mathcal{A}_{FBFB}$
are depicted in figure~\ref{FBFB}.

\begin{figure}[tb]
  \begin{center}
          \includegraphics[clip, width=10cm]{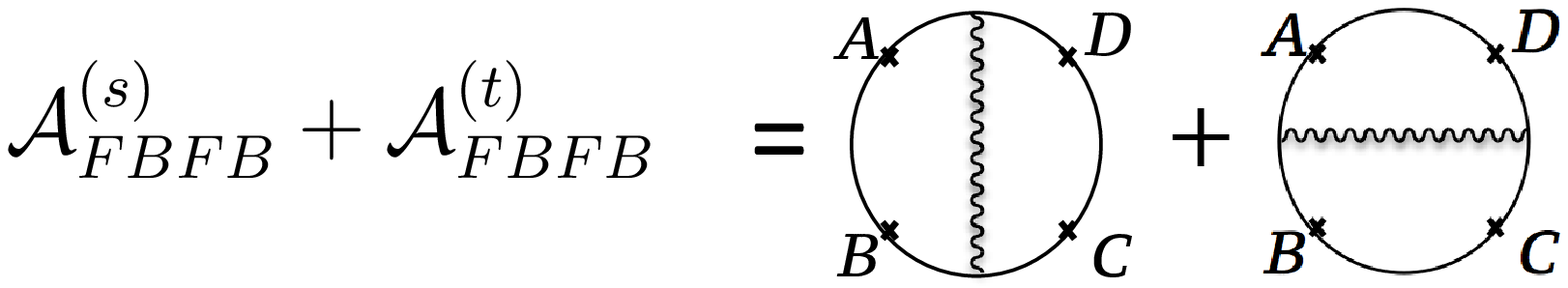}
          \caption{The integration contour of $X$ in $\mathcal{A}_{FBFB}^{(s)}$ and $\mathcal{A}_{FBFB}^{(t)}$.} 
          \label{FBFB-st}
  \end{center}
\end{figure}

Each of the two diagrams in figure~\ref{FBFB} consists
of two cubic vertices from $S_R^{(1)}$ and one Ramond propagator.
The $s$-channel contribution~$\mathcal{A}_{FBFB}^{(s)}$ is calculated as
\begin{align}
 \mathcal{A}_{FBFB}^{(s)}\ 
 =&\ \bigl\langle\,\Psi_A\, \phi_B\ \,
 \overbracket[0.5pt]{\!\! \Psi \,\bigr\rangle\big(-\bigl\langle\,\Psi\!\!}\ \,
\Psi_C\,\phi_D\,\bigr\rangle\big)
=\ \bigl\langle\,\Psi_A\,\phi_B\,\frac{b_0 X \eta}{L_0}\,(\,\Psi_C\,\phi_D\,)\,\bigr\rangle.
\end{align}
Similarly, the $t$-channel contribution~$\mathcal{A}_{FBFB}^{(t)}$ is
\begin{align}
 \mathcal{A}_{FBFB}^{(t)}\ 
 =&\ \bigl\langle\,\phi_B\,\Psi_C\,\frac{b_0 X \eta}{L_0}\,(\phi_D\,\Psi_A)\,\bigr\rangle.
\end{align}
We express $\mathcal{A}_{FBFB}^{(s)}$ and $\mathcal{A}_{FBFB}^{(t)}$
in terms of $\Phi_B$ and $\Phi_D$ using the BPZ inner product in the small Hilbert space.
We find
\begin{equation}
 \mathcal{A}_{FBFB}^{(s)}+ \mathcal{A}_{FBFB}^{(t)}
 = \ \llangle\,\Psi_A\,\Phi_B\,\frac{b_0 X}{L_0}\,(\Psi_C\,\Phi_D)\,\rrangle
 +\llangle\,\Phi_B\,\Psi_C\,\frac{b_0 X }{L_0}\,(\Phi_D\,\Psi_A)\,\rrangle. \label{A_FBFB^(s)+A_FBFB^(t)} 
\end{equation}
In each of the two terms on the right-hand side of~\eqref{A_FBFB^(s)+A_FBFB^(t)},
the integration of the odd modulus is implemented by the operator $X$ in the Ramond propagator.
However, the integration contour of $X$ is different in the $s$-channel and in the $t$-channel.
See figure~\ref{FBFB-st} for illustration.

Let us transmute $X$ in each channel to $X_0$ acting $\Phi_B$
in a way similar to~\eqref{X-to-X_0}.
For $\mathcal{A}_{FBFB}^{(s)}$ we have
\begin{align} 
\mathcal{A}_{FBFB}^{(s)}
=&\ \llangle\,\Psi_A\,\Phi_B\,\frac{b_0 X}{L_0}\,(\Psi_C\,\Phi_D)\,\rrangle
= {} -\bigl\langle\,\Psi_A\, \xi_0 \Phi_B \,\frac{b_0 \{\,Q,\Xi\,\} }{L_0}\,(\Psi_C\, \Phi_D)\,\bigr\rangle
\nonumber\\
=&\ \llangle\,\Psi_A\, X_0 \Phi_B\,\frac{b_0}{L_0}\,(\Psi_C\,\Phi_D)\,\rrangle
- \bigl\langle\,\Psi_A\, \xi_0 \Phi_B \,\Xi\,(\Psi_C\,\Phi_D)\,\bigr\rangle\,.
\label{X-to-X_0-2}
\end{align}
For $\mathcal{A}_{FBFB}^{(t)}$ we have
\begin{align}
 \mathcal{A}_{FBFB}^{(t)}\ 
 =& \llangle\,\Phi_B\,\Psi_C\,\frac{b_0 X }{L_0}\,(\Phi_D\,\Psi_A)\,\rrangle
= \bigl\langle\, \xi_0 \Phi_B \,\Psi_C\,\frac{b_0 \{\,Q,\Xi\,\}}{L_0}\,(\Phi_D\,\Psi_A)\,\bigr\rangle
\nonumber\\
=&\ \llangle\, X_0 \Phi_B\,\Psi_C\,\frac{b_0}{L_0} \, (\Phi_D\, \Psi_A)\,\rrangle
+\bigl\langle\,\xi_0 \Phi_B\,\Psi_C\,\Xi\,(\Phi_D\,\Psi_A)\,\bigr\rangle\,.
\label{X-to-X_0-3}
\end{align}
The total contribution $\mathcal{A}_{FBFB}^{(s)}+\mathcal{A}_{FFBB}^{(t)}$ from Feynman diagrams
with two cubic vertices and one propagator is
\begin{equation}\begin{split} 
\mathcal{A}_{FBFB}^{(s)}+\mathcal{A}_{FBFB}^{(t)} 
=&
\llangle\,\Psi_A\, X_0 \Phi_B\,\frac{b_0}{L_0}\,(\Psi_C\,\Phi_D)\,\rrangle
+\llangle\, X_0 \Phi_B\,\Psi_C\,\frac{b_0}{L_0} \, (\Phi_D\, \Psi_A)\,\rrangle \\
&- \bigl\langle\,\Psi_A\, \xi_0 \Phi_B\,\Xi\,(\Psi_C\,\Phi_D)\,\bigr\rangle\, 
+\bigl\langle\,\xi_0 \Phi_B\,\Psi_C\,\Xi\,(\Phi_D\,\Psi_A)\,\bigr\rangle . \label{FBFB-1p}
\end{split}
\end{equation}

Compared with the amplitude~$\mathcal{A}_{FBFB}^{{\rm WS}}$ in the world-sheet theory,
the difference $\Delta \mathcal{A}_{FBFB}$ is given by
\begin{equation}
\begin{split}
\Delta \mathcal{A}_{FBFB} 
& \equiv \mathcal{A}_{FBFB}^{(s)}+\mathcal{A}_{FBFB}^{(t)}-\mathcal{A}_{FBFB}^{{\rm WS}} \\
& = {}-\bigl\langle\,\Psi_A\, \xi_0 \Phi_B\,\Xi\,(\Psi_C\,\Phi_D)\,\bigr\rangle\, 
+\bigl\langle\,\xi_0 \Phi_B\,\Psi_C\,\Xi\,(\Phi_D\,\Psi_A)\,\bigr\rangle .
\end{split}
\end{equation}
As in the case of the fermion-fermion-boson-boson ordering,
the difference~$\Delta \mathcal{A}_{FBFB}$
also consists of terms without the propagator $b_0 / L_0$
and is localized
at the boundary of the moduli space
between the region covered by the $s$-channel contribution
and the region covered by the $t$-channel contribution.

\begin{figure}[t]
  \begin{center}
          \includegraphics[clip, width=2.5cm]{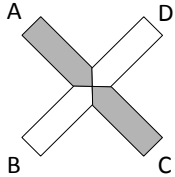}        
    \caption{Feynman diagram with a quartic vertex for $\mathcal{A}_{FBFB}$.}
    \label{FBFB4}
  \end{center}
\end{figure}

Let us now include an additional contribution to~$\mathcal{A}_{FBFB}$ from the Feynman diagram
with a quartic vertex depicted in figure~\ref{FBFB4}.
The contribution from this diagram is denoted by~$\mathcal{A}_{FBFB}^{(4)}$
and is calculated as
\begin{align}
 \mathcal{A}_{FBFB}^{(4)}\ =&\
-\frac{1}{2}\,\bigl\langle\,\phi_B\,\Psi_C\,\Xi\,(\eta\phi_D\,\Psi_A)\,\bigr\rangle
-\frac{1}{2}\,\bigl\langle\,\phi_D\,\Psi_A\,\Xi\,(\eta\phi_B\,\Psi_C)\,\bigr\rangle
\nonumber\\
&
+ \frac{1}{2}\,\bigl\langle\,\Psi_A\,\phi_B\,\Xi\,(\Psi_C\,\eta\phi_D)\,\bigr\rangle
+ \frac{1}{2}\,\bigl\langle\,\Psi_C\,\phi_D\,\Xi\,(\Psi_A\,\eta\phi_B)\,\bigr\rangle.
\end{align}
To compare this with the difference $\Delta \mathcal{A}_{FBFB}$,
we write it in terms of $\Phi_B$ and $\Phi_D$ in the following way:
\begin{align}
 \mathcal{A}_{FBFB}^{(4)}\ =&\
-\frac{1}{2}\,\bigl\langle\,\xi_0 \Phi_B\,\Psi_C\,\Xi\,(\Phi_D\,\Psi_A)\,\bigr\rangle
-\frac{1}{2}\,\bigl\langle\,\xi_0 \Phi_D\,\Psi_A\,\Xi\,(\eta \xi_0 \Phi_B\,\Psi_C)\,\bigr\rangle
\nonumber\\
&
+ \frac{1}{2}\,\bigl\langle\,\Psi_A\,\xi_0 \Phi_B\,\Xi\,(\Psi_C\,\Phi_D)\,\bigr\rangle
+ \frac{1}{2}\,\bigl\langle\,\Psi_C\,\xi_0 \Phi_D\,\Xi\,(\Psi_A\,\eta \xi_0 \Phi_B)\,\bigr\rangle.
\end{align}
We then move $\eta$ to obtain
\begin{align}
 \mathcal{A}_{FBFB}^{(4)}\ 
 =&\
\bigl\langle\,\Psi_A\,\xi_0 \Phi_B\,\Xi\,(\Psi_C\,\Phi_D)\,\bigr\rangle
- \bigl\langle\,\xi_0 \Phi_B\,\Psi_C\,\Xi\,(\Phi_D\,\Psi_A)\,\bigr\rangle\,. \label{FBFB-np}
\end{align}
This precisely cancels the difference~$\Delta \mathcal{A}_{FBFB}$,
so the amplitude~$\mathcal{A}_{FBFB}$ in open superstring field theory,
\begin{equation}
\mathcal{A}_{FBFB} = \mathcal{A}_{FBFB}^{(s)} + \mathcal{A}_{FBFB}^{(t)} + \mathcal{A}_{FBFB}^{(4)} \,,
\end{equation}
coincides with the amplitude~$\mathcal{A}_{FBFB}^{{\rm WS}}$ in the world-sheet theory:
\begin{equation}
\mathcal{A}_{FBFB} = \mathcal{A}_{FBFB}^{{\rm WS}} \,.
\end{equation}

\begin{figure}[t]
  \begin{center}
    \begin{tabular}{c}    
      \begin{minipage}{0.40\hsize}
        \begin{center}
          \includegraphics[clip, width=6cm]{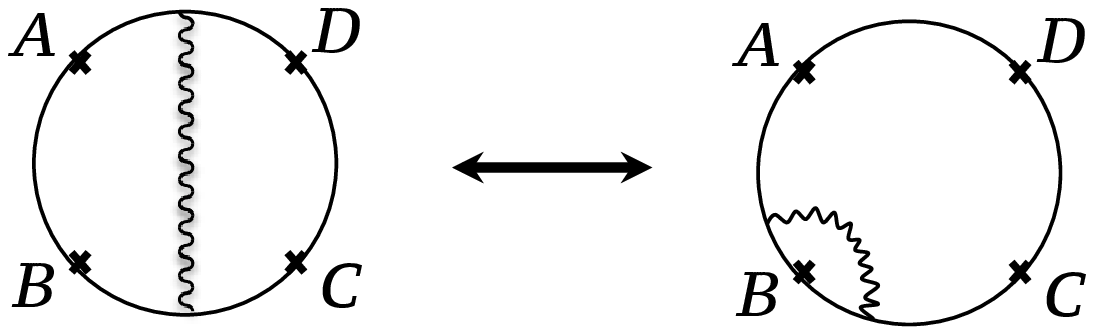}
          \caption{The first term on the right-hand side of~(\ref{FBFB-np}) contains $\Xi$
          acting on the $s$-channel intermediate state in the Ramond sector
          and $\xi_0$ acting on $\Phi_B$.
          This term exchanges $X$ in the $s$-channel and $X_0$ acting on $\Phi_B$ 
          as seen in~\eqref{X-to-X_0-2}.} 
          \label{ABtoB}
        \end{center}
      \end{minipage}
      \hspace{0.1\hsize}
      \begin{minipage}{0.40\hsize}
        \begin{center}
          \includegraphics[clip, width=6cm]{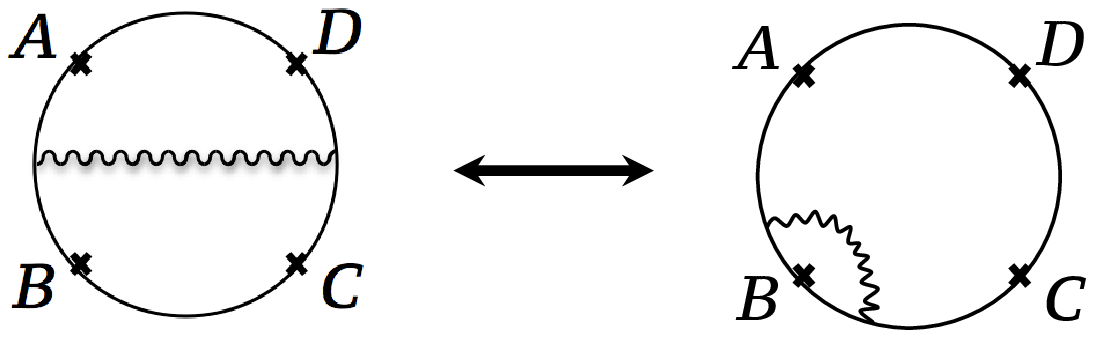}
          \caption{The second term on the right-hand side of~(\ref{FBFB-np}) contains $\Xi$
          acting on the $t$-channel intermediate state in the Ramond sector
          and $\xi_0$ acting on $\Phi_B$.
          This term exchanges $X$ in the $t$-channel and $X_0$ acting on $\Phi_B$ 
          as seen in~\eqref{X-to-X_0-3}.} 
          \label{BCtoB}
        \end{center}
      \end{minipage}    
    \end{tabular}
  \end{center}
\end{figure}

As in the fermion-fermion-boson-boson ordering,
we have again seen that the contribution~$\mathcal{A}_{FBFB}^{(4)}$ from the quartic interaction
adjusted the different assignment of picture-changing operators in the $s$-channel and in the $t$-channel.
The role of each term of $\mathcal{A}_{FBFB}^{(4)}$ in~\eqref{FBFB-np}
is illustrated in figures~\ref{ABtoB} and~\ref{BCtoB}.

\section{Five-point amplitudes}
\label{five-point-amplitudes}
\setcounter{equation}{0}
Before the construction of the gauge-invariant actions
including the Ramond sector~\cite{Kunitomo:2015usa, Sen:2015uaa},
scattering amplitudes involving external fermions in open superstring field theory
were calculated based on the approach
where a constraint is imposed on the equation of motion
after it was derived from an action.
As we mentioned in the introduction,
the set of Feynman rules proposed in~\cite{Michishita:2004by}
based on this approach
correctly reproduced four-point amplitudes in the world-sheet theory,
but it was reported that five-point amplitudes
were not correctly reproduced~\cite{Michishita:Riken}.
This issue was later resolved by a refined set of Feynman rules proposed in~\cite{Kunitomo:2014qla}.
The two sets of Feynman rules generally give different results
for diagrams containing at least two Ramond propagators
or for diagrams involving a vertex of two fermions and an even number of bosons
with at least one of the fermions being contracted with a propagator,
and these diagrams first appear in five-point amplitudes
when we increase the number of external states.

This motivated us to calculate five-point amplitudes
based on the Feynman rules derived from
the gauge-invariant actions constructed in~\cite{Kunitomo:2015usa, Sen:2015uaa},
and in this section
we present the calculations of five-point amplitudes involving two external fermions
with the fermion-fermion-boson-boson-boson ordering.
We present the calculations of five-point amplitudes involving two external fermions
with the fermion-boson-fermion-boson-boson ordering
and five-point amplitudes involving four external fermions
in appendix~\ref{appendix}.

\subsection{Five-point amplitudes in the world-sheet theory} 
Let us first present an expression for five-point amplitudes in the world-sheet theory.
It is known that the moduli space of disks with an arbitrary number of punctures
on the boundary is covered by contributions from Feynman diagrams
which consist of Witten's cubic vertices and propagators~\cite{Zwiebach:1990az}.
For the open superstring, we also need to consider fermionic directions of the supermoduli space,
and we insert an appropriate number of picture-changing operators
{\it on the same set of external states for all Feynman diagrams}.
We consider five-point amplitudes of on-shell external states denoted by
$\Psi_A$, $\Psi_B$, $\Phi_C$, $\Phi_D$, and $\Phi_E$,
where $\Psi_A$ and $\Psi_B$ are in the Ramond sector
and $\Phi_C$, $\Phi_D$, and $\Phi_E$ are in the NS sector.
In this case we need to insert two picture-changing operators
and we insert them on $\Phi_C$ and $\Phi_D$ for all diagrams.
The scattering amplitude $\mathcal{A}_{FFBBB}^{{\rm WS}}$ in the world-sheet theory
is then written as
\begin{align}
\mathcal{A}_{FFBBB}^{{\rm WS}}
=&\
-\,\llangle\,\Psi_A\,\Psi_B\,\frac{b_0}{L_0}\,\Bigl( \,X_0\Phi_C\,\frac{b_0}{L_0}\,(
X_0\Phi_D\,\Phi_E) \Bigr)\,\rrangle
- \llangle\,\Psi_B\,X_0\Phi_C\,\frac{b_0}{L_0}\,\Bigl( X_0\Phi_D\,
\frac{b_0}{L_0}\,(\Phi_E\,\Psi_A)\Bigr)\,\rrangle
\nonumber\\
&\
- \llangle\, X_0\Phi_C\,X_0\Phi_D\,
\frac{b_0}{L_0}\,\Bigl( \Phi_E\,\frac{b_0}{L_0}\,(\Psi_A\,\Psi_B) \Bigr) \,\rrangle
- \llangle\, X_0\Phi_D\,\Phi_E
\frac{b_0}{L_0}\,\Bigl( \Psi_A\,\frac{b_0}{L_0}\,(\Psi_B\,X_0\Phi_C)\Bigr)\,\rrangle
\nonumber\\
&\
- \llangle\,\Phi_E\,\Psi_A\,\frac{b_0}{L_0}\,
\Bigl( \Psi_B\,\frac{b_0}{L_0}\,(X_0\Phi_C\,X_0\Phi_D) \Bigr)\,\rrangle\,. \label{A_FFBBB^WS}
\end{align}
The goal of this section is to reproduce this expression
by the calculation of the scattering amplitude in open superstring field theory.

\subsection{Contributions from diagrams with two propagators}
\label{2p}

\begin{figure}[h]
  \begin{center}
    \begin{tabular}{c}
    
      \begin{minipage}{0.30\hsize}
        \begin{center}
          \includegraphics[clip, width=3.5cm]{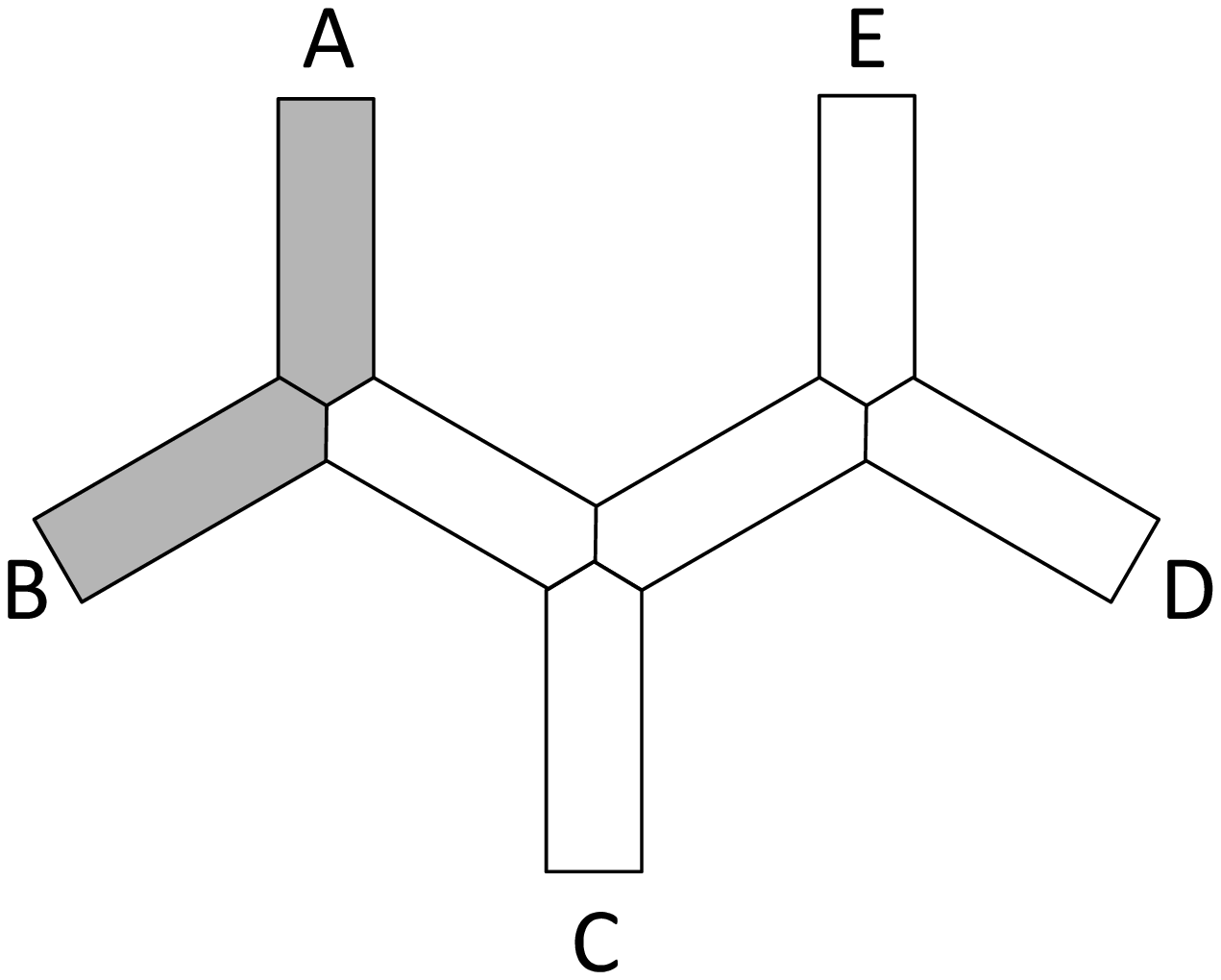}
          \hspace{1.6cm} $(2P)(a)$
                  \end{center}
      \end{minipage}
      \begin{minipage}{0.30\hsize}
        \begin{center}
          \includegraphics[clip, width=3.5cm]{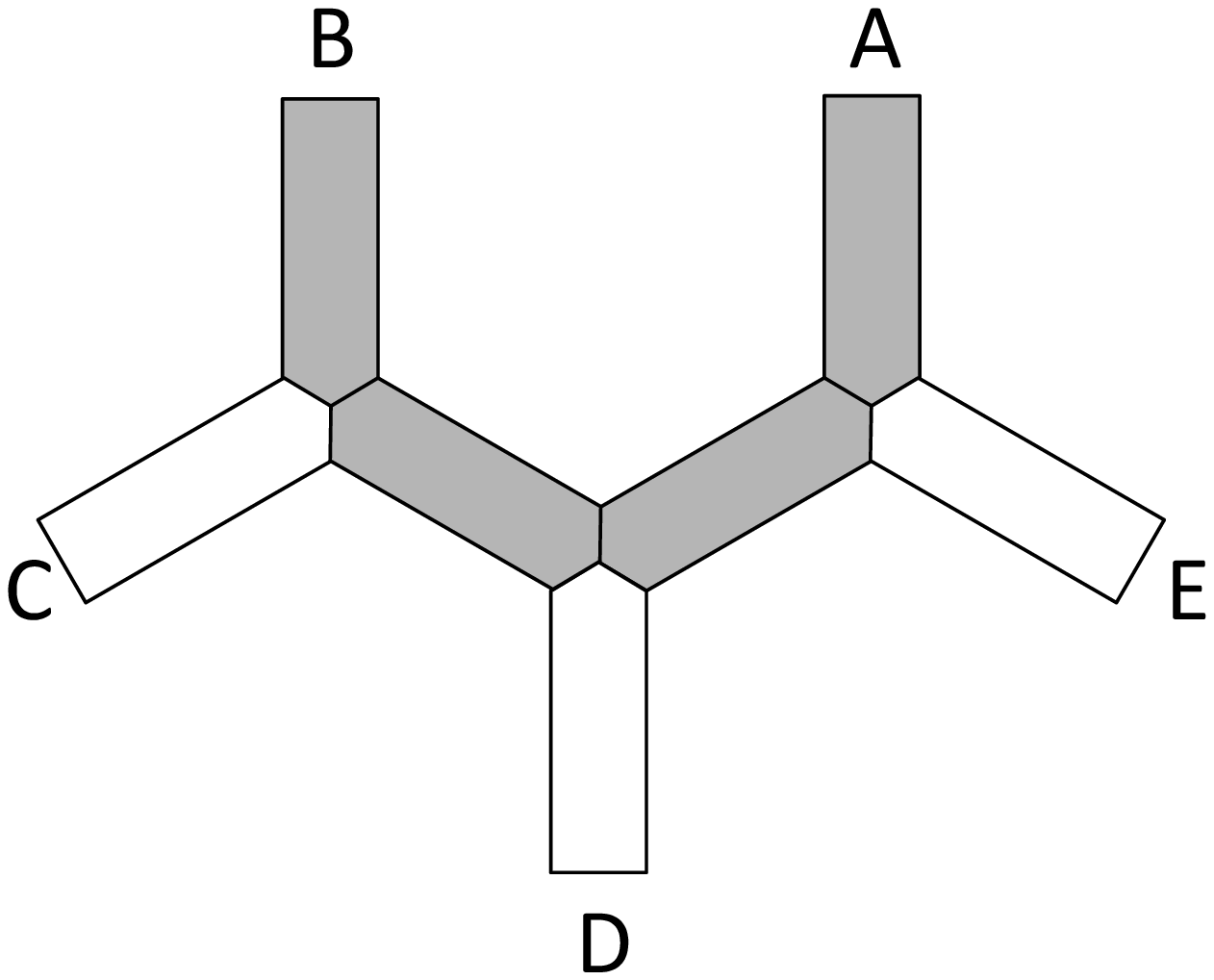}
          \hspace{1.6cm}$(2P)(b)$
        \end{center}
      \end{minipage}
      \begin{minipage}{0.30\hsize}
        \begin{center}
          \includegraphics[clip, width=3.5cm]{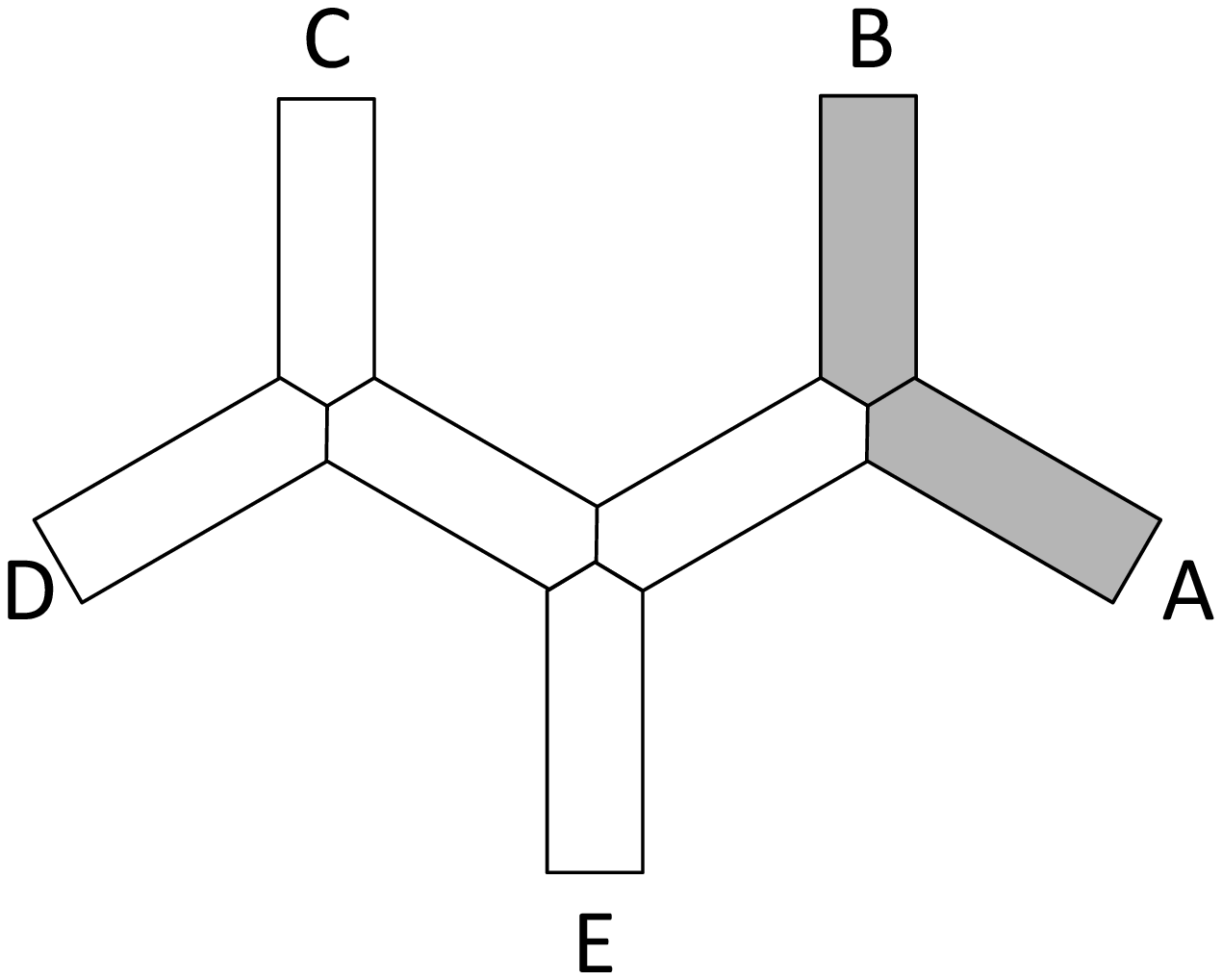}
          \hspace{1.6cm} $(2P)(c)$
        \end{center}
      \end{minipage}
      \\
      \begin{minipage}{0.30\hsize}
        \begin{center}
          \includegraphics[clip, width=3.5cm]{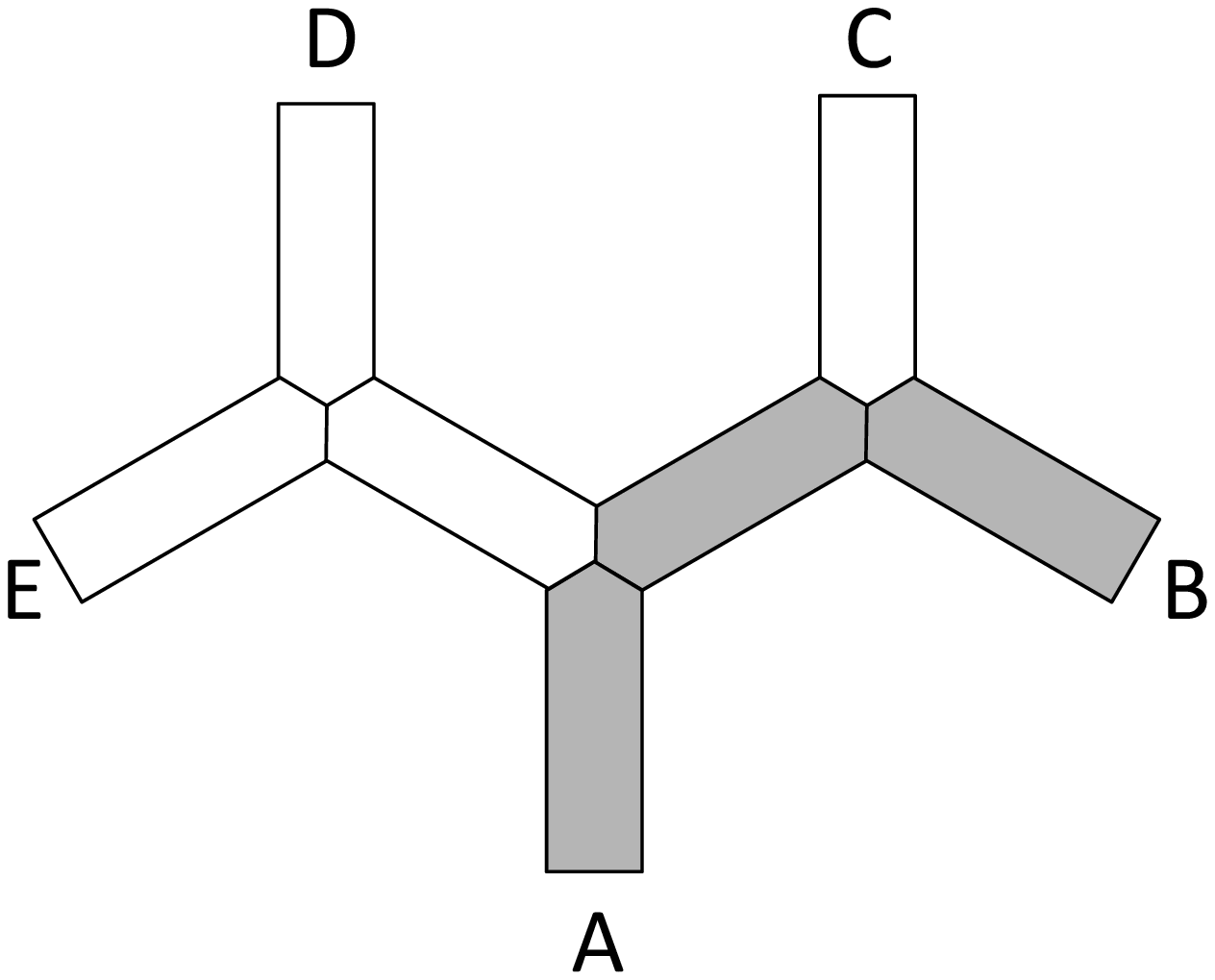}
          \hspace{1.6cm} $(2P)(d)$
        \end{center}
      \end{minipage}
      \begin{minipage}{0.30\hsize}
        \begin{center}
          \includegraphics[clip, width=3.5cm]{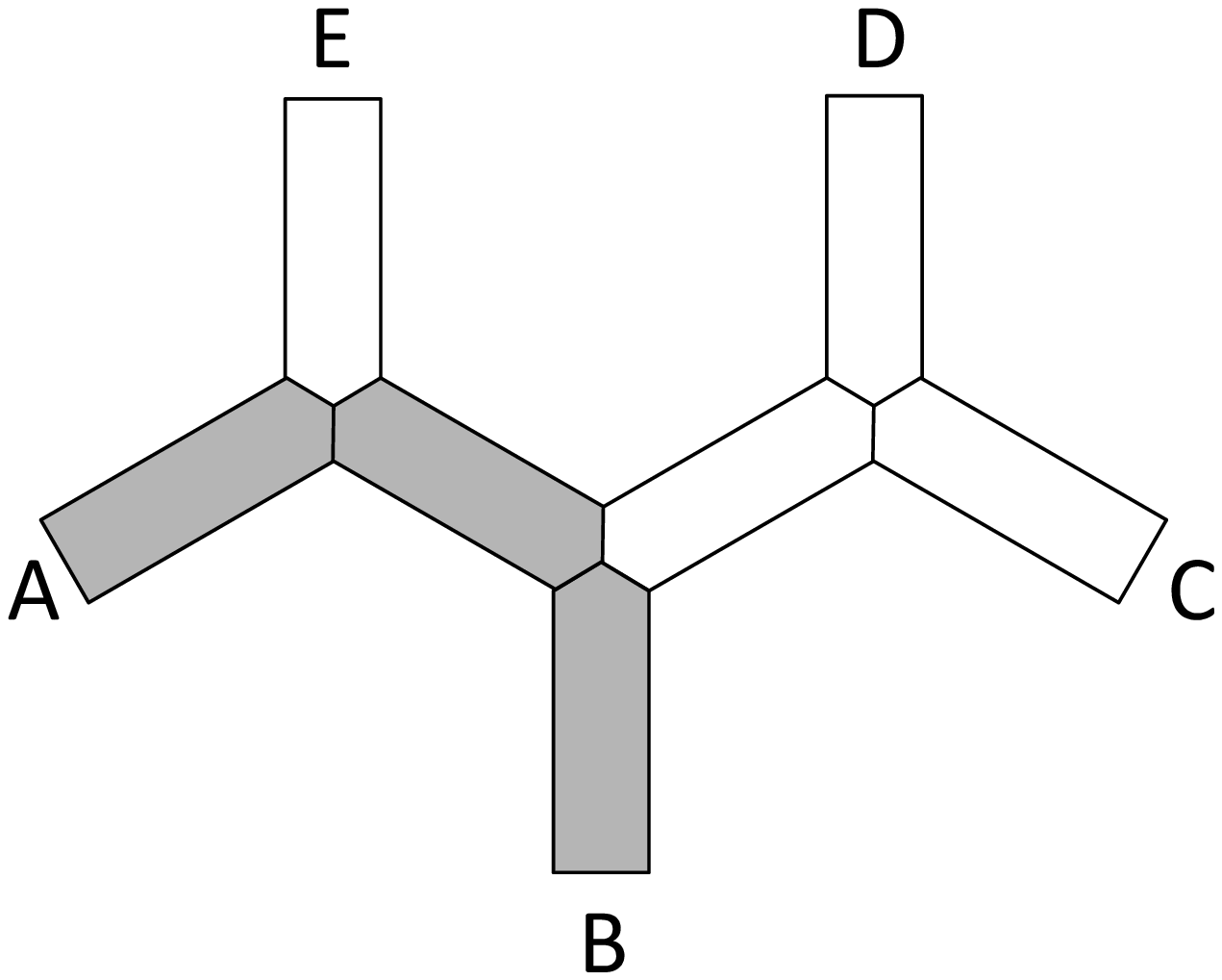}
          \hspace{1.6cm} $(2P)(e)$
        \end{center}
      \end{minipage}
      
    \end{tabular}
    \caption{Feynman diagrams with three cubic vertices and two propagators for $\mathcal{A}_{FFBBB}$.}
    \label{5point-2propagator}
  \end{center}
\end{figure}
We denote the corresponding amplitude in open superstring field theory by $\mathcal{A}_{FFBBB}$.
In this subsection we calculate contributions to $\mathcal{A}_{FFBBB}$
from Feynman diagrams with three cubic vertices and two propagators.
Such diagrams are depicted in figure~\ref{5point-2propagator}.

The diagram $(2P)(a)$ in figure~\ref{5point-2propagator}
consists of two cubic vertices from $S_{NS}^{(1)}$, one cubic vertex from $S_{R}^{(1)}$,
and two NS propagators. It is calculated as follows:
\begin{align}
\mathcal{A}_{FFBBB}^{(2P)(a)}\ =&\
 \big(- \bigl\langle\,\Psi_A\,\Psi_B\ \,
\overbracket[0.5pt]{\!\! \phi \,\bigr\rangle\big) \big(\, \frac{1}{2}\bigl\langle\, Q\phi\!\!}\ \,
\phi_C \  \eta \, \overbracket[0.5pt]{\!\!\phi \,\bigr\rangle\big) 
\Big(- \frac{1}{2}\bigl\langle\,\phi\!\!}\, \ (Q\phi_D\,\eta\phi_E + 
\eta\phi_D\,Q\phi_E)\,\bigr\rangle\Big)
\nonumber\\
&
+ \big(- \bigl\langle\,\Psi_A\,\Psi_B\ \, 
\overbracket[0.5pt]{\!\! \phi \,\bigr\rangle\big) \big(\, \frac{1}{2}\bigl\langle\, \eta\phi\!\!}\, \
\phi_C \ Q \, \overbracket[0.5pt]{\!\! \phi \,\bigr\rangle\big) 
\Big(- \frac{1}{2}\bigl\langle\,\phi\!\!}\ \, (Q\phi_D\,\eta\phi_E + 
\eta\phi_D\,Q\phi_E)\,\bigr\rangle\Big)
\nonumber\\
=&\                            
-\frac{1}{4}\,\bigl\langle\,\Psi_A\,\Psi_B\,\frac{\xi_0 b_0 Q}{L_0}\, \Bigl(
\phi_C\,\frac{\eta \xi_0 b_0}{L_0}\,(Q\phi_D\,\eta\phi_E + \eta\phi_D\,Q\phi_E) \Bigr)\,\bigr\rangle
\nonumber\\
&\
-\frac{1}{4}\,\bigl\langle\,\Psi_A\,\Psi_B\,\frac{\xi_0 b_0 \eta}{L_0}\,\Bigl(
\phi_C\,\frac{Q \xi_0 b_0}{L_0}\,(Q\phi_D\,\eta\phi_E + \eta\phi_D\,Q\phi_E) \Bigr)\,\bigr\rangle.
 \label{(2p)(a)} 
\end{align}
The corresponding term of the amplitude~$\mathcal{A}_{FFBBB}^{{\rm WS}}$ in the world-sheet theory is
\begin{equation}
{}-\llangle \, \Psi_A \, \Psi_B \, \frac{b_0}{L_0} \, \Bigl( \, X_0\Phi_C \, \frac{b_0}{L_0} \,(\, X_0\Phi_D \,  \Phi_E)\, \Bigr) \, \rrangle,
\end{equation}
which can be translated into the language of the large Hilbert space as
\begin{equation}
{}-\llangle \, \Psi_A \, \Psi_B \, \frac{b_0}{L_0} \, \Bigl( \, X_0\Phi_C \, \frac{b_0}{L_0} \,(\, X_0\Phi_D \,  \Phi_E)\, \Bigr) \, \rrangle
= {}-\bigl\langle \, \Psi_A \, \Psi_B \, \frac{b_0}{L_0} \, \Bigl( \, Q\phi_C \, \frac{b_0}{L_0} \,(\, Q\phi_D \,  \phi_E)\, \Bigr) \, \bigr\rangle.
\label{(2p)(a)-WS}
\end{equation}
In this section and in appendix~\ref{appendix}, we mainly work in the large Hilbert space.
Note that our choice of inserting picture-changing operators on $\Phi_C$ and $\Phi_D$
in the small Hilbert space corresponds to inserting the BRST operators on $\phi_C$ and $\phi_D$
in the large Hilbert space.
Let us therefore move the BRST operators in~\eqref{(2p)(a)}
to act on $\phi_C$ and $\phi_D$,
as we did in~\eqref{D-to-C} for four-point amplitudes.

For example,
the second term in the first line on the right-hand side of~\eqref{(2p)(a)}
is calculated as follows. 
We first move $\eta$ to delete $\xi_0$ in the second NS propagator as
\begin{align}
-\bigl\langle \, \Psi_A \, \Psi_B \, \frac{\xi_0 b_0 Q}{L_0}\, 
\Bigl(\, \phi_C \, 
\frac{\eta \xi_0 b_0 }{L_0} \, (\, \eta\phi_D \, Q\phi_E \, )\, \Bigr) \, \bigr\rangle 
= -\bigl\langle \, \Psi_A \, \Psi_B \, \frac{\xi_0 b_0 Q}{L_0} \, 
\Bigl( \, \phi_C \, \frac{b_0}{L_0} (\, \eta\phi_D \, Q \phi_E \, ) \,\Bigr) \, \bigr\rangle,
\end{align}
then we can move the first BRST operator to obtain
\begin{equation}
\begin{split}
& -\bigl\langle \, \Psi_A \, \Psi_B \, \frac{\xi_0 b_0 Q}{L_0} \, 
\Bigl( \, \phi_C \, \frac{b_0}{L_0} (\, \eta\phi_D \, Q \phi_E \, ) \,\Bigr) \, \bigr\rangle \\
& = {}- \bigl\langle \, \Psi_A \, \Psi_B \, \frac{\xi_0 b_0}{L_0} \,\Bigl(\, Q\phi_C \, \frac{b_0}{L_0}\, (\, \eta\phi_D \, Q\phi_E \,) \, \Bigr) \, \bigr\rangle
- \bigl\langle \, \Psi_A \, \Psi_B \, \frac{\xi_0 b_0}{L_0} (\,\phi_C \, \eta\phi_D \, Q\phi_E \, ) \, \bigr\rangle \,.
\end{split}
\end{equation}
The second term on the right-hand side was generated
when the BRST operator anticommutes with one of the propagators,
and the number of the propagators is reduced in this term.
In the first term on the right-hand side with two propagators,
one of the BRST operators acts on $\phi_C$,
but the other one acts on $\phi_E$.
Let us move this to act on $\phi_D$.
However, we first need to remove $\eta$ acting on $\phi_D$ as follows:
\begin{equation}
\begin{split}
{}- \bigl\langle \, \Psi_A \, \Psi_B \, \frac{\xi_0 b_0}{L_0} \,\Bigl(\, Q\phi_C \, \frac{b_0}{L_0}\, (\, \eta\phi_D \, Q\phi_E \,) \, \Bigr) \, \bigr\rangle
= \bigl\langle \, \Psi_A \, \Psi_B \, \frac{b_0}{L_0} \, \Bigl( \, Q\phi_C \, \frac{b_0}{L_0} \,(\, \phi_D \, Q \phi_E)\, \Bigr) \, \bigr\rangle.
\end{split}
\end{equation}
We can now move the BRST operator from $\phi_E$ to $\phi_D$:
\begin{align}
&\bigl\langle \, \Psi_A \, \Psi_B \, \frac{b_0}{L_0} \, \Bigl( \, Q\phi_C \, \frac{b_0}{L_0} \,(\, \phi_D \, Q \phi_E)\, \Bigr) \, \bigr\rangle \nonumber\\
=& -\bigl\langle \, \Psi_A \, \Psi_B \, \frac{b_0}{L_0} \, \Bigl( \, Q\phi_C \, \frac{b_0}{L_0} \,(\, Q\phi_D \,  \phi_E)\, \Bigr) \, \bigr\rangle \nonumber \\
& +\bigl\langle \, \Psi_A \, \Psi_B \, \frac{b_0}{L_0} \, ( \, Q\phi_C \,\phi_D \, \phi_E)\, ) \, \bigr\rangle
+\bigl\langle \, \Psi_A \, \Psi_B \,  Q\phi_C \, \frac{b_0}{L_0} \,(\, \phi_D \, \phi_E) \, \bigr\rangle.
\label{(2p)(a)-example}
\end{align}
Note that two more terms with one propagator are generated.

We similarly transform each term on the right-hand side of~\eqref{(2p)(a)}
and find that the contribution $\mathcal{A}_{FFBBB}^{(2P)(a)}$ from the diagram $(2P)(a)$
in figure~\ref{5point-2propagator} can be written as
\begin{align}
\mathcal{A}_{FFBBB}^{(2P)(a)}
=&\, -\,\bigl\langle\,\Psi_A\,\Psi_B\,\frac{b_0}{L_0}\,\Bigl( Q\phi_C\,\frac{b_0}{L_0}\,
( \, Q\phi_D\,\phi_E \, ) \, \Bigr) \,\bigr\rangle
\nonumber\\
&\
-\frac{1}{4}\,\bigl\langle\,\Psi_A\,\Psi_B\,\frac{\xi_0 b_0}{L_0}\, (
\phi_C\,(Q\phi_D\,\eta\phi_E + \eta\phi_D\,Q\phi_E) ) \,\bigr\rangle
+ \frac{1}{2}\,\bigl\langle\,\Psi_A\,\Psi_B\,\frac{\xi_0 b_0}{L_0}\, (
Q\phi_C\,\eta(\phi_D\,\phi_E) ) \,\bigr\rangle
\nonumber\\
&\
+\frac{1}{4}\,\bigl\langle\,(Q\phi_D\,\eta\phi_E + \eta\phi_D\,Q\phi_E)\,
\frac{\xi_0 b_0}{L_0}\,(\Psi_A\,\Psi_B\,\phi_C )\,\bigr\rangle
+\frac{1}{2}\,\bigl\langle\,\eta(\phi_D\,\phi_E)\,\frac{\xi_0 b_0}{L_0}\,
(\Psi_A\,\Psi_B\,Q\phi_C)\,\bigr\rangle\,.
\label{(2P)(a)-FFBBB}
\end{align}
The first term on the right-hand side which contains two propagators
reproduces the term~\eqref{(2p)(a)-WS} in the world-sheet theory.
All other terms, which were generated when we moved the BRST operator, contain one propagator.
The moduli space of disks with five punctures on the boundary are two dimensional,
and the contribution from the term with two propagators cover a two-dimensional region
of the moduli space.
On the other hand, the contributions from the terms with one propagator
have support in a one-dimensional region of the moduli space
and they can be regarded as surface terms of the two-dimensional integral.
We will see if these surface terms cancel when we combine contributions from all Feynman diagrams.

The diagram $(2P)(b)$ in figure~\ref{5point-2propagator}
consists of three cubic vertices from $S_{R}^{(1)}$
and two Ramond propagators. It is calculated as follows:
\begin{align}
 \mathcal{A}_{FFBBB}^{(2P)(b)}\ =&\
\bigl\langle\,\Psi_B\,\phi_C\ \,
\overbracket[0.5pt]{\!\! \Psi\,\bigr\rangle \bigl\langle\, \Psi \!\!}\ \,
\phi_D\, 
\overbracket[0.5pt]{\!\! \Psi\,\bigr\rangle \bigl\langle\,\Psi \!\!}\ \
\phi_E\,\Psi_A\,\bigr\rangle
\nonumber\\
=&\
\bigl\langle\,\Psi_B\,\phi_C\,\frac{b_0 X \eta}{L_0}\, \Bigl( \phi_D\,
\frac{b_0 X \eta}{L_0}\,(\phi_E\,\Psi_A ) \Bigr) \,\bigr\rangle .
\end{align}
We move two insertions of $\eta$ to find
\begin{equation}
\bigl\langle\,\Psi_B\,\phi_C\,\frac{b_0 X \eta}{L_0}\, \Bigl( \phi_D\,
\frac{b_0 X \eta}{L_0}\,(\phi_E\,\Psi_A ) \Bigr) \,\bigr\rangle
=\bigl\langle \, \Psi_B \, \phi_C \, \frac{b_0 X}{L_0} \, \Bigl( \, \eta\phi_D \, \frac{b_0 X}{L_0}\, (\, \eta\phi_E \, \Psi_A \, ) \, \Bigr) \, \bigr\rangle.
\end{equation}
The operator $X$ can be written as $X = \{ \, Q , \, \Xi \, \}$
so that we have two BRST operators.
We want to move them to act on $\phi_C$ and $\phi_D$.
We first write the operator $X$ in the second Ramond propagator as $\{ \, Q , \, \Xi \, \}$
and move $Q$ to $\phi_C$:
\begin{equation}
\begin{split}
& \bigl\langle \, \Psi_B \, \phi_C \, \frac{b_0 X}{L_0} \, \Bigl( \, \eta\phi_D \, \frac{b_0 \{ \, Q , \, \Xi \, \}}{L_0}\, (\, \eta\phi_E \, \Psi_A \, ) \, \Bigr) \, \bigr\rangle \\
& = \bigl\langle \, \Psi_B \, Q \phi_C \, \frac{b_0 X}{L_0} \, \Bigl( \, \eta\phi_D \, \frac{b_0 \Xi}{L_0} \, (\, \eta\phi_E \,\Psi_A \,) \, \Bigr) \, \bigr\rangle \\
& \quad~
+ \bigl\langle\,\eta\phi_E\,\Psi_A\,\frac{b_0 \Xi}{L_0}\,( X\,(\Psi_B\,\phi_C)\,\eta\phi_D ) \,\bigr\rangle
+ \bigl\langle\,\Psi_B\,\phi_C\,\frac{b_0 X}{L_0}\,( \eta\phi_D\,\Xi\,(\eta\phi_E\,\Psi_A) ) \,\bigr\rangle \,.
\end{split}
\end{equation} 
Before we move the other BRST operator to $\phi_D$ in the first term on the right-hand side,
we need to remove $\eta$ acting on $\phi_D$:
\begin{equation}
\bigl\langle \, \Psi_B \, Q \phi_C \, \frac{b_0 X}{L_0} \, \Bigl( \, \eta\phi_D \, \frac{b_0 \Xi}{L_0} \, (\, \eta\phi_E \,\Psi_A \,) \Bigr) \, \bigr\rangle 
=
\bigl\langle \, \Psi_B \, Q \phi_C \, \frac{b_0 X}{L_0} \, \Bigl( \, \phi_D \, \frac{b_0}{L_0} \, (\, \eta\phi_E \,\Psi_A \,) \, \Bigr) \,\bigr\rangle .
\end{equation} 
We then write the operator $X$ in the first Ramond propagator as $\{ \, Q , \, \Xi \, \}$
and move $Q$ to $\phi_D$:
\begin{equation}
\begin{split}
& \bigl\langle \, \Psi_B \, Q \phi_C \, \frac{b_0 \{ \, Q , \, \Xi \, \}}{L_0} \, \Bigl( \, \phi_D \, \frac{b_0}{L_0} \, (\, \eta\phi_E \,\Psi_A \,) \, \Bigr) \,\bigr\rangle \\
& = \bigl\langle \, \Psi_B \, Q \phi_C \, \frac{b_0 \Xi}{L_0} \, \Bigl( \, Q\phi_D \, \frac{b_0}{L_0} \, (\, \eta\phi_E \,\Psi_A \,) \, \Bigr) \,\bigr\rangle \\
& \quad~
+ \bigl\langle\,\Psi_B\,Q\phi_C\,\frac{b_0 \Xi}{L_0}\,(\phi_D\,
\eta\phi_E\,\Psi_A)\,\bigr\rangle
- \bigl\langle\,\eta\phi_E\,\Psi_A\,\frac{b_0}{L_0}\,( \Xi\,(\Psi_B\,Q\phi_C)\,\phi_D ) \,\bigr\rangle \,.
\end{split}
\end{equation}
Therefore, the contribution $\mathcal{A}_{FFBBB}^{(2P)(b)}$ from the diagram $(2P)(b)$
in figure~\ref{5point-2propagator} can be written as
\begin{align}
\mathcal{A}_{FFBBB}^{(2P)(b)}=&\
\bigl\langle\,\Psi_B\,Q\phi_C\,\frac{b_0 \Xi}{L_0}\,\Bigl( Q\phi_D\,
\frac{b_0}{L_0}\,(\eta\phi_E\,\Psi_A ) \Bigr) \,\bigr\rangle
\nonumber\\
&\
+ \bigl\langle\,\Psi_B\,Q\phi_C\,\frac{b_0 \Xi}{L_0}\,(\phi_D\,
\eta\phi_E\,\Psi_A)\,\bigr\rangle
+ \bigl\langle\,\Psi_B\,\phi_C\,\frac{b_0 X}{L_0}\,( \eta\phi_D\,\Xi\,(\eta\phi_E\,\Psi_A) ) \,\bigr\rangle
\nonumber\\
&\
- \bigl\langle\,\eta\phi_E\,\Psi_A\,\frac{b_0}{L_0}\,( \Xi\,(\Psi_B\,Q\phi_C)\,\phi_D ) \,\bigr\rangle
+ \bigl\langle\,\eta\phi_E\,\Psi_A\,\frac{b_0 \Xi}{L_0}\,( X\,(\Psi_B\,\phi_C)\,\eta\phi_D ) \,\bigr\rangle\,.
\label{(2P)(b)-FFBBB}
\end{align}
The first term on the right-hand side reproduces
the corresponding term in the world-sheet theory,
\begin{equation}
\bigl\langle\,\Psi_B\,Q\phi_C\,\frac{b_0 \Xi}{L_0}\,\Bigl( Q\phi_D\,
\frac{b_0}{L_0}\,(\eta\phi_E\,\Psi_A ) \Bigr) \,\bigr\rangle
= {}-\llangle \, \Psi_B \, X_0 \Phi_C \, \frac{b_0}{L_0} \, \Bigl( \, X_0\Phi_D \, \frac{b_0}{L_0} \, (\, \Phi_E \,\Psi_A \,) \, \Bigr) \,\rrangle \,,
\end{equation}
and four terms containing one propagator have been generated.

The diagram $(2P)(c)$ in figure~\ref{5point-2propagator}
consists of two cubic vertices from $S_{NS}^{(1)}$, one cubic vertex from $S_{R}^{(1)}$,
and two NS propagators. It is calculated as follows:
\begin{equation}
\begin{split}
\mathcal{A}_{FFBBB}^{(2P)(c)}
& = \Big(-\frac{1}{2}\,\bigl\langle\,(Q\phi_C\,\eta\phi_D + \eta\phi_C\,Q\phi_D)\ \
\overbracket[0.5pt]{\!\!\! \phi\,\bigr\rangle\Big)
\Big(\,\frac{1}{2}\, \bigl\langle\, Q\phi\!\!\!}\ \
\phi_E\ \
\overbracket[0.5pt]{\!\!\! \eta\phi\,\bigr\rangle\Big) ( -\, \bigl\langle\, \phi\!\!\!}\ \
\Psi_A\,\Psi_B\,\bigr\rangle) \\
& \quad~
+ \Big(-\frac{1}{2}\,\bigl\langle\,(Q\phi_C\,\eta\phi_D + \eta\phi_C\,Q\phi_D)\ \
\overbracket[0.5pt]{\!\!\! \phi\,\bigr\rangle\Big)
\Big(\,\frac{1}{2}\, \bigl\langle\, \eta\phi\!\!\!}\ \
\phi_E\ \
\overbracket[0.5pt]{\!\!\! Q\phi\,\bigr\rangle\Big) ( -\, \bigl\langle\, \phi\!\!\!}\ \
\Psi_A\,\Psi_B\,\bigr\rangle) \,.
\end{split}
\end{equation}
This can be transformed as in the case of the diagram~$(2P)(a)$:
\begin{align}
&\mathcal{A}_{FFBBB}^{(2P)(c)}
= -\frac{1}{4}\,\bigl\langle\,(Q\phi_C\,\eta\phi_D + \eta\phi_C\,Q\phi_D)\,
\frac{\xi_0 b_0 Q}{L_0}\,\Bigl( \phi_E\,\frac{\eta \xi_0 b_0}{L_0}\,( \Psi_A\,\Psi_B) \Bigr)\,\bigr\rangle
\nonumber\\
& \qquad \qquad \quad
-\frac{1}{4}\,\bigl\langle\,(Q\phi_C\,\eta\phi_D + \eta\phi_C\,Q\phi_D)\,
\frac{\xi_0 b_0 \eta}{L_0}\,\Bigl( \phi_E\,\frac{Q \xi_0 b_0}{L_0}\,(\Psi_A\,\Psi_B) \Bigr)\,\bigr\rangle
\nonumber\\
=&\
- \bigl\langle\, Q\phi_C\,Q\phi_D\,
\frac{b_0}{L_0}\, \Bigl( \eta\phi_E\,\frac{\xi_0 b_0}{L_0}\,( \Psi_A\,\Psi_B) \Bigr) \,\bigr\rangle
\nonumber\\
&\
+ \frac{1}{2}\,\bigl\langle\,(Q\phi_C\,\phi_D - \phi_C\,Q\phi_D)\,
\frac{\xi_0 b_0}{L_0}\,(\eta\phi_E\,\Psi_A\,\Psi_B)\,\bigr\rangle
+ \frac{1}{4}\,\bigl\langle\,(Q\phi_C\,\eta\phi_D + \eta\phi_C\,Q\phi_D)\,
\frac{\xi_0 b_0}{L_0}\,(\phi_E\,\Psi_A\,\Psi_B)\,\bigr\rangle
\nonumber\\
&\
- \frac{1}{2}\,\bigl\langle\,\Psi_A\,\Psi_B\,
\frac{\xi_0 b_0}{L_0}\, (
(Q\phi_C\,\phi_D - \phi_C\,Q\phi_D)\,\eta\phi_E ) \,\bigr\rangle
-\frac{1}{4}\,\bigl\langle\,\Psi_A\,\Psi_B\,\frac{\xi_0 b_0}{L_0}\, (
(Q\phi_C\,\eta\phi_D + \eta\phi_C\,Q\phi_D)\,\phi_E ) \,\bigr\rangle\,. 
\label{(2P)(c)-FFBBB}
\end{align}

The diagram $(2P)(d)$ in figure~\ref{5point-2propagator}
consists of one cubic vertex from $S_{NS}^{(1)}$, two cubic vertices from $S_{R}^{(1)}$,
one NS propagator, and one Ramond propagators. It is calculated as follows:
\begin{align}
 \mathcal{A}_{FFBBB}^{(2P)(d)}\ =&\
\Big(-\,\frac{1}{2}\, \bigl\langle\,(Q\phi_D\,\eta\phi_E + \eta\phi_D\,Q\phi_E)\ \
\overbracket[0.5pt]{\!\!\! \phi\,\bigr\rangle\Big)
(-\, \bigl\langle\, \phi\!\!\!}\ \
\Psi_A\ \
\overbracket[0.5pt]{\!\!\! \Psi\,\bigr\rangle)
(\,-\,\bigl\langle\, \Psi\!\!\!}\ \
\Psi_B\,\phi_C\,\bigr\rangle)
\nonumber\\
=&\
\frac{1}{2}\,\bigl\langle\,(Q\phi_D\,\eta\phi_E + \eta\phi_D\,Q\phi_E)\,
\frac{\xi_0 b_0}{L_0}\, \Bigl( 
\Psi_A\,\frac{b_0 X \eta}{L_0}\,(\Psi_B\,\phi_C) \Bigr) \,\bigr\rangle. \label{(2p)(d)}
\end{align}
We have one BRST operator in the cubic vertex from $S_{NS}^{(1)}$,
and the other BRST operator is from the operator $X$.
We want to move them to act on $\phi_C$ and $\phi_D$.
Let us first move $\eta$ in the Ramond propagator to find
\begin{equation}
\begin{split}
& \frac{1}{2}\,\bigl\langle\,(Q\phi_D\,\eta\phi_E + \eta\phi_D\,Q\phi_E)\,
\frac{\xi_0 b_0}{L_0}\, \Bigl( 
\Psi_A\,\frac{b_0 X \eta}{L_0}\,(\Psi_B\,\phi_C) \Bigr) \,\bigr\rangle \\
& = {}-\frac{1}{2}\,\bigl\langle\,(Q\phi_D\,\eta\phi_E + \eta\phi_D\,Q\phi_E)\,
\frac{b_0}{L_0}\, \Bigl( 
\Psi_A\,\frac{b_0 X}{L_0}\,(\Psi_B\,\phi_C) \Bigr) \,\bigr\rangle
\end{split}
\end{equation}
For the first term on the right-hand side,
we already have one BRST operator acting on $\phi_D$
so that we move the BRST operator from $X$ to act on $\phi_C$.
For the second term on the right-hand side,
we move the BRST operator from $X$ to act on $\phi_C$.
Then we move $\eta$ acting on $\phi_D$
and move the other BRST operator from $\phi_E$ to $\phi_D$.
The upshot is
\begin{align}
\mathcal{A}_{FFBBB}^{(2P)(d)}=&\
\bigl\langle\, Q\phi_D\,\eta\phi_E
\frac{b_0}{L_0}\, \Bigl( \Psi_A\,\frac{b_0 \Xi}{L_0}\,(\Psi_B\,Q\phi_C) \Bigr) \,\bigr\rangle
\nonumber\\
&
+\frac{1}{2}\,\bigl\langle\,\phi_D\,\phi_E\,\frac{b_0}{L_0}\,
(\Psi_A\,\Psi_B\,Q\phi_C)\,\bigr\rangle
-\frac{1}{2}\,\bigl\langle\,(Q\phi_D\,\eta\phi_E + \eta\phi_D\,Q\phi_E)\,
\frac{b_0}{L_0}\,( \Psi_A\,\Xi\,(\Psi_B\,\phi_C) ) \,\bigr\rangle
\nonumber\\
&
-\frac{1}{2}\,\bigl\langle\,\Psi_B\,Q\phi_C\,\frac{b_0}{L_0}\,
( \phi_D\,\phi_E\,\Psi_A)\,\bigr\rangle
+\frac{1}{2}\,\bigl\langle\,\Psi_B\,\phi_C\,\frac{b_0 \Xi}{L_0}\,( 
(Q\phi_D\,\eta\phi_E + \eta\phi_D\,Q\phi_E)\,\Psi_A ) \,\bigr\rangle\,.
\label{(2P)(d)-FFBBB}
\end{align}

The diagram $(2P)(e)$ in figure~\ref{5point-2propagator}
consists of one cubic vertex from $S_{NS}^{(1)}$, two cubic vertices from $S_{R}^{(1)}$,
one NS propagator, and one Ramond propagators. It is calculated as follows:
\begin{equation}
\begin{split}
 \mathcal{A}_{FFBBB}^{(2P)(e)} & =
(-\, \bigl\langle\,\phi_E\,\Psi_A\ \ 
\overbracket[0.5pt]{\!\!\! \Psi\,\bigr\rangle)
(-\, \bigl\langle\, \Psi\!\!\!}\ \ \Psi_B\ \ 
\overbracket[0.5pt]{\!\!\! \phi\,\bigr\rangle)
\Big(-\,\frac{1}{2}\, \bigl\langle\, \phi\!\!\!}\ \
(Q\phi_C\,\eta\phi_D + \eta\phi_C\,Q\phi_D)\,\bigr\rangle
\Big) \\
& = \frac{1}{2}\,\bigl\langle\,\phi_E\,\Psi_A\,\frac{b_0 X \eta}{L_0}\, \Bigl( 
\Psi_B\,\frac{\xi_0 b_0}{L_0}\,
(Q\phi_C\,\eta\phi_D + \eta\phi_C\,Q\phi_D) \Bigr) \,\bigr\rangle \,.
\end{split}
\end{equation}
The structure of this diagram is similar to that of the diagram $(2P)(d)$,
and we can transform the contribution $\mathcal{A}_{FFBBB}^{(2P)(e)}$ from this diagram
as follows:
\begin{align}
\mathcal{A}_{FFBBB}^{(2P)(e)}\
=&\
\bigl\langle\,\eta\phi_E\,\Psi_A\,\frac{b_0 \Xi}{L_0}\, \Bigl(
\Psi_B\,\frac{b_0}{L_0}\,(Q\phi_C\,Q\phi_D) \Bigr) \,\bigr\rangle
\nonumber\\
&\
+ \frac{1}{2}\,\bigl\langle\,\eta\phi_E\,\Psi_A\,\frac{b_0 \Xi}{L_0}\, ( 
\Psi_B\,(Q\phi_C\,\phi_D - \phi_C\,Q\phi_D)) \,\bigr\rangle
\nonumber\\
&\
+ \frac{1}{2}\,\bigl\langle\,(Q\phi_C\,\phi_D - \phi_C\,Q\phi_D)\,
\frac{b_0}{L_0}\, ( 
\Xi\,(\eta\phi_E\,\Psi_A)\,\Psi_B ) \,\bigr\rangle\,.
\label{(2P)(e)-FFBBB}
\end{align}

\subsection{Contributions from diagrams with one propagator}
\label{1p}
\begin{figure}[htb]\label{5p1p}
  \begin{center}
    \begin{tabular}{c}
    
      \begin{minipage}{0.30\hsize}
        \begin{center}
          \includegraphics[clip, width=3.5cm]{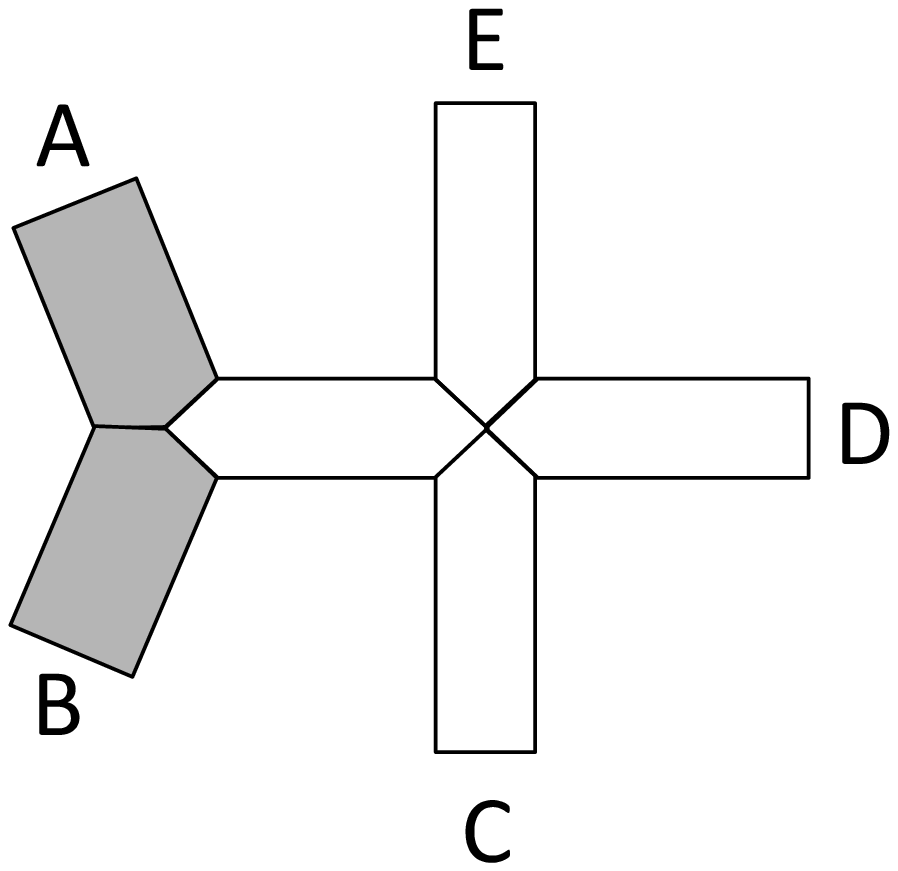}
          \hspace{1.6cm} $(1P)(a)$
                  \end{center}
      \end{minipage}
      \begin{minipage}{0.30\hsize}
        \begin{center}
          \includegraphics[clip, width=3.5cm]{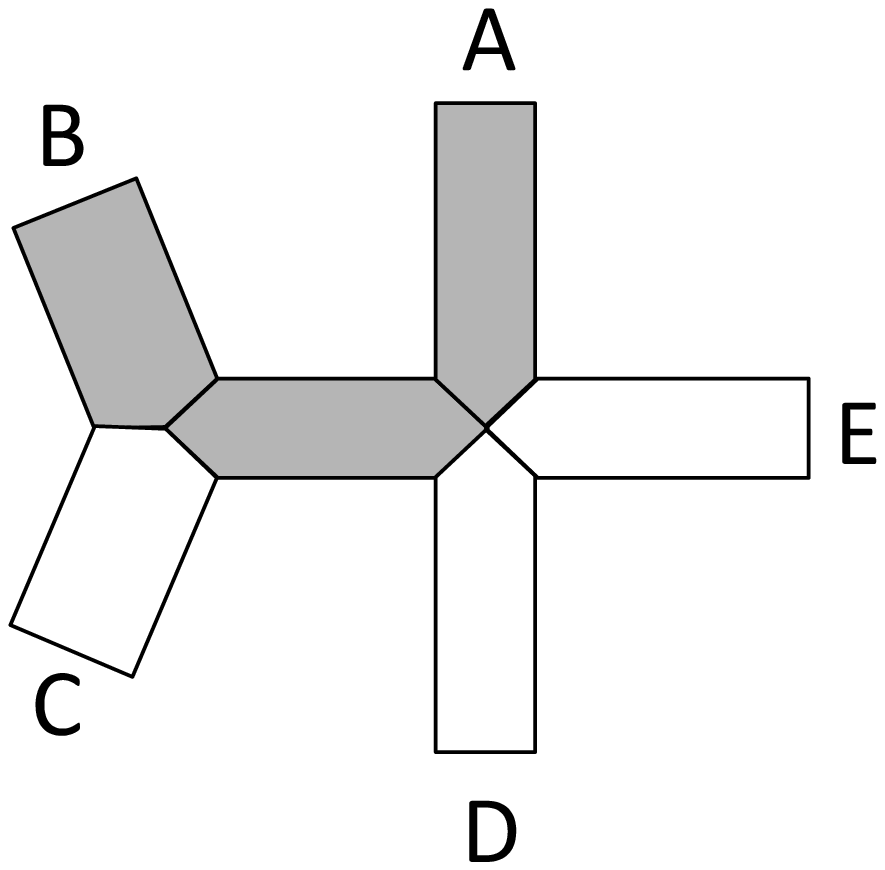}
          \hspace{1.6cm}$(1P)(b)$
        \end{center}
      \end{minipage}
      \begin{minipage}{0.30\hsize}
        \begin{center}
          \includegraphics[clip, width=3.5cm]{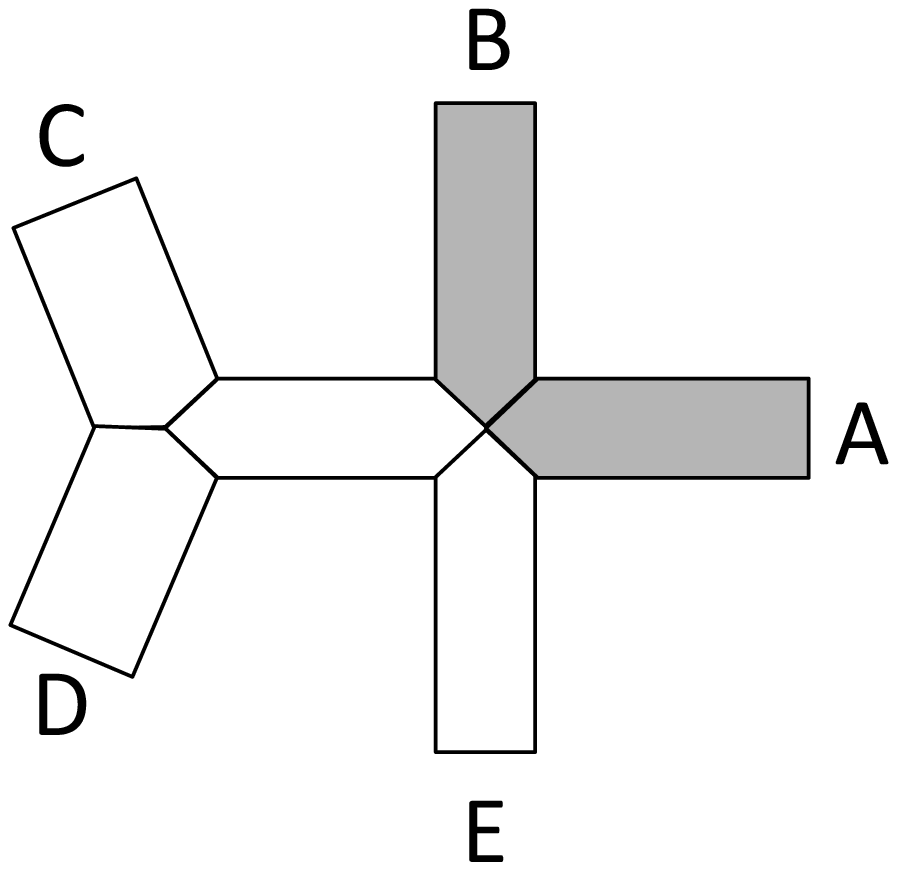}
          \hspace{1.6cm} $(1P)(c)$
        \end{center}
      \end{minipage}
      \\
      \begin{minipage}{0.30\hsize}
        \begin{center}
          \includegraphics[clip, width=3.5cm]{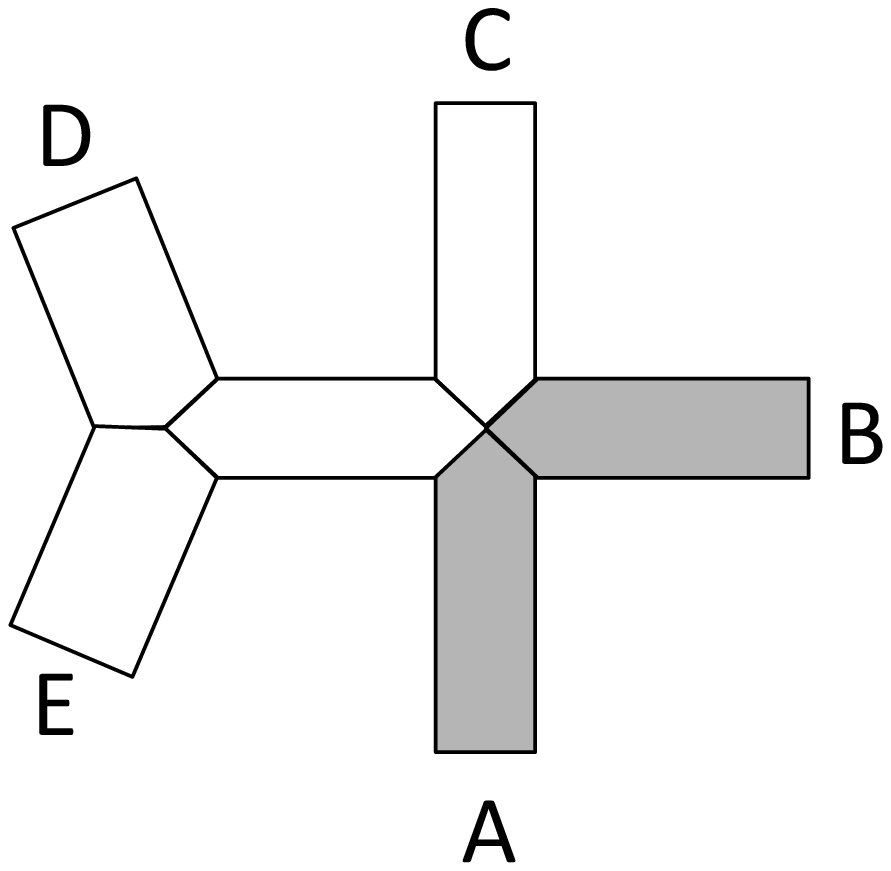}
          \hspace{1.6cm} $(1P)(d)$
        \end{center}
      \end{minipage}
      \begin{minipage}{0.30\hsize}
        \begin{center}
          \includegraphics[clip, width=3.5cm]{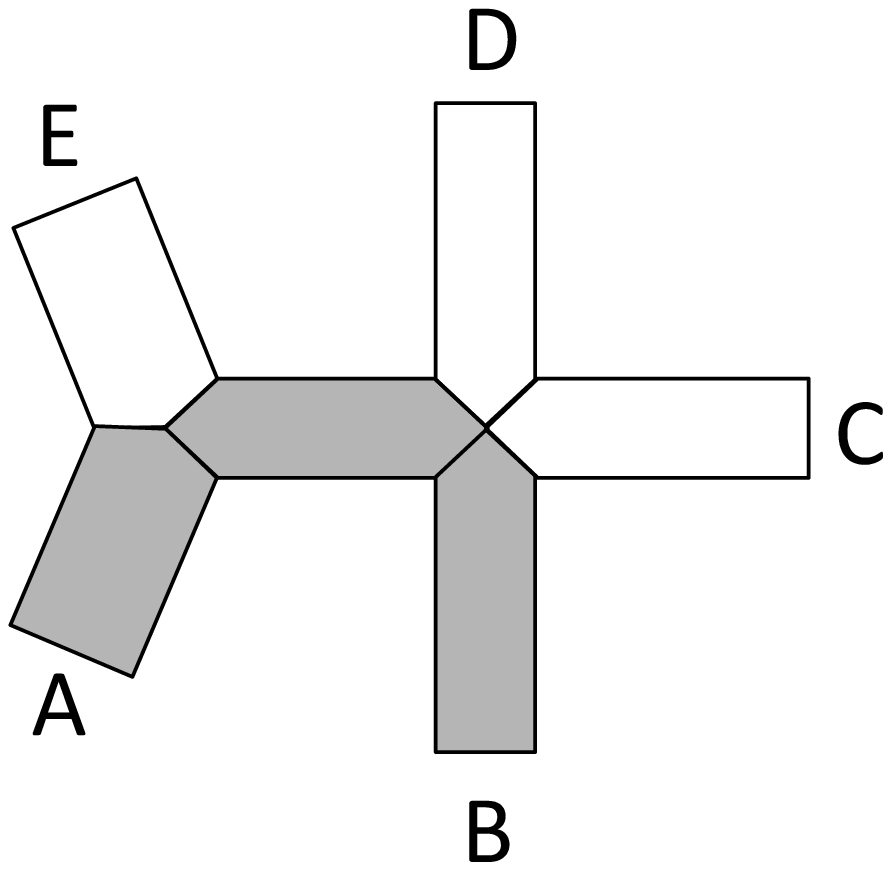}
          \hspace{1.6cm} $(1P)(e)$
        \end{center}
      \end{minipage}
      
    \end{tabular}
    \caption{Feynman diagrams with one quartic vertex, one cubic vertex, and one propagator for $\mathcal{A}_{FFBBB}$.}
    \label{5point-1propagator}
  \end{center}
\end{figure}
In this subsection we calculate contributions to $\mathcal{A}_{FFBBB}$
from Feynman diagrams with one quartic vertex, one cubic vertex, and one propagator.
Such diagrams are depicted in figure~\ref{5point-1propagator}.

The diagram $(1P)(a)$ in figure~\ref{5point-1propagator}
consists of one quartic vertex from $S_{NS}^{(2)}$,
one cubic vertex from $S_{R}^{(1)}$,
and one NS propagator.
Note that the quartic interaction $S_{NS}^{(2)}$ has the following property:
\begin{align}
 \delta S_{NS}^{(2)}\ =&\
\frac{1}{12}\,\bigl\langle\,\delta\phi\, (
\phi\,(Q\phi\,\eta\phi - \eta\phi\,Q\phi)
+
(Q\phi\,\eta\phi - \eta\phi\,Q\phi)\,\phi
-2\,(Q\phi\,\phi\,\eta\phi - \eta\phi\,\phi\,Q\phi)
\nonumber\\
&\hspace{15mm}
-(Q\eta\phi)\,\phi^2 + 2\phi\,(Q\eta\phi)\,\phi 
- \phi^2\,(Q\eta\phi))\,\bigr\rangle\,.
\end{align}
Using this property, the contribution $\mathcal{A}_{FFBBB}^{(1P)(a)}$ from the diagram $(1P)(a)$
is calculated as
\begin{align}
 \mathcal{A}_{FFBBB}^{(1P)(a)}\ =&\
(-\, \bigl\langle\, \Psi_A\,\Psi_B\ \ 
\overbracket[0.5pt]{\!\!\! \phi\,\bigr\rangle)
\Big(-\,\frac{1}{12}\,\bigl\langle\,\phi\!\!\!}\ \ 
(\phi_C\,(Q\phi_D\,\eta\phi_E - \eta\phi_D\,Q\phi_E)
+(Q\phi_C\,\eta\phi_D - \eta\phi_C\,Q\phi_D)\,\phi_E
\nonumber\\
&\hspace{60mm}
-2\,(Q\phi_C\,\phi_D\,\eta\phi_E - \eta\phi_C\,\phi_D\,Q\phi_E)
)\,\bigr\rangle\Big)
\nonumber\\
=&\
-\frac{1}{12}\,\bigl\langle\,\Psi_A\,\Psi_B\,\frac{\xi_0 b_0}{L_0}\,
(\phi_C\,(Q\phi_D\,\eta\phi_E - \eta\phi_D\,Q\phi_E) 
+ (Q\phi_C\,\eta\phi_D - \eta\phi_C\,Q\phi_D)\,\phi_E
\nonumber\\
&\hspace{60mm}
-2\,(Q\phi_C\,\phi_D\,\eta\phi_E - \eta\phi_C\,\phi_D\,Q\phi_E)
)\,\bigr\rangle\,.
\label{(1P)(a)-FFBBB}
\end{align}

The diagram $(1P)(b)$ in figure~\ref{5point-1propagator}
consists of one quartic vertex with the fermion-fermion-boson-boson ordering from $S_R^{(2)}$,
one cubic vertex from $S_{R}^{(1)}$,
and one Ramond propagator.
It is calculated as
\begin{align}
\mathcal{A}_{FFBBB}^{(1P)(b)}\ =&\
\bigl\langle\,\Psi_B\,\phi_C\ \ 
\overbracket[0.5pt]{\!\!\! \Psi\,\bigr\rangle
\Big(\,\frac{1}{2}\bigl\langle\,\Psi\!\!\!}\ \ 
(\phi_D\,\Xi\,(\eta\phi_E\,\Psi_A) 
+ \eta\phi_D\,\Xi\,(\phi_E\,\Psi_A) )
\,\bigr\rangle\Big)
\nonumber\\
=&\ -\frac{1}{2} \, \bigl\langle\,\Psi_B\,\phi_C \,
\frac{b_0 X \eta}{L_0} \, (\phi_D\,\Xi\,(\eta\phi_E\,\Psi_A) 
+ \eta\phi_D\,\Xi\,(\phi_E\,\Psi_A) )
\,\bigr\rangle\Big) \,.
\end{align}
We move $\eta$ in the Ramond propagator to find
\begin{align}
\mathcal{A}_{FFBBB}^{(1P)(b)}\ =&\
-\,\bigl\langle\,\Psi_B\,\phi_C\,\frac{b_0 X}{L_0}\, (
\eta\phi_D\,\Xi\,(\eta\phi_E\,\Psi_A) ) \,\bigr\rangle
\nonumber\\
&\
-\frac{1}{2}\,\bigl\langle\,\Psi_B\,\phi_C\,\frac{b_0 X}{L_0}\, (
(\phi_D\,\eta\phi_E - \eta\phi_D\,\phi_E)\,\Psi_A ) \,\bigr\rangle \,.
\label{(1P)(b)-FFBBB}
\end{align}

The diagram $(1P)(c)$ in figure~\ref{5point-1propagator}
consists of one quartic vertex with the fermion-fermion-boson-boson ordering from $S_R^{(2)}$,
one cubic vertex from $S_{NS}^{(1)}$,
and one NS propagator.
It is given by
\begin{align}
 \mathcal{A}_{FFBBB}^{(1P)(c)}\ =&\
\Big(\,-\,\frac{1}{2}\, \bigl\langle\, (Q\phi_C\,\eta\phi_D + \eta\phi_C\,Q\phi_D)\ \ 
\overbracket[0.5pt]{\!\!\! \phi\,\bigr\rangle\Big)
\Big(\,-\,\frac{1}{2}\bigl\langle\,\phi\!\!\!}\ \ 
\Xi\,(\eta\phi_E\,\Psi_A)\,\Psi_B \,\bigr\rangle\Big)
\nonumber\\
&\
+ \Big(\,-\,\frac{1}{2}\,\bigl\langle\, (Q\phi_C\,\eta\phi_D + \eta\phi_C\,Q\phi_D)\ \ 
\overbracket[0.5pt]{\!\!\! \phi\,\bigr\rangle\Big)
\Big(\,-\,\frac{1}{2}\bigl\langle\,\eta\phi\!\!\!}\ \ 
\Xi\,(\phi_E\,\Psi_A)\,\Psi_B \,\bigr\rangle\Big) \,.
\end{align}
For the second term on the right-hand side,
we move $\eta$ acting on $\phi$ in the propagator to find
\begin{align}
 \mathcal{A}_{FFBBB}^{(1P)(c)}\ =&\
\frac{1}{2}\,\bigl\langle\,(Q\phi_C\,\eta\phi_D + \eta\phi_C\,Q\phi_D)\,
\frac{\xi_0 b_0}{L_0}\,( \Xi\,(\eta\phi_E\,\Psi_A)\,\Psi_B ) \,\bigr\rangle
\nonumber\\
&\
- \frac{1}{4}\,\bigl\langle\,(Q\phi_C\,\eta\phi_D + \eta\phi_C\,Q\phi_D)\,
\frac{\xi_0 b_0}{L_0}\,( \phi_E\,\Psi_A\,\Psi_B ) \,\bigr\rangle\,.
\label{(1P)(c)-FFBBB}
\end{align}

The diagram $(1P)(d)$ in figure~\ref{5point-1propagator}
consists of one quartic vertex with the fermion-fermion-boson-boson ordering from $S_R^{(2)}$,
one cubic vertex from $S_{NS}^{(1)}$,
and one NS propagator.
It is calculated as in the case of the diagram $(1P)(c)$:
\begin{align}
 \mathcal{A}_{FFBBB}^{(1P)(d)}\ =&\
\Big(-\,\frac{1}{2}\, \bigl\langle\, (Q\phi_D\,\eta\phi_E + \eta\phi_D\,Q\phi_E)\ \
\overbracket[0.5pt]{\!\!\! \phi\,\bigr\rangle\Big)
\Big(\,-\,\frac{1}{2}\bigl\langle\,\phi\!\!\!}\ \ 
\Psi_A\,\Xi\,(\Psi_B\,\eta\phi_C)\,\bigr\rangle\Big)
\nonumber\\
&\
+ \Big(-\,\frac{1}{2}\,\bigl\langle\, (Q\phi_D\,\eta\phi_E + \eta\phi_D\,Q\phi_E)\ \
\overbracket[0.5pt]{\!\!\! \phi\,\bigr\rangle\Big)
\Big(\,-\,\frac{1}{2}\bigl\langle\,\eta\phi\!\!\!}\ \ 
\Psi_A\,\Xi\,(\Psi_B\,\phi_C)\,\bigr\rangle\Big)
\nonumber\\
=&\
\frac{1}{2}\,\bigl\langle\,(Q\phi_D\,\eta\phi_E + \eta\phi_D\,Q\phi_E)\,
\frac{\xi_0 b_0}{L_0}\,( \Psi_A\,\Xi(\Psi_B\,\eta\phi_C) ) \,\bigr\rangle \nonumber \\
&+\frac{1}{4}\,\bigl\langle\,(Q\phi_D\,\eta\phi_E + \eta\phi_D\,Q\phi_E)\,
\frac{\xi_0 b_0}{L_0}\,( \Psi_A\,\Psi_B\,\phi_C) \,\bigr\rangle.
\label{(1P)(d)-FFBBB}
\end{align}

The diagram $(1P)(e)$ in figure~\ref{5point-1propagator}
consists of one quartic vertex with the fermion-fermion-boson-boson ordering from $S_R^{(2)}$,
one cubic vertex from $S_{R}^{(1)}$,
and one Ramond propagator.
It is calculated as in the case of the diagram $(1P)(b)$:
\begin{align}
 \mathcal{A}_{FFBBB}^{(1P)(e)}\ =&\
(-\,\bigl\langle\,\phi_E\,\Psi_A\ \ 
\overbracket[0.5pt]{\!\!\! \Psi\,\bigr\rangle)
\Big(\,-\,\frac{1}{2}\bigl\langle\,\Psi\!\!\!}\ \ 
\Xi\,(\Psi_B\,\eta\phi_C)\,\phi_D\,\bigr\rangle\Big)
\nonumber\\
&\
+(-\,\bigl\langle\,\phi_E\,\Psi_A\ \ 
\overbracket[0.5pt]{\!\!\! \Psi\,\bigr\rangle)
\Big(\,\frac{1}{2}\bigl\langle\,\Psi\!\!\!}\ \ 
\Xi\,(\Psi_B\,\phi_C)\,\eta\phi_D\,\bigr\rangle\Big)
\nonumber\\
=&\
-\,\bigl\langle\,\eta\phi_E\,\Psi_A\,\frac{b_0 X}{L_0}\,( 
\Xi\,(\Psi_B\,\phi_C)\,\eta\phi_D ) \,\bigr\rangle
-\,\frac{1}{2}\,\bigl\langle\,\eta\phi_E\,\Psi_A\,\frac{b_0 X}{L_0}\,
(\Psi_B\,\phi_C\,\phi_D)\,\bigr\rangle\,.
\label{(1P)(e)-FFBBB}
\end{align}

\subsection{Contributions from diagrams without propagators}\label{np}

Let us finally consider contributions to~$\mathcal{A}_{FFBBB}$ from Feynman diagrams
with one quintic vertex.
We decompose the quintic interaction~$S_{R}^{(3)}$ as follows:
\begin{equation}
S_{R}^{(3)}=S_{FFBBB}+S_{FBFBB} \,,
\end{equation}
where $S_{FFBBB}$ consists of terms with the fermion-fermion-boson-boson-boson ordering 
given by
\begin{align}
S_{FFBBB}
=& -\frac{1}{3}\,\bigl\langle \, \phi\,\Psi\,\Xi(\,\Xi(\,\Psi\,\eta\phi\,)\,\eta\phi\,) \, \bigr\rangle
-\frac{1}{3}\,\bigl\langle \,\phi\,\Xi(\,\eta\phi\,\Xi(\,\eta\phi\,\Psi\,)\,)\,\Psi \, \bigr\rangle
-\frac{1}{3}\,\bigl\langle \, \phi\,\Xi(\,\eta\phi\,\Psi\,)\,\Xi(\,\Psi\,\eta\phi\,)\, \bigr\rangle \nonumber \\
&+\frac{1}{6}\,\bigl\langle \, \phi\,\Psi\,\Xi(\,\Psi\,\eta\phi\,\phi\,)\, \bigr\rangle
-\frac{1}{6}\,\bigl\langle \,\phi\,\Psi\,\Xi(\,\Psi\,\phi\,\eta\phi\,) \, \bigr\rangle \nonumber \\
&+\frac{1}{6}\bigl\langle \,\phi\,\Xi(\,\eta\phi\,\phi\,\Psi)\,\Psi \, \bigr\rangle 
-\frac{1}{6}\bigl\langle \phi\,\Xi(\,\phi\,\eta\phi\,\Psi)\,\Psi \, \bigr\rangle
\end{align}
and $S_{FBFBB}$ consists of terms with the fermion-boson-fermion-boson-boson ordering 
given by
\begin{align}
S_{FBFBB}
=&-\frac{1}{3}\, \bigl\langle\, \phi\,\Psi\,\Xi(\, \eta\phi\,\Xi(\,\eta\phi\,\Psi\,)\,)\, \bigr\rangle
-\frac{1}{3}\, \bigl\langle\,\phi\,\Psi\,\Xi(\,\eta\phi\,\Xi(\,\Psi\,\eta\phi\,)\,) \,\bigr\rangle
-\frac{1}{3}\, \bigl\langle\,\phi\,\Psi\,\Xi(\,\Xi(\eta\phi\,\Psi\,)\,\eta\phi\,) \,\bigr\rangle \nonumber\\
&-\frac{1}{3}\, \bigl\langle\,\phi\,\Xi(\,\eta\phi\,(\Xi(\,\Psi\,\eta\phi\,)\,)\,\Psi \,\bigr\rangle
-\frac{1}{3}\, \bigl\langle\, \phi\,\Xi(\,\Xi(\,\eta\phi\,\Psi\,)\,\eta\phi\,)\,\Psi\,\bigr\rangle
-\frac{1}{3}\, \bigl\langle\,\phi\,\Xi(\, \Xi(\,\Psi\,\eta\phi\,)\,\eta\phi\,) \,\Psi\,\bigr\rangle \nonumber \\
&-\frac{1}{3}\, \bigl\langle\,\phi\,\Xi(\,\eta\phi\,\Psi\,)\,\Xi(\eta\phi\Psi\,)\,\,\bigr\rangle
-\frac{1}{3}\, \bigl\langle\, \phi\,\Xi(\,\Psi\,\eta\phi\,)\,\Xi(\,\eta\phi\,\Psi\,) \,\bigr\rangle
-\frac{1}{3}\, \bigl\langle\, \phi\,\Xi(\,\Psi\,\eta\phi\,)\,\Xi(\,\Psi\,\eta\phi\,)\,\bigr\rangle \nonumber \\
&+\frac{1}{6}\, \bigl\langle\,\phi\,\Psi\,\Xi(\,\eta\phi\,\phi\,\Psi\,) \,\bigr\rangle
-\frac{1}{6}\, \bigl\langle\,\phi\,\Psi\,\Xi(\,\phi\,\eta\phi\,\Psi\,) \,\bigr\rangle \nonumber \\
&+\frac{1}{6}\, \bigl\langle\,\phi\,\Xi(\,\Psi\,\eta\phi\,\phi\,)\,\Psi \,\bigr\rangle 
-\frac{1}{6}\, \bigl\langle\,\phi\,\Xi(\,\Psi\,\phi\,\eta\phi\,)\,\Psi \,\bigr\rangle \, . \label{quintic-FBFBB}
\end{align}

\begin{figure}[H]
  \begin{center}
    \begin{tabular}{c}
    
      \begin{minipage}{0.33\hsize}
        \begin{center}
          \includegraphics[clip, width=3.5cm]{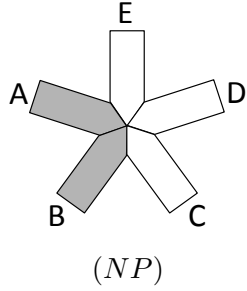}
          \hspace{1.6cm} $(NP)$
                  \end{center}
      \end{minipage}
        \end{tabular}
    \caption{Feynman diagram with a quintic vertex for $\mathcal{A}_{FFBBB}$.}
    \label{5point-nopropagator}
  \end{center}
\end{figure}
The Feynman diagram made of a quintic vertex with the fermion-fermion-boson-boson-boson ordering
is depicted in figure~\ref{5point-nopropagator}. 
The contribution from this diagram denoted by~$\mathcal{A}_{FFBBB}^{(NP)}$ is calculated from~$S_{FFBBB}$
as follows:
\begin{align}
\mathcal{A}_{FFBBB}^{(NP)} 
=& -\frac{1}{3}\,\bigl\langle \, \phi_E\,\Psi_A\,\Xi(\,\Xi(\,\Psi_B\,\eta\phi_C\,)\,\eta\phi_D\,) \, \bigr\rangle \nonumber \\
&-\frac{1}{3}\,\bigl\langle \,\phi_C\,\Xi(\,\eta\phi_D\,\Xi(\,\eta\phi_E\,\Psi_A\,)\,)\,\Psi_B \, \bigr\rangle
-\frac{1}{3}\,\bigl\langle \, \phi_D\,\Xi(\,\eta\phi_E\,\Psi_A\,)\,\Xi(\,\Psi_B\,\eta\phi_C\,)\, \bigr\rangle \nonumber \\
&+\frac{1}{6}\,\bigl\langle \, \phi_E\,\Psi_A\,\Xi(\,\Psi_B\,\eta\phi_C\,\phi_D\,)\, \bigr\rangle
-\frac{1}{6}\,\bigl\langle \,\phi_E\,\Psi_A\,\Xi(\,\Psi_B\,\phi_C\,\eta\phi_D\,) \, \bigr\rangle \nonumber \\
&+\frac{1}{6}\bigl\langle \,\phi_C\,\Xi(\,\eta\phi_D\,\phi_E\,\Psi_A\,)\,\Psi_B \, \bigr\rangle 
-\frac{1}{6}\bigl\langle \phi_C\,\Xi(\,\phi_D\,\eta\phi_E\,\Psi_A\,)\,\Psi_B \, \bigr\rangle.
\end{align}
We transform each term such that the first string field is $\Psi_A$
and move $\eta$ acting on $\phi_C$ to find
\begin{align}
\mathcal{A}_{FFBBB}^{(NP)} 
=&\
-\,\frac{1}{6}\,\bigl\langle\,\Psi_A\,\Psi_B\,
\phi_C\,\phi_D\,\phi_E\,\bigr\rangle
+\,\frac{1}{2}\,\bigl\langle\,\Psi_A\,\Xi\,(\Psi_B\,\phi_C)\,
(\phi_D\,\eta\phi_E - \eta\phi_D\,\phi_E)\,\bigr\rangle
\nonumber\\
&\
-\,\frac{1}{2}\,\bigl\langle\,\Psi_A\,\Xi\,(\Psi_B\,\phi_C\,
\phi_D)\,\eta\phi_E\,\bigr\rangle
-\,\bigl\langle\,\Psi_A\,\Xi\,(\Xi\,(\Psi_B\,\phi_C)\,
\eta\phi_D)\,\eta\phi_E\,\bigr\rangle\,.
\label{FFBBB NP}
\end{align}

\subsection{Contributions from all diagrams}
\label{cancellation}

In subsection~\ref{2p} we calculated contributions to the amplitude~$\mathcal{A}_{FFBBB}$ from diagrams
with two propagators shown in figure~\ref{5point-2propagator}.
We moved two BRST operators to act on $\phi_C$ and $\phi_D$
so that we obtain $X_0 \Phi_C$ and $X_0 \Phi_D$ of the amplitude~$\mathcal{A}_{FFBBB}^{\rm WS}$
in the world-sheet theory when we reduce the expressions to the small Hilbert space.
We found that all the terms of~$\mathcal{A}_{FFBBB}^{\rm WS}$ in~\eqref{A_FFBBB^WS}
are reproduced,
but we also found extra terms with one propagator are generated
when we moved the BRST operators.
In subsection~\ref{1p} we calculated contributions to the amplitude~$\mathcal{A}_{FFBBB}$ from diagrams
with one propagator shown in figure~\ref{5point-1propagator}.
We combine the extra terms with one propagator from the diagrams with two propagators
and the contributions from the diagram with one propagator
and then decompose them according to the location of the propagator as follows:
\begin{equation}
\begin{split}
& ( \mathcal{A}_{FFBBB}^{(2P)} -\mathcal{A}_{FFBBB}^{\rm WS} )
+\mathcal{A}_{FFBBB}^{(1P)}
 \\
& = \mathcal{A}_{FFBBB}^{(AB|CDE)}
+\mathcal{A}_{FFBBB}^{(BC|DEA)}
+\mathcal{A}_{FFBBB}^{(CD|EAB)}
+\mathcal{A}_{FFBBB}^{(DE|ABC)}
+\mathcal{A}_{FFBBB}^{(EA|BCD)} \,,
\end{split}
\end{equation}
where
\begin{equation}
\begin{split}
\mathcal{A}_{FFBBB}^{(2P)}
& = \mathcal{A}_{FFBBB}^{(2P)(a)}
+\mathcal{A}_{FFBBB}^{(2P)(b)}
+\mathcal{A}_{FFBBB}^{(2P)(c)}
+\mathcal{A}_{FFBBB}^{(2P)(d)}
+\mathcal{A}_{FFBBB}^{(2P)(e)} \,, \\
\mathcal{A}_{FFBBB}^{(1P)}
& = \mathcal{A}_{FFBBB}^{(1P)(a)}
+\mathcal{A}_{FFBBB}^{(1P)(b)}
+\mathcal{A}_{FFBBB}^{(1P)(c)}
+\mathcal{A}_{FFBBB}^{(1P)(d)}
+\mathcal{A}_{FFBBB}^{(1P)(e)} \,.
\end{split}
\end{equation}
We denoted the location of the propagator by a vertical slash
in the superscripts such as $(AB|CDE)$.

Let us first consider the contribution~$\mathcal{A}_{FFBBB}^{(AB|CDE)}$.
The sources of~$\mathcal{A}_{FFBBB}^{(AB|CDE)}$
are two of the four terms with one propagator generated from the diagram $(2P)(a)$,
two of the four terms with one propagator generated from the diagram $(2P)(c)$,
and the whole contribution from the one-propagator diagram $(1P)(a)$.
We collect
the third and fourth terms of~$\mathcal{A}_{FFBBB}^{(2P)(a)}$ in~\eqref{(2P)(a)-FFBBB}, 
the fourth and fifth terms of~$\mathcal{A}_{FFBBB}^{(2P)(c)}$ in~\eqref{(2P)(c)-FFBBB},
and $\mathcal{A}_{FFBBB}^{(1P)(a)}$ in~\eqref{(1P)(a)-FFBBB}
to obtain
\begin{align}
& \mathcal{A}_{FFBBB}^{(AB|CDE)} \nonumber \\
= &\
-\frac{1}{4}\,\bigl\langle\,\Psi_A\,\Psi_B\,\frac{\xi_0 b_0}{L_0}\, ( 
\phi_C\,(Q\phi_D\,\eta\phi_E + \eta\phi_D\,Q\phi_E) ) \,\bigr\rangle
+ \frac{1}{2}\,\bigl\langle\,\Psi_A\,\Psi_B\,\frac{\xi_0 b_0}{L_0}\, ( 
Q\phi_C\,\eta(\phi_D\,\phi_E) ) \,\bigr\rangle
\nonumber\\
&\
- \frac{1}{2}\,\bigl\langle\,\Psi_A\,\Psi_B\,
\frac{\xi_0 b_0}{L_0}\, ( 
(Q\phi_C\,\phi_D - \phi_C\,Q\phi_D)\,\eta\phi_E ) \,\bigr\rangle
-\frac{1}{4}\,\bigl\langle\,\Psi_A\,\Psi_B\,\frac{\xi_0 b_0}{L_0}\, ( 
(Q\phi_C\,\eta\phi_D + \eta\phi_C\,Q\phi_D)\,\phi_E ) \,\bigr\rangle
\nonumber\\
&\
-\frac{1}{12}\,\bigl\langle\,\Psi_A\,\Psi_B\,\frac{\xi_0 b_0}{L_0}\,
(\phi_C\,(Q\phi_D\,\eta\phi_E - \eta\phi_D\,Q\phi_E) 
+ (Q\phi_C\,\eta\phi_D - \eta\phi_C\,Q\phi_D)\,\phi_E
\nonumber\\
&\hspace{100mm}
-2\,(Q\phi_C\,\phi_D\,\eta\phi_E - \eta\phi_C\,\phi_D\,Q\phi_E)
)\,\bigr\rangle.
\end{align} 
The sum of these terms does not vanish,
but it reduces to a form with no propagators:
\begin{equation}
\begin{split}
\mathcal{A}_{FFBBB}^{(AB|CDE)}
& =\ \frac{1}{6}\,\bigl\langle\,\Psi_A\,\Psi_B\,
\frac{\xi_0 b_0}{L_0}\, Q \eta ( \phi_C\,\phi_D\,\phi_E ) \,\bigr\rangle \\
& =\ \frac{1}{6}\,\bigl\langle\,\Psi_A\,\Psi_B\,\phi_C\,\phi_D\,\phi_E\,\bigr\rangle\,. 
\end{split}
\end{equation}

The sources of the contribution~$\mathcal{A}_{FFBBB}^{(BC|DEA)}$
are the diagrams~$(2P)(b)$, $(2P)(d)$, and~$(1P)(b)$.
We collect relevant terms in~\eqref{(2P)(b)-FFBBB}, \eqref{(2P)(d)-FFBBB}, and~\eqref{(1P)(b)-FFBBB}
to find
\begin{align}
\mathcal{A}_{FFBBB}^{(BC|DEA)} = &\ 
\bigl\langle\,\Psi_B\,Q\phi_C\,\frac{b_0 \Xi}{L_0}\,(\phi_D\,
\eta\phi_E\,\Psi_A)\,\bigr\rangle
+ \bigl\langle\,\Psi_B\,\phi_C\,\frac{b_0 X}{L_0}\,( \eta\phi_D\,\Xi\,(\eta\phi_E\,\Psi_A)) \,\bigr\rangle
\nonumber\\
&\
-\frac{1}{2}\,\bigl\langle\,\Psi_B\,Q\phi_C\,\frac{b_0}{L_0}\,
(\phi_D\,\phi_E\,\Psi_A)\,\bigr\rangle
+\frac{1}{2}\,\bigl\langle\,\Psi_B\,\phi_C\,\frac{b_0 \Xi}{L_0}\, ( 
(Q\phi_D\,\eta\phi_E + \eta\phi_D\,Q\phi_E)\,\Psi_A ) \,\bigr\rangle\,
\nonumber\\
&\
-\,\bigl\langle\,\Psi_B\,\phi_C\,\frac{b_0 X}{L_0}\,( 
\eta\phi_D\,\Xi\,(\eta\phi_E\,\Psi_A) ) \,\bigr\rangle
-\frac{1}{2}\,\bigl\langle\,\Psi_B\,\phi_C\,\frac{b_0 X}{L_0}\, ( 
(\phi_D\,\eta\phi_E - \eta\phi_D\,\phi_E)\,\Psi_A ) \,\bigr\rangle \,.
\end{align}
Again the sum of these terms does not vanish,
but it reduces to a form with no propagators:
\begin{align}
\mathcal{A}_{FFBBB}^{(BC|DEA)} = 
-\frac{1}{2}\,\bigl\langle\,\Psi_A\,\Xi\,(\Psi_B\,\phi_C)\,
(\phi_D\,\eta\phi_E - \eta\phi_D\,\phi_E)\,\bigr\rangle\,.
\end{align}

The sources of the contribution~$\mathcal{A}_{FFBBB}^{(CD|EAB)}$
are the diagrams~$(2P)(c)$, $(2P)(e)$, and~$(1P)(c)$.
We collect relevant terms in~\eqref{(2P)(c)-FFBBB}, \eqref{(2P)(e)-FFBBB}, and~\eqref{(1P)(c)-FFBBB}
to find
\begin{align}
\mathcal{A}_{FFBBB}^{(CD|EAB)} = &\ 
\frac{1}{2}\,\bigl\langle\,(Q\phi_C\,\phi_D - \phi_C\,Q\phi_D)\,
\frac{\xi_0 b_0}{L_0}\,(\eta\phi_E\,\Psi_A\,\Psi_B)\,\bigr\rangle
\nonumber\\ 
&\ 
+ \frac{1}{4}\,\bigl\langle\,(Q\phi_C\,\eta\phi_D + \eta\phi_C\,Q\phi_D)\,
\frac{\xi_0 b_0}{L_0}\,(\phi_E\,\Psi_A\,\Psi_B)\,\bigr\rangle
\nonumber\\
&\
+ \frac{1}{2}\,\bigl\langle\,(Q\phi_C\,\phi_D - \phi_C\,Q\phi_D)\,
\frac{b_0}{L_0}\, ( 
\Xi\,(\eta\phi_E\,\Psi_A)\,\Psi_B ) \,\bigr\rangle
\nonumber\\
&\
+ \frac{1}{2}\,\bigl\langle\,(Q\phi_C\,\eta\phi_D + \eta\phi_C\,Q\phi_D)\,
\frac{\xi_0 b_0}{L_0}\,( \Xi\,(\eta\phi_E\,\Psi_A)\,\Psi_B ) \,\bigr\rangle
\nonumber\\
&\
- \frac{1}{4}\,\bigl\langle\,(Q\phi_C\,\eta\phi_D + \eta\phi_C\,Q\phi_D)\,
\frac{\xi_0 b_0}{L_0}\,(\phi_E\,\Psi_A\,\Psi_B)\,\bigr\rangle \,.
\end{align}
This time the sum of these terms turns out to vanish:
\begin{equation}
\mathcal{A}_{FFBBB}^{(CD|EAB)} = 0 \,.
\end{equation}

The sources of the contribution~$\mathcal{A}_{FFBBB}^{(DE|ABC)}$
are the diagrams~$(2P)(a)$, $(2P)(d)$, and~$(1P)(d)$.
We collect relevant terms in~\eqref{(2P)(a)-FFBBB}, \eqref{(2P)(d)-FFBBB}, and~\eqref{(1P)(d)-FFBBB}
to find
\begin{align}
\mathcal{A}_{FFBBB}^{(DE|ABC)} = &\ 
\frac{1}{2}\,\bigl\langle\,\phi_D\,\phi_E\,\frac{b_0}{L_0}\,
(\Psi_A\,\Psi_B\,Q\phi_C)\,\bigr\rangle
-\frac{1}{2}\,\bigl\langle\,(Q\phi_D\,\eta\phi_E + \eta\phi_D\,Q\phi_E)\,
\frac{b_0}{L_0}\,(
\Psi_A\,\Xi\,(\Psi_B\,\phi_C)) \,\bigr\rangle
\nonumber\\
&\
+\frac{1}{4}\,\bigl\langle\,(Q\phi_D\,\eta\phi_E + \eta\phi_D\,Q\phi_E)\,
\frac{\xi_0 b_0}{L_0}\,(\Psi_A\,\Psi_B\,\phi_C)\,\bigr\rangle
+\frac{1}{2}\,\bigl\langle\,\eta(\phi_D\,\phi_E)\,\frac{\xi_0 b_0}{L_0}\,
(\Psi_A\,\Psi_B\,Q\phi_C)\,\bigr\rangle
\nonumber\\
&\
+ \frac{1}{2}\,\bigl\langle\,(Q\phi_D\,\eta\phi_E + \eta\phi_D\,Q\phi_E)\,
\frac{\xi_0 b_0}{L_0}\,( \Psi_A\,\Xi(\Psi_B\,\eta\phi_C) ) \,\bigr\rangle
\nonumber\\
&\
+ \frac{1}{4}\,\bigl\langle\,(Q\phi_D\,\eta\phi_E + \eta\phi_D\,Q\phi_E)\,
\frac{\xi_0 b_0}{L_0}\,(\Psi_A\,\Psi_B\,\phi_C)\,\bigr\rangle \,.
\end{align}
The sum of these terms again turns out to vanish:
\begin{equation}
\mathcal{A}_{FFBBB}^{(DE|ABC)} = 0 \,.
\end{equation}

The sources of the contribution~$\mathcal{A}_{FFBBB}^{(EA|BCD)}$
are the diagrams~$(2P)(b)$, $(2P)(e)$, and~$(1P)(e)$.
We collect relevant terms in~\eqref{(2P)(b)-FFBBB}, \eqref{(2P)(e)-FFBBB}, and~\eqref{(1P)(e)-FFBBB}
to find
\begin{align}
\mathcal{A}_{FFBBB}^{(EA|BCD)} = &\ 
  \frac{1}{2}\,\bigl\langle\,\eta\phi_E\,\Psi_A\,\frac{b_0 \Xi}{L_0}\, (
\Psi_B\,(Q\phi_C\,\phi_D - \phi_C\,Q\phi_D) )\,\bigr\rangle
\nonumber\\
&\
- \bigl\langle\,\eta\phi_E\,\Psi_A\,\frac{b_0}{L_0}\,( \Xi\,(\Psi_B\,Q\phi_C)\,\phi_D )\,\bigr\rangle
+ \bigl\langle\,\eta\phi_E\,\Psi_A\,\frac{b_0 \Xi}{L_0}\,( X\,(\Psi_B\,\phi_C)\,\eta\phi_D ) \,\bigr\rangle
\nonumber\\
&\
-\,\bigl\langle\,\eta\phi_E\,\Psi_A\,\frac{b_0 X}{L_0}\, (
\Xi\,(\Psi_B\,\phi_C)\,\eta\phi_D ) \,\bigr\rangle
-\,\frac{1}{2}\,\bigl\langle\,\eta\phi_E\,\Psi_A\,\frac{b_0 X}{L_0}\,
(\Psi_B\,\phi_C\,\phi_D)\,\bigr\rangle\,.
\end{align}
The sum of these terms does not vanish,
but it reduces to a form with no propagators:
\begin{equation}
\mathcal{A}_{FFBBB}^{(EA|BCD)}
= \frac{1}{2}\,\bigl\langle\,\Psi_A\,\Xi\,(\Psi_B\,\phi_C\,\phi_D)\,\eta\phi_E\,\bigr\rangle
+\,\bigl\langle\,\Psi_A\,\Xi\,\left(\Xi\,(\Psi_B\,\phi_C)\,\eta\phi_D\right)\eta\phi_E\,\bigr\rangle\,.
\end{equation}

To summarize,
we found that the contributions $\mathcal{A}_{FFBBB}^{(CD|EAB)}$
and $\mathcal{A}_{FFBBB}^{(DE|ABC)}$ vanish,
and each of the nonvanishing contributions
$\mathcal{A}_{FFBBB}^{(AB|CDE)}$, $\mathcal{A}_{FFBBB}^{(BC|DEA)}$,
and $\mathcal{A}_{FFBBB}^{(EA|BCD)}$
reduces to a form with no propagators.
The sum of all the terms is given by
\begin{align}
& \mathcal{A}_{FFBBB}^{(AB|CDE)}
+\mathcal{A}_{FFBBB}^{(BC|DEA)}
+\mathcal{A}_{FFBBB}^{(CD|EAB)}
+\mathcal{A}_{FFBBB}^{(DE|ABC)}
+\mathcal{A}_{FFBBB}^{(EA|BCD)}
\nonumber \\
& =
\frac{1}{6}\,\bigl\langle\,\Psi_A\,\Psi_B\,\phi_C\,\phi_D\,\phi_E\,\bigr\rangle
-\frac{1}{2}\,\bigl\langle\,\Psi_A\,\Xi\,(\Psi_B\,\phi_C)\,
(\phi_D\,\eta\phi_E - \eta\phi_D\,\phi_E)\,\bigr\rangle
\nonumber\\
& \quad~
+\,\frac{1}{2}\,\bigl\langle\,\Psi_A\,\Xi\,(\Psi_B\,\phi_C\,\phi_D)\,\eta\phi_E\,\bigr\rangle
+\,\bigl\langle\,\Psi_A\,\Xi\,\left(\Xi\,(\Psi_B\,\phi_C)\,\eta\phi_D\right)\eta\phi_E\,\bigr\rangle\,. 
\end{align}
We find that the sum of these remaining contributions is precisely canceled
by the contribution $\mathcal{A}_{FFBBB}^{(NP)}$ from the diagram without the propagator
shown in figure~\ref{5point-nopropagator}:
\begin{equation}
\mathcal{A}_{FFBBB}^{(AB|CDE)}
+\mathcal{A}_{FFBBB}^{(BC|DEA)}
+\mathcal{A}_{FFBBB}^{(CD|EAB)}
+\mathcal{A}_{FFBBB}^{(DE|ABC)}
+\mathcal{A}_{FFBBB}^{(EA|BCD)}
= {}-\mathcal{A}_{FFBBB}^{(NP)} \,.
\end{equation}
We thus conclude that the amplitude~$\mathcal{A}_{FFBBB}^{\rm WS}$ in the world-sheet theory
is correctly reproduced by the amplitude~$\mathcal{A}_{FFBBB}$
calculated in open superstring field theory:
\begin{equation}
\mathcal{A}_{FFBBB} =\mathcal{A}_{FFBBB}^{\rm WS} \,,
\end{equation}
where
\begin{equation}
\mathcal{A}_{FFBBB}
=\mathcal{A}_{FFBBB}^{(2P)} +\mathcal{A}_{FFBBB}^{(1P)} +\mathcal{A}_{FFBBB}^{(NP)} \,.
\end{equation}
\section{Conclusions and discussion}
We calculated the four-point amplitudes~$\mathcal{A}_{FFBB}$, $\mathcal{A}_{FBFB}$, and $\mathcal{A}_{FFFF}$
and the five-point amplitudes~$\mathcal{A}_{FFBBB}$, $\mathcal{A}_{FBFBB}$, and $\mathcal{A}_{FFFFB}$
and confirmed that the corresponding amplitudes in the world-sheet theory are reproduced.
Our calculations can be interpreted as those for the complete action constructed in~\cite{Kunitomo:2015usa}
or as those for the covariant formulation developed by Sen~\cite{Sen:2015uaa}
with spurious free fields.

For the four-point amplitudes, we elucidated the role of the quartic interaction
in reproducing the amplitudes in the world-sheet theory.
For example, the location of the picture-changing operator is different
in the $s$-channel contribution~$\mathcal{A}_{FFBB}^{(s)}$
and the $t$-channel contribution~$\mathcal{A}_{FFBB}^{(t)}$ from
Feynman diagrams with two cubic vertices and one propagator,
and we found that the contribution~$\mathcal{A}_{FFBB}^{(4)}$ compensates the difference.
This is the generalization of the results for the NS sector in~\cite{Iimori:2013kha}.
A new ingredient is the operator $X$ in the Ramond propagator,
and we found that the quartic interaction $S_R^{(2)}$
not only moves the operator $X_0$ acting on an external state as in figure~\ref{FFBB-xi-1}
but also transmutes $X$ to $X_0$ as in figure~\ref{FFBB-xi-2}. 

For the five-point amplitudes, the correct amplitudes were reproduced
in a more intricate way via the quartic and quintic interactions.
As we mentioned in the introduction,
we do not yet understand why the use of the large Hilbert space is so successful,
and understanding of the mechanism of reproducing the correct amplitudes
in our calculations can be a step forward to this direction.

Another important direction for the generalization
would be to consider one-loop amplitudes.
Since states in the Ramond sector also propagate in the loop,
we need an action including the Ramond sector.
We are now at the starting point for the calculation of loop diagrams
in open superstring field theory.
Since ghosts also propagate in the loop,
gauge fixing is necessary
and we use the Batalin-Vilkovisky formalism
for gauge fixing in string field theory.
The Batalin-Vilkovisky quantization for the WZW-like action in the NS sector is discussed in \cite{Kroyter:2012ni,Torii:2011zz,Torii:2012nj,Berkovits:2012np,Iimori:2015aea,Berkovits}, and it  has turned out to be extremely complicated.
The action constructed in~\cite{Kunitomo:2015usa} inherits the same difficulty
because it uses the WZW-like action in the NS sector.
On the other hand, it is known that the Batalin-Vilkovisky quantization is straightforward
when the action has an $A_\infty$ structure.
Therefore, the action constructed in~\cite{Erler:2016ybs, Konopka:2016grr} has an advantage,
although the action itself is complicated for higher-order interactions.
The issue of the covering of the supermoduli space of super-Riemann surfaces
is even more important for one-loop amplitudes
in the context of open superstring field theory
because it is related to the fundamental question
of whether open superstring field theory is complete as a quantum theory
or whether independent degrees of freedom such as closed string fields
are necessary at the quantum level.

\section*{Acknowledgments}
The work of Y.O. was supported in part by a Grant-in-Aid for Scientific Research (B) No. 25287049 and a Grant-in-Aid for Scientific Research (C) No. 24540254 from the Japan Society for the Promotion of Science (JSPS).
\appendix
\setcounter{equation}{0}
\section{Other five-point amplitudes}
\label{appendix}
In this appendix we present calculations of five-point amplitudes
involving four fermions
and of five-point amplitudes involving two fermions
with the fermion-boson-fermion-boson-boson ordering.
\subsection{Fermion-fermion-fermion-fermion-boson amplitudes}
We begin with five-point amplitudes of four fermions and one boson.
Since disks with four Ramond punctures and one NS puncture have two even moduli and one odd modulus,
we need one insertion of the picture-changing operator.
We choose the operator $X_0$ to act on the NS state
for the amplitude~$\mathcal{A}_{FFFFB}^{{\rm WS}}$ in the world-sheet theory,
and then it is given by
\begin{align}
\mathcal{A}_{FFFFB}^{{\rm WS}}
= &-\llangle \, \Psi_A \, \Psi_B \,  \frac{b_0}{L_0} \, \Bigl( \, \Psi_C \, \frac{b_0}{L_0} \, (\, \Psi_D \, X_0 \Phi_E \, )  \Bigr)\rrangle
-\llangle \, \Psi_B \, \Psi_C \,  \frac{b_0}{L_0} \, \Bigl( \, \Psi_D \, \frac{b_0}{L_0} \, (\, X_0\Phi_E \, \Psi_A \, )  \Bigr)\rrangle \nonumber\\
&-\llangle \, \Psi_C \, \Psi_D \,  \frac{b_0}{L_0} \, \Bigl( \, X_0\Psi_E \, \frac{b_0}{L_0} \, (\, \Psi_A \, \Phi_B \, )  \Bigr)\rrangle 
-\llangle \, \Psi_D \, X_0\Phi_E \,  \frac{b_0}{L_0} \, \Bigl( \, \Psi_A \, \frac{b_0}{L_0} \, (\, \Psi_B \, \Phi_C \, )  \Bigr)\rrangle \nonumber \\
&-\llangle \, X_0\Phi_E \, \Psi_A \,  \frac{b_0}{L_0} \, \Bigl( \, \Psi_B \, \frac{b_0}{L_0} \, (\, \Psi_C \, \Psi_D \, )  \Bigr)\rrangle. \label{WS-FFFFB}
\end{align} 

\subsubsection{Contributions from diagrams with two propagators} \label{FFFFB-2P-diagrams}
\begin{figure}[htbp]
  \begin{center}
    \begin{tabular}{c}
    
      \begin{minipage}{0.30\hsize}
        \begin{center}
          \includegraphics[clip, width=3.5cm]{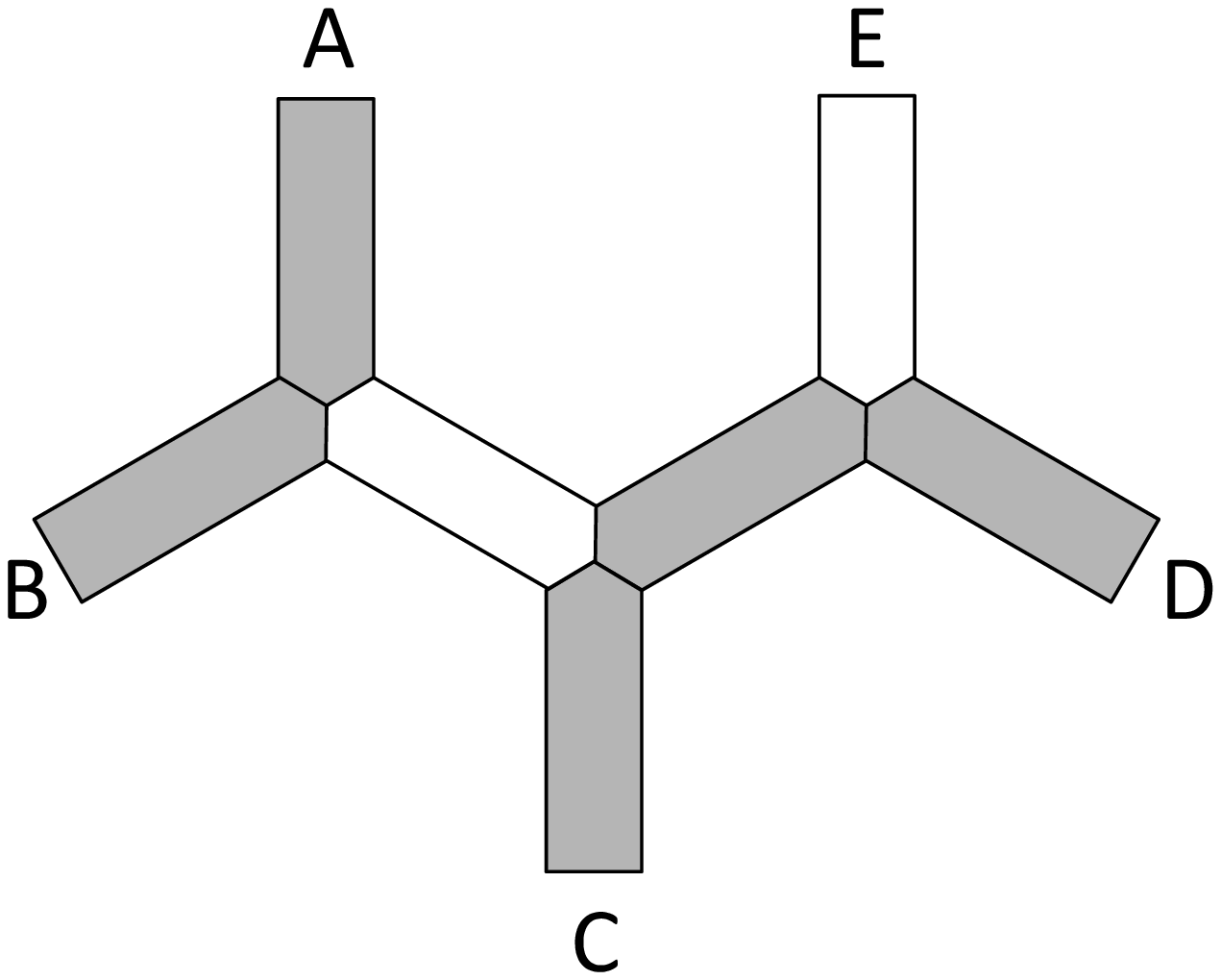}
          \hspace{1.6cm} $(2P)(a)$
                  \end{center}
      \end{minipage}
      \begin{minipage}{0.30\hsize}
        \begin{center}
          \includegraphics[clip, width=3.5cm]{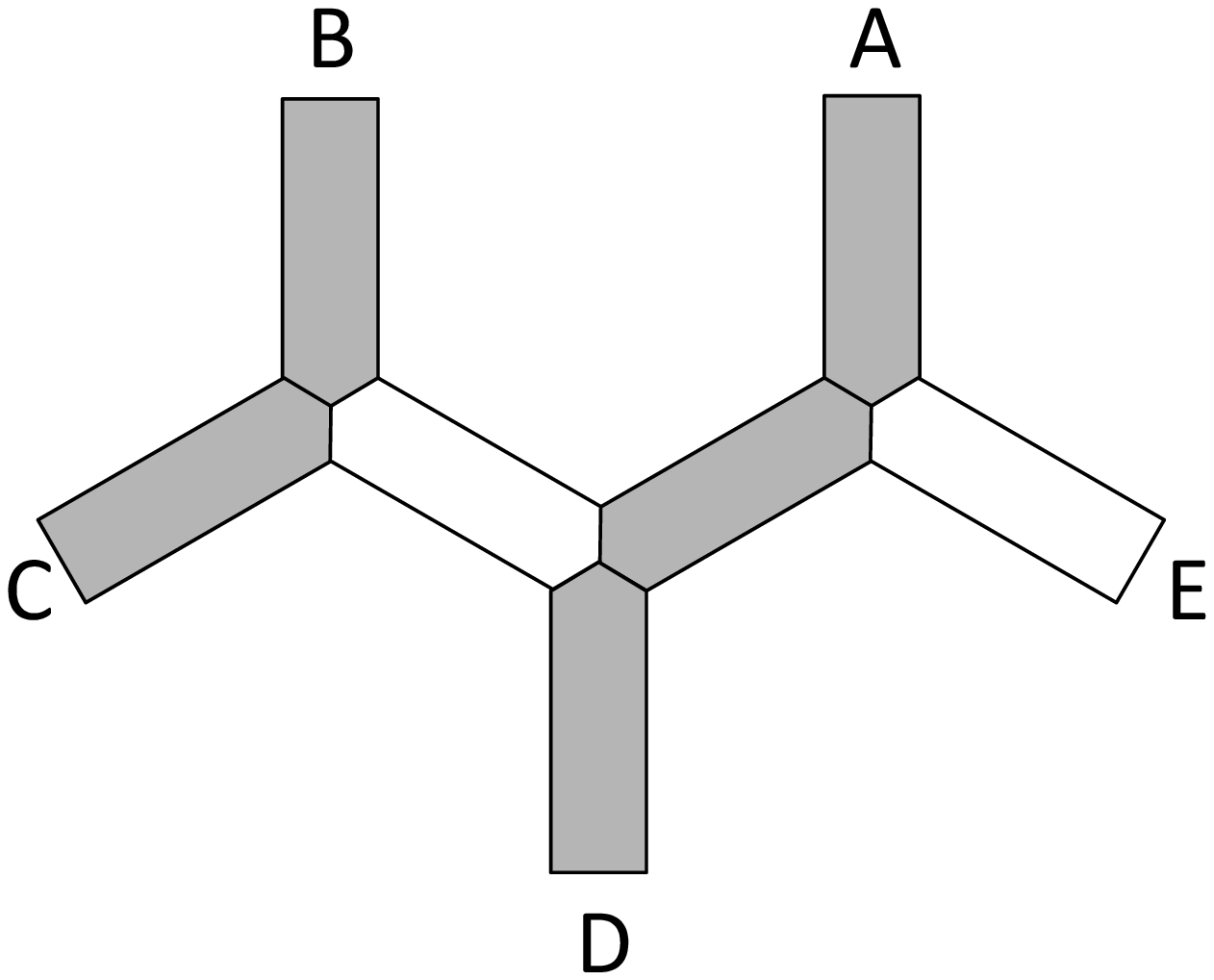}
          \hspace{1.6cm}$(2P)(b)$
        \end{center}
      \end{minipage}
      \begin{minipage}{0.30\hsize}
        \begin{center}
          \includegraphics[clip, width=3.5cm]{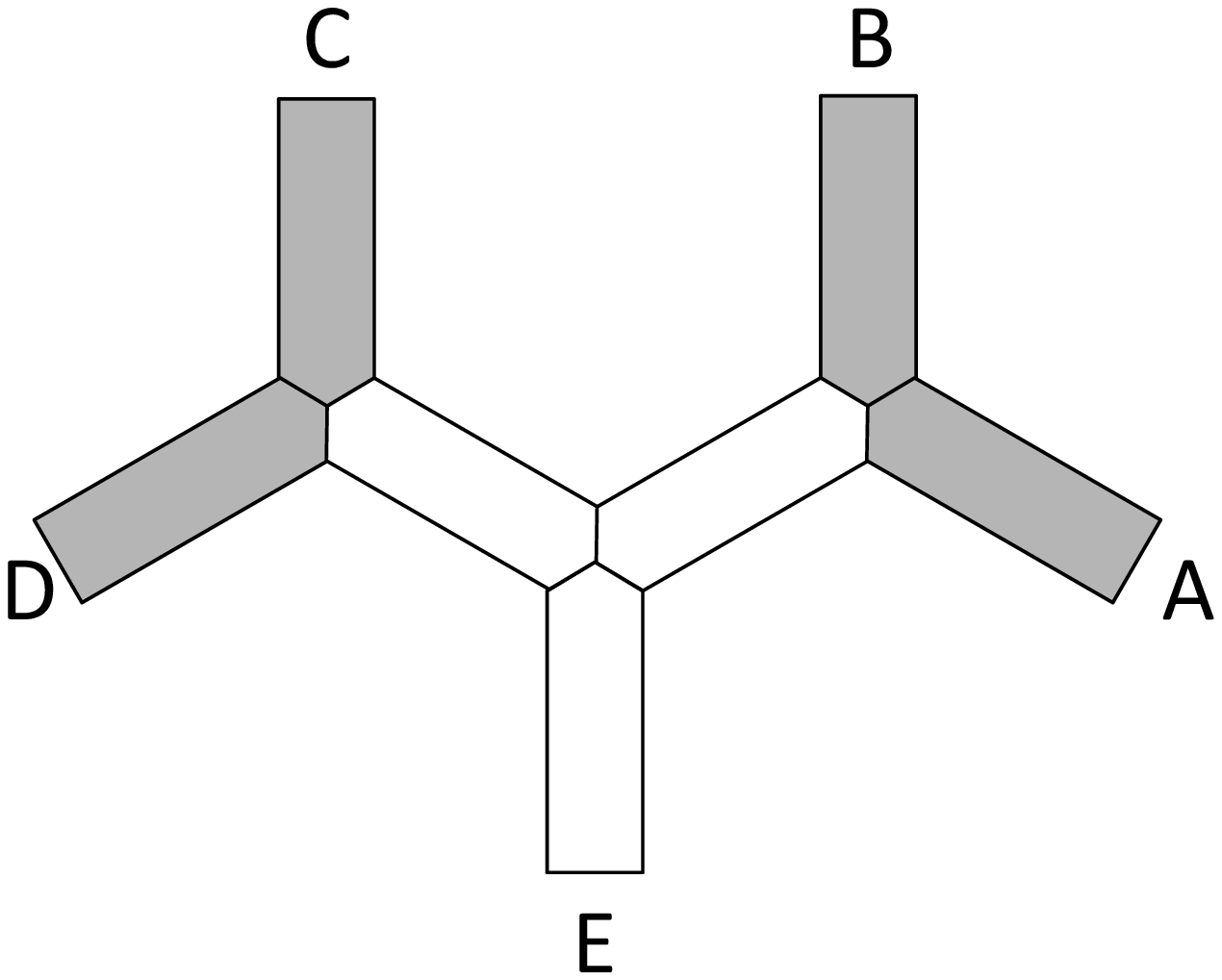}
          \hspace{1.6cm} $(2P)(c)$
        \end{center}
      \end{minipage}
      \\
      \begin{minipage}{0.30\hsize}
        \begin{center}
          \includegraphics[clip, width=3.5cm]{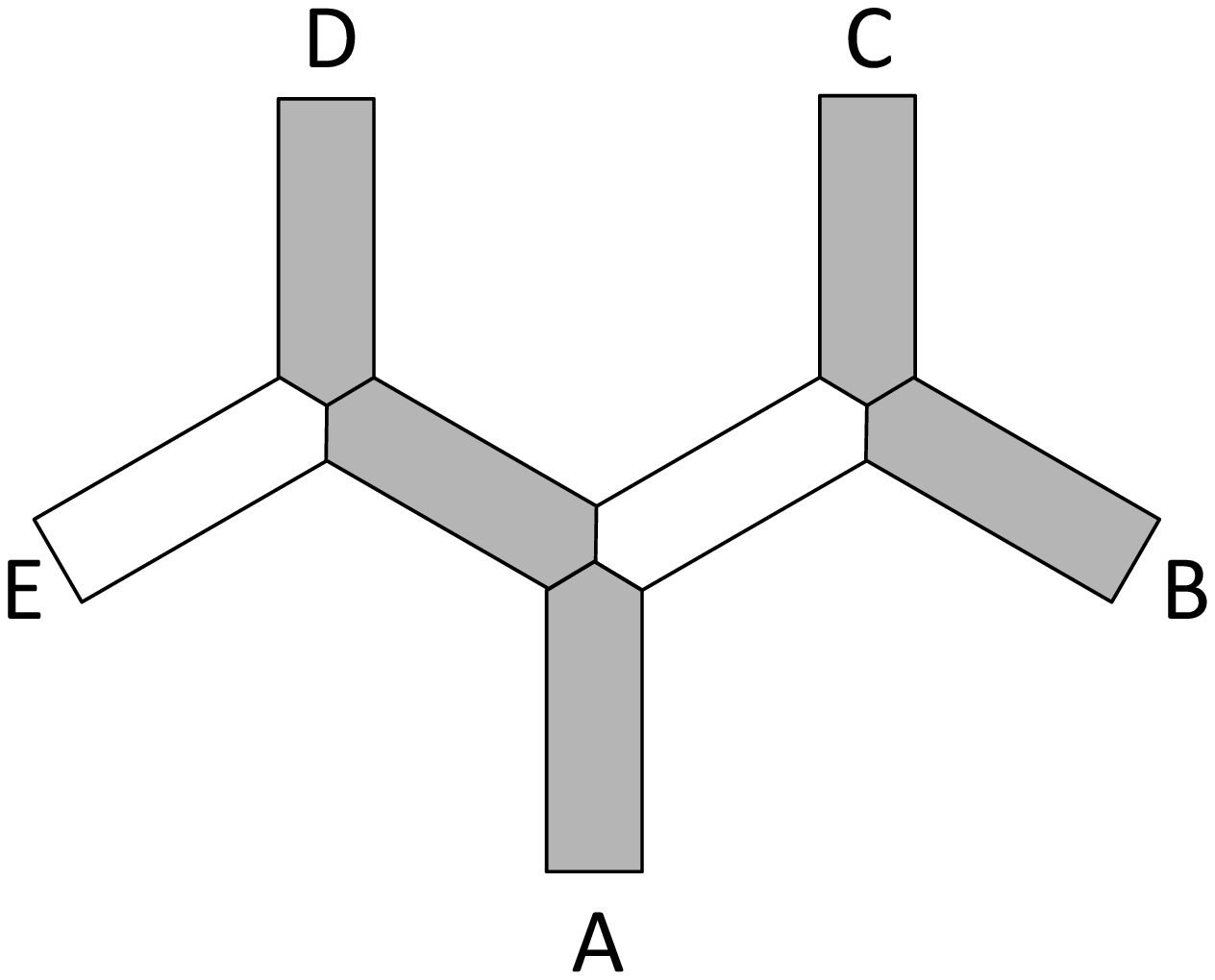}
          \hspace{1.6cm} $(2P)(d)$
        \end{center}
      \end{minipage}
      \begin{minipage}{0.30\hsize}
        \begin{center}
          \includegraphics[clip, width=3.5cm]{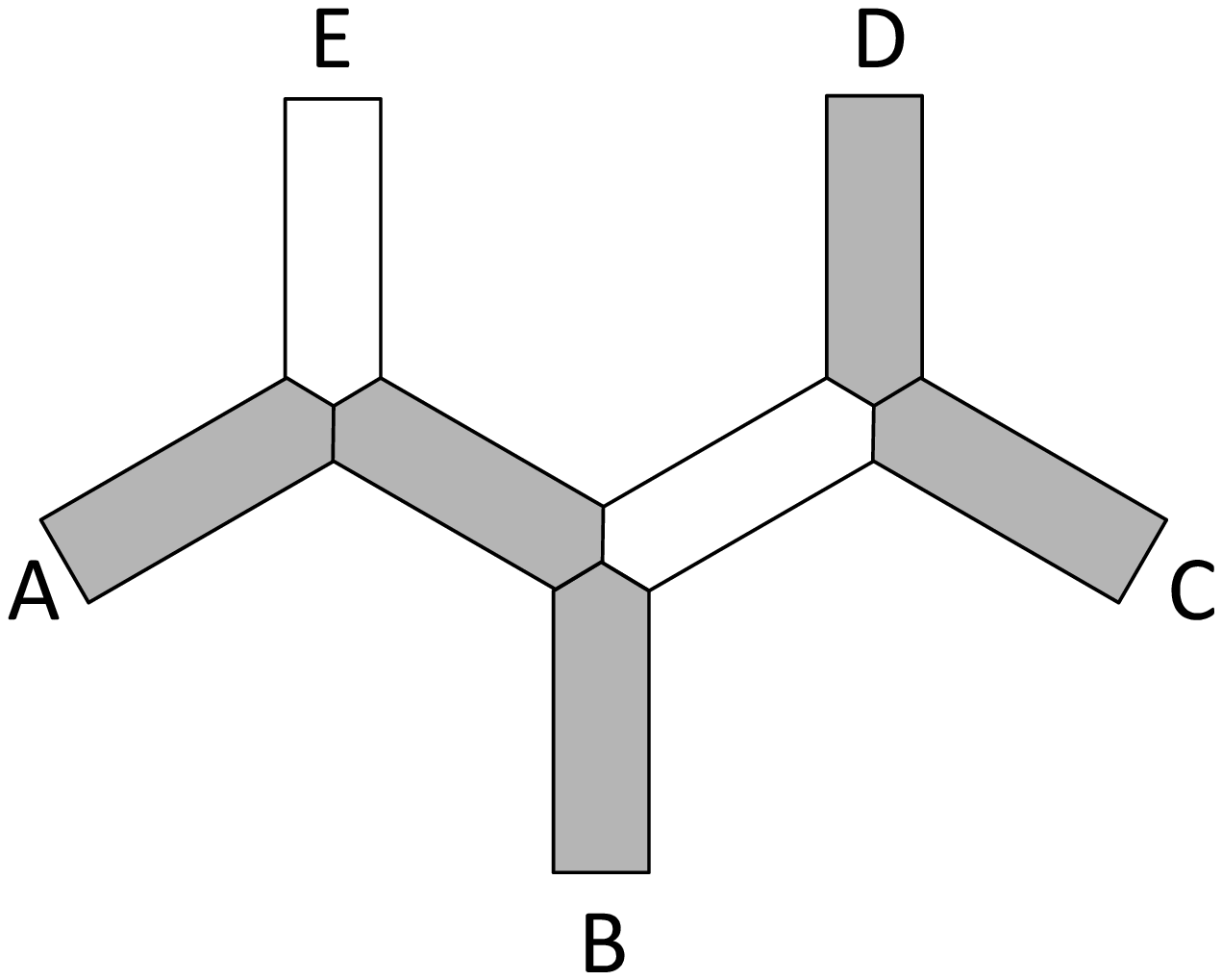}
          \hspace{1.6cm} $(2P)(e)$
        \end{center}
      \end{minipage}
      
    \end{tabular}
    \caption{Feynman diagrams with three cubic vertices and two propagators for $\mathcal{A}_{FFFFB}$.}
    \label{FFFFB-2p}
  \end{center}
\end{figure}

We denote the corresponding amplitude in open string field theory by $\mathcal{A}_{FFFFB}$.
In this subsection we calculate contributions from Feynman diagrams with three cubic vertices
and two propagators.
Such diagrams are depicted in figure~\ref{FFFFB-2p}.

The diagram $(2P)(a)$ in figure \ref{FFFFB-2p} consists of three cubic vertices from $S_{R}^{(1)}$, one NS propagator, and one Ramond propagator.
The contribution~$\mathcal{A}_{FFFFB}^{(2P)(a)}$ from the diagram $(2P)(a)$ is calculated as
\begin{align}
\mathcal{A}_{FFFFB}^{(2P)(a)}\ =&\
\big(- \bigl\langle\,\Psi_A\,\Psi_B\ \ 
\overbracket[0.5pt]{\!\!\! \phi\ \,\bigr\rangle\big) \big(-\bigl\langle\,\ \phi\!\!\!}\ \
\Psi_C\ \
\overbracket[0.5pt]{\!\!\! \Psi\ \,\bigr\rangle\big) \big(-\bigl\langle\,\ \Psi\!\!\!}\ \
\Psi_D\,\phi_E\,\bigr\rangle\big) 
\nonumber\\
=&\
\bigl\langle\,\Psi_A\,\Psi_B\,\frac{\xi_0 b_0}{L_0}\,\Bigl( \,\Psi_C\,\frac{b_0 X \eta}{L_0}\,
( \, \Psi_D\,\phi_E\,) \, \Bigr) \, \bigr\rangle.
\end{align}
In order to compare with the amplitude~$\mathcal{A}_{FFFFB}^{\rm WS}$ in the world-sheet theory,
we want to transmute $X$ in the Ramond propagator to $X_0$ acting on $\phi_E$.
Following the method in \eqref{X-to-X_0},
we transform the contribution~$\mathcal{A}_{FFFFB}^{(2P)(a)}$ as follows:
\begin{align}
\mathcal{A}_{FFFFB}^{(2P)(a)}=&{}-\llangle \, \Psi_A \, \Psi_B \, \frac{b_0}{L_0} \, \Bigl( \, \Psi_C\,\frac{b_0
}{L_0}\,( \, \Psi_D\,X_0\Phi_E\,) \, \Bigr) \, \rrangle \nonumber \\
&{}- \bigl\langle\,\Psi_A\,\Psi_B\,\frac{b_0}{L_0}\,( \, \Psi_C\,\Xi\,(\Psi_D\,\phi_E)\,) \, \bigr\rangle
+ \bigl\langle\,\Psi_D\,\phi_E\,\frac{b_0 \Xi}{L_0}\,(\,\Psi_A\,\Psi_B\,\Psi_C\,)\,\bigr\rangle\,.
\end{align}
Note that two terms with one propagator are generated in this procedure.
We fix the form of such terms with one propagator
by the condition that the combination $\eta \phi_E$ does not appear.
When we encounter $\eta \phi_E$, we move this $\eta$ acting on $\phi_E$
to satisfy the condition.

The diagram $(2P)(b)$  in figure \ref{FFFFB-2p} consists of  three cubic vertices from $S_{R}^{(1)}$, one NS propagator, and one Ramond propagator.
The contribution~$\mathcal{A}_{FFFFB}^{(2P)(b)}$ from this diagram is calculated as
\begin{align}
 \mathcal{A}_{FFFFB}^{(2P)(b)}\ =&\
\big(- \bigl\langle\,\Psi_B\,\Psi_C\ \ 
\overbracket[0.5pt]{\!\!\! \phi\ \,\bigr\rangle\big) \big(-\bigl\langle\,\ \phi\!\!\!}\ \
\Psi_D\ \
\overbracket[0.5pt]{\!\!\! \Psi\ \,\bigr\rangle\big) \bigl\langle\,\ \Psi\!\!\!}\ \
\phi_E\,\Psi_A\,\bigr\rangle
\nonumber\\
=&\
- \bigl\langle\,\Psi_B\,\Psi_C\,\frac{\xi_0 b_0}{L_0}\,\Bigl( \, \Psi_D\,\frac{b_0 X \eta}{L_0}\,
( \, \phi_E\,\Psi_A\,) \, \Bigr) \, \bigr\rangle.
\end{align}
Following the method in~\eqref{X-to-X_0}, it is transformed as follows.
\begin{align}
\mathcal{A}_{FFFFB}^{(2P)(b)}=& {}- \llangle\,\Psi_B\,\Psi_C\,\frac{b_0}{L_0}\,\Bigl( \, \Psi_D\,\frac{b_0}{L_0}\,( \, X_0\Phi_E\,\Psi_A\,) \, \Bigr) \, \rrangle
\nonumber\\
&
+ \bigl\langle\,\Psi_B\,\Psi_C\,\frac{b_0}{L_0}\,( \, \Psi_D\,\Xi\,(\phi_E\,\Psi_A)\,) \, \bigr\rangle
- \bigl\langle\,\phi_E\,\Psi_A\,\frac{b_0\, \Xi}{L_0}\,(\,\Psi_B\,\Psi_C\,\Psi_D\,)\,\bigr\rangle\,.
\end{align}

The diagram $(2P)(c)$ in figure~\ref{FFFFB-2p} consists of two cubic vertices from $S_{R}^{(1)}$, one cubic vertex from $S_{NS}^{(1)}$, and two NS propagators.
It is calculated as
\begin{align}
 \mathcal{A}_{FFFFB}^{(2P)(c)}\ =&\
\big(- \bigl\langle\,\Psi_C\,\Psi_D\ \ 
\overbracket[0.5pt]{\!\!\! \phi\ \,\bigr\rangle\big) \Big(\frac{1}{2}\,\bigl\langle\,\ Q\phi\!\!\!}\ \
\phi_E\ \
\overbracket[0.5pt]{\!\!\! \eta\phi\ \,\bigr\rangle\Big)\big(- \bigl\langle\,\ \phi\!\!\!}\ \
\Psi_A\,\Psi_B\,\bigr\rangle\big)
\nonumber\\
&
+
\big(- \bigl\langle\,\Psi_C\,\Psi_D\ \ 
\overbracket[0.5pt]{\!\!\! \phi\ \,\bigr\rangle\big) \Big(\frac{1}{2}\,\bigl\langle\,\ \eta\phi\!\!\!}\ \
\phi_E\ \
\overbracket[0.5pt]{\!\!\! Q\phi\ \,\bigr\rangle\Big)\big(- \bigl\langle\,\ \phi\!\!\!}\ \
\Psi_A\,\Psi_B\,\bigr\rangle\big)
\nonumber\\
=&\
-\frac{1}{2}\,\bigl\langle\,\Psi_C\,\Psi_D\,\frac{\xi_0 b_0 Q}{L_0}\,\Bigl( \, \phi_E\,
\frac{\eta \xi_0 b_0}{L_0}\,( \, \Psi_A\,\Psi_B\,) \, \Bigr) \,\bigr\rangle \nonumber \\
&\ -\frac{1}{2}\,\bigl\langle\,\Psi_C\,\Psi_D\,\frac{\xi_0 b_0 \eta}{L_0}\,\Bigl( \, \phi_E\,
\frac{Q \xi_0 b_0}{L_0}\,( \, \Psi_A\,\Psi_B\,) \, \Bigr) \, \bigr\rangle.
\end{align}
We move $\eta$ in the direction to avoid the combination $\eta \phi_E$
and then move the BRST operator to act on $\phi_E$:
\begin{align}
\mathcal{A}_{FFFFB}^{(2P)(c)}\ =&\
{}- \llangle\,\Psi_C\,\Psi_D\,\frac{ b_0}{L_0}\,\Bigl( \, X_0\Phi_E\,
\frac{b_0}{L_0}\,( \, \Psi_A\,\Psi_B\,) \, \Bigr) \, \rrangle
\nonumber\\
&
- \frac{1}{2}\,\bigl\langle\,\Psi_C\,\Psi_D\,\frac{\xi_0 b_0}{L_0}\,( \, \phi_E\,\Psi_A\,\Psi_B\,) \, \bigr\rangle
+ \frac{1}{2}\,\bigl\langle\,\Psi_A\,\Psi_B\,\frac{\xi_0 b_0}{L_0}\,( \, \Psi_C\,\Psi_D\,\phi_E\,) \, \bigr\rangle\,.
\end{align} 

The calculations for the diagrams~$(2P)(d)$ and $(2P)(e)$ are similar
to those for in the diagrams $(2P)(a)$ and $(2P)(b)$, respectively.
The contribution~$\mathcal{A}_{FFFFB}^{(2P)(d)}$ from the diagram $(2P)(d)$
and the contribution~$\mathcal{A}_{FFFFB}^{(2P)(e)}$ from the diagram $(2P)(e)$
are given by
\begin{align}
 \mathcal{A}_{FFFFB}^{(2P)(d)}\ 
=&\
-\llangle\,\Psi_D\,X_0\Phi_E\,\frac{b_0}{L_0}\,\Bigl( \, \Psi_A\,\frac{b_0}{L_0}\,
( \, \Psi_B\,\Psi_C\,) \, \Bigr) \, \rrangle
\nonumber\\
&
- \bigl\langle\,\Psi_D\,\phi_E\,\frac{b_0 \Xi}{L_0}\,(\, \Psi_A\,\Psi_B\,\Psi_C\,) \, \bigr\rangle
+ \bigl\langle\,\Psi_B\,\Psi_C\,\frac{b_0}{L_0}\,( \, \Xi\,(\Psi_D\,\phi_E)\,\Psi_A\,) \, \bigr\rangle\,, \\
 \mathcal{A}_{FFFFB}^{(2P)(e)}\ 
=&\
-\llangle\, X_0\Phi_E\,\Psi_A\,\frac{b_0}{L_0}\,\Bigl( \, \Psi_B\,\frac{b_0}{L_0}\,
( \, \Psi_C\,\Psi_D\,) \, \Bigr) \, \rrangle
\nonumber\\
&
+ \bigl\langle\,\phi_E\,\Psi_A\,\frac{b_0 \Xi}{L_0}\,( \, \Psi_B\,\Psi_C\,\Psi_D\,) \, \bigr\rangle
-\bigl\langle\,\Psi_C\,\Psi_D\,\frac{b_0}{L_0}\,( \, \Xi\,(\phi_E\,\Psi_A)\,\Psi_B\,) \, \bigr\rangle\,.
\end{align}

\subsubsection{Contributions from diagrams with one propagator}
\begin{figure}[htb]
  \begin{center}
    \begin{tabular}{c}
    
      \begin{minipage}{0.30\hsize}
        \begin{center}
          \includegraphics[clip, width=3.5cm]{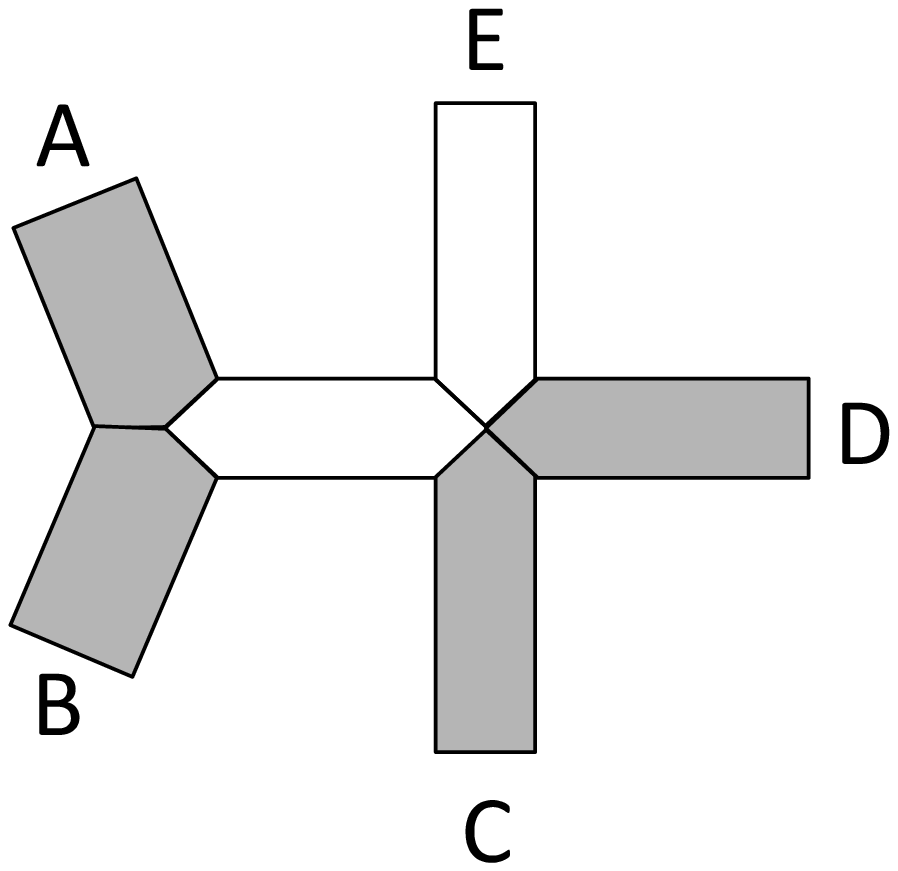}
          \hspace{1.6cm} $(1P)(a)$
                  \end{center}
      \end{minipage}
      \begin{minipage}{0.30\hsize}
        \begin{center}
          \includegraphics[clip, width=3.5cm]{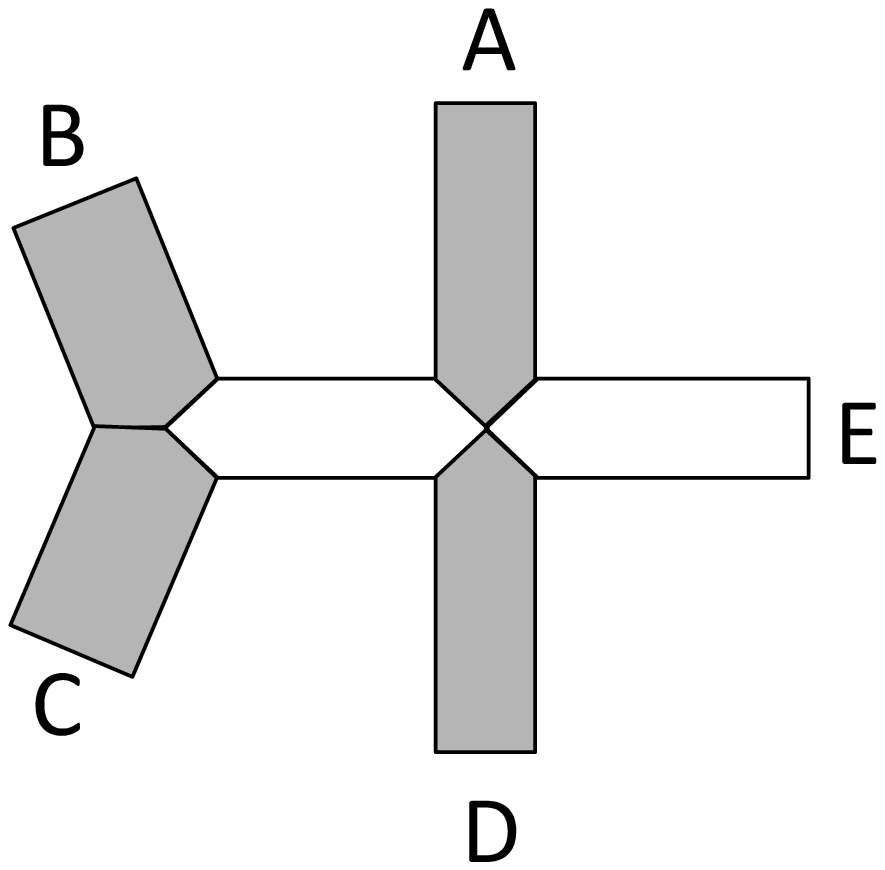}
          \hspace{1.6cm}$(1P)(b)$
        \end{center}
      \end{minipage}
      \begin{minipage}{0.30\hsize}
        \begin{center}
          \includegraphics[clip, width=3.5cm]{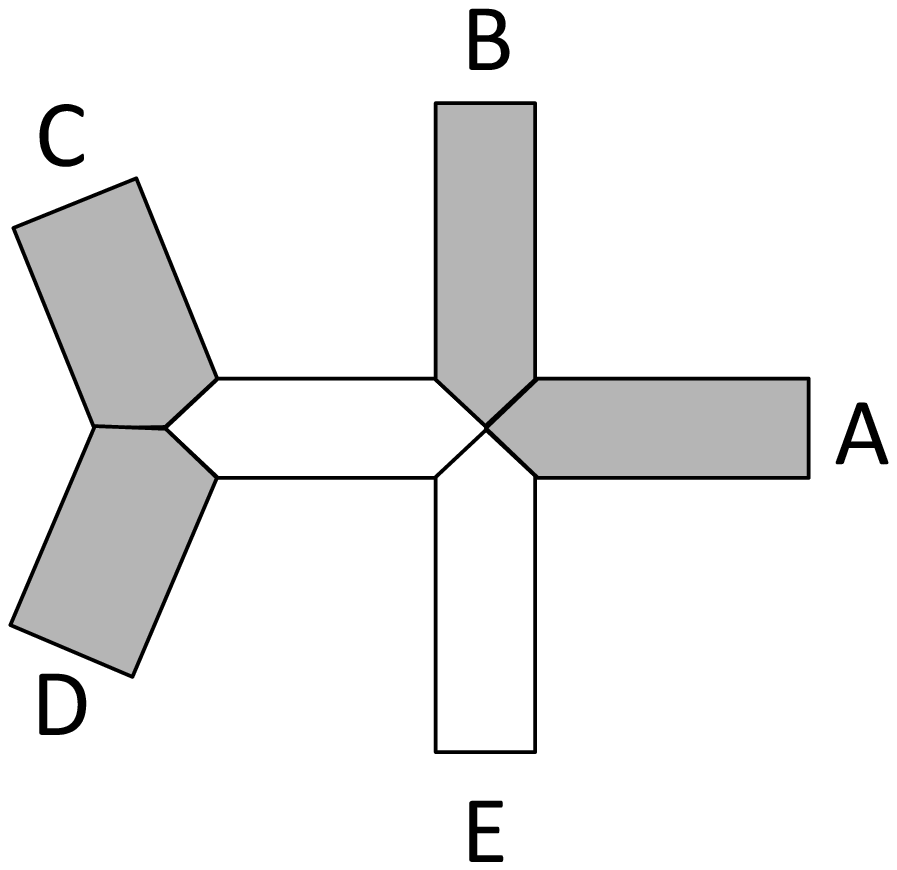}
          \hspace{1.6cm} $(1P)(c)$
        \end{center}
      \end{minipage}
         \end{tabular}
    \caption{Feynman diagrams with one cubic vertex, one quartic vertex, and one propagator for $\mathcal{A}_{FFFFB}$.}
    \label{FFFFB-1p}
  \end{center}
\end{figure}

Let us next calculate contributions from Feynman diagrams
with one cubic vertex, one quartic vertex, and one propagator for $\mathcal{A}_{FFFFB}$.
Such Feynman diagrams are depicted in figure~\ref{FFFFB-1p}.

The diagram $(1P)(a)$ in figure \ref{FFFFB-1p} consists of one cubic vertex from $S_{R}^{(1)}$, one quartic vertex with the ordering fermion-fermion-boson-boson from $S_{R}^{(2)}$, and one NS propagator.
The contribution~$\mathcal{A}_{FFFFB}^{(1P)(a)}$ from the diagram~$(1P)(a)$ is calculated as
\begin{align}
 \mathcal{A}_{FFFFB}^{(1P)(a)}\ =&\
\big(- \bigl\langle\,\Psi_A\,\Psi_B\ \ 
\overbracket[0.5pt]{\!\!\! \phi\ \,\bigr\rangle\big) \Big(- \frac{1}{2}\bigl\langle\,\ \phi\!\!\!}\ \
\Psi_C\,\Xi\,(\Psi_D\,\eta\phi_E)\,\bigr\rangle\Big)
\nonumber\\
&
+ \big(- \bigl\langle\,\Psi_A\,\Psi_B\ \ 
\overbracket[0.5pt]{\!\!\! \phi\ \,\bigr\rangle\big) \Big(- \frac{1}{2}\bigl\langle\,\ \eta\phi\!\!\!}\ \
\Psi_C\,\Xi\,(\Psi_D\,\phi_E)\,\bigr\rangle\Big)
\nonumber\\
=&\
\frac{1}{2}\,\bigl\langle\,\Psi_A\,\Psi_B\,\frac{\xi_0 b_0}{L_0}\,( \, \Psi_C\,
\Xi\,(\Psi_D\,\eta\phi_E)\,) \,\bigr\rangle
-\frac{1}{2}\,\bigl\langle\,\Psi_A\,\Psi_B\,\frac{\xi_0 b_0 \eta}{L_0}\,
( \, \Psi_C\,\Xi\,(\Psi_D\,\phi_E)\,) \, \bigr\rangle.
\end{align}
We move $\eta$ to avoid the combination $\eta \phi_E$ to find
\begin{align}
&\
\bigl\langle\,\Psi_A\,\Psi_B\,\frac{b_0}{L_0}\,(\,\Psi_C\,\Xi\,(\Psi_D\,\phi_E)\,)\,\bigr\rangle
- \frac{1}{2}\,\bigl\langle\,\Psi_A\,\Psi_B\,\frac{\xi_0 b_0}{L_0}\,
(\,\Psi_C\,\Psi_D\,\phi_E\,)\,\bigr\rangle\,.
\end{align}

The diagram $(1P)(b)$ in figure \ref{FFFFB-1p} consists of one cubic vertex $S_{R}^{(1)}$, one quartic vertex with the ordering fermion-boson-fermion-boson from $S_{R}^{(2)}$, and one NS propagator.
The contribution~$\mathcal{A}_{FFFFB}^{(1P)(b)}$ from the diagram~$(1P)(b)$ is given by
\begin{align}
& \mathcal{A}_{FFFFB}^{(1P)(b)}\  \nonumber \\
 =&\
\big(- \bigl\langle\,\Psi_B\,\Psi_C\ \ 
\overbracket[0.5pt]{\!\!\! \phi\ \,\bigr\rangle\big) \Big(- \frac{1}{2}\bigl\langle\,\ \phi\!\!\!}\ \
\Psi_D\,\Xi\,(\eta\phi_E\,\Psi_A)\,\bigr\rangle\Big)
+\big(- \bigl\langle\,\Psi_B\,\Psi_C\ \ 
\overbracket[0.5pt]{\!\!\! \phi\ \,\bigr\rangle\big) \Big(- \frac{1}{2}\bigl\langle\,\ \phi\!\!\!}\ \
\Xi\,(\Psi_D\,\eta\phi_E)\,\Psi_A\,\bigr\rangle\Big)
\nonumber\\
&
+ \big(- \bigl\langle\,\Psi_B\,\Psi_C\ \ 
\overbracket[0.5pt]{\!\!\! \phi\ \,\bigr\rangle\big) \Big(\ \frac{1}{2}\bigl\langle\,\ \eta\phi\!\!\!}\ \
\Psi_D\,\Xi\,(\phi_E\,\Psi_A)\,\bigr\rangle\Big)
+ \big(- \bigl\langle\,\Psi_B\,\Psi_C\ \ 
\overbracket[0.5pt]{\!\!\! \phi\ \,\bigr\rangle\big) \Big(\ \frac{1}{2}\bigl\langle\,\ \eta\phi\!\!\!}\ \
\Xi\,(\Psi_D\,\phi_E)\,\Psi_A\,\bigr\rangle\Big)
\nonumber\\
=&\
\frac{1}{2}\,\bigl\langle\,\Psi_B\,\Psi_C\,\frac{\xi_0 b_0}{L_0}\,( \, \Psi_D\,
\Xi\,(\eta\phi_E\,\Psi_A)\,) \, \bigr\rangle
+\frac{1}{2}\,\bigl\langle\,\Psi_B\,\Psi_C\,\frac{\xi_0 b_0}{L_0}\,
( \, \Xi\,(\Psi_D\,\eta\phi_E)\,\Psi_A\,) \, \bigr\rangle
\nonumber\\
&
+\frac{1}{2}\,\bigl\langle\,\Psi_B\,\Psi_C\,\frac{\xi_0 b_0 \eta}{L_0}\,( \, \Psi_D\,
\Xi\,(\phi_E\,\Psi_A)\,) \, \bigr\rangle
+\frac{1}{2}\,\bigl\langle\,\Psi_B\,\Psi_C\,\frac{\xi_0 b_0 \eta}{L_0}\,
( \, \Xi\,(\Psi_D\,\phi_E)\,\Psi_A\,) \, \bigr\rangle.
\end{align}
We move $\eta$ to avoid the combination $\eta\phi_E$ and find
\begin{align}
\mathcal{A}_{FFFFB}^{(1P)(b)}\ =&\
- \bigl\langle\,\Psi_B\,\Psi_C\,\frac{b_0}{L_0}\,( \, \Psi_D\,\Xi\,(\phi_E\,\Psi_A)\,) \, \bigr\rangle
-\bigl\langle\,\Psi_B\,\Psi_C\,\frac{b_0}{L_0}\,( \, \Xi\,(\Psi_D\,\phi_E)\,\Psi_A\,) \, \bigr\rangle\,.
\end{align}

The calculation for the diagram $(1P)(c)$ in figure \ref{FFFFB-1p} is similar
to that for the diagram $(1P)(a)$.
The contribution~$\mathcal{A}_{FFFFB}^{(1P)(c)}$ for the diagram~$(1P)(c)$ is given by
\begin{align}
 \mathcal{A}_{FFFFB}^{(1P)(c)}\ 
=&\
\bigl\langle\,\Psi_C\,\Psi_D\,\frac{b_0}{L_0}\,( \, \Xi\,(\phi_E\,\Psi_A)\,\Psi_B\,) \, \bigr\rangle
+\frac{1}{2}\,\bigl\langle\,\Psi_C\,\Psi_D\,\frac{\xi_0 b_0}{L_0}\,( \, \phi_E\,\Psi_A\,\Psi_B\,) \, \bigr\rangle\,.
\end{align}

\subsubsection{Contributions from all diagrams}
In the calculation of the contributions from diagrams with two propagators in subsection~\ref{FFFFB-2P-diagrams},
we found that all the terms of~$\mathcal{A}_{FFFFB}^{{\rm WS}}$ in the world-sheet theory are reproduced,
but we also found that extra terms with one propagator are generated. 
We denote the sum of those extra terms as
\begin{equation}
\Delta\mathcal{A}_{FFFFB} \equiv \mathcal{A}_{FFFFB}^{(2P)}-\mathcal{A}_{FFFFB}^{{\rm WS}},
\end{equation}
where
\begin{equation}
\mathcal{A}_{FFFFB}^{(2P)} 
= \mathcal{A}_{FFFFB}^{(2P)(a)}
+\mathcal{A}_{FFFFB}^{(2P)(b)}
+\mathcal{A}_{FFFFB}^{(2P)(c)}
+\mathcal{A}_{FFFFB}^{(2P)(d)}
+\mathcal{A}_{FFFFB}^{(2P)(e)},
\end{equation}
and it is given by
\begin{align}
\Delta\mathcal{A}_{FFFFB}
=&
-\bigl\langle\,\Psi_A\,\Psi_B\,\frac{b_0}{L_0}\,(\,\Psi_C\,\Xi\,(\Psi_D\,\phi_E)\,)\,\bigr\rangle
+ \frac{1}{2}\,\bigl\langle\,\Psi_A\,\Psi_B\,\frac{\xi_0 b_0}{L_0}\,(\,\Psi_C\,\Psi_D\,\phi_E\,)\,\bigr\rangle \nonumber \\
&+ \bigl\langle\,\Psi_B\,\Psi_C\,\frac{b_0}{L_0}\,( \, \Psi_D\,\Xi\,(\phi_E\,\Psi_A)\,) \, \bigr\rangle
+\bigl\langle\,\Psi_B\,\Psi_C\,\frac{b_0}{L_0}\,( \, \Xi\,(\Psi_D\,\phi_E)\,\Psi_A\,) \, \bigr\rangle \nonumber \\
&
-\bigl\langle\,\Psi_C\,\Psi_D\,\frac{b_0}{L_0}\,( \, \Xi\,(\phi_E\,\Psi_A)\,\Psi_B\,) \, \bigr\rangle
-\frac{1}{2}\,\bigl\langle\,\Psi_C\,\Psi_D\,\frac{\xi_0 b_0}{L_0}\,( \, \phi_E\,\Psi_A\,\Psi_B\,) \, \bigr\rangle\,.
\end{align}
On the other hand, the sum of the amplitudes from diagrams with one propagator is written as
 \begin{align}
\mathcal{A}_{FFFFB}^{(1P)}
=&\mathcal{A}_{FFFFB}^{(1P)(a)}+\mathcal{A}_{FFFFB}^{(1P)(b)}+\mathcal{A}_{FFFFB}^{(1P)(c)} \nonumber \\
=&
\bigl\langle\,\Psi_A\,\Psi_B\,\frac{b_0}{L_0}\,(\,\Psi_C\,\Xi\,(\Psi_D\,\phi_E)\,)\,\bigr\rangle
- \frac{1}{2}\,\bigl\langle\,\Psi_A\,\Psi_B\,\frac{\xi_0 b_0}{L_0}\,(\,\Psi_C\,\Psi_D\,\phi_E\,)\,\bigr\rangle \nonumber \\
&- \bigl\langle\,\Psi_B\,\Psi_C\,\frac{b_0}{L_0}\,( \, \Psi_D\,\Xi\,(\phi_E\,\Psi_A)\,) \, \bigr\rangle
-\bigl\langle\,\Psi_B\,\Psi_C\,\frac{b_0}{L_0}\,( \, \Xi\,(\Psi_D\,\phi_E)\,\Psi_A\,) \, \bigr\rangle \nonumber \\
&
+\bigl\langle\,\Psi_C\,\Psi_D\,\frac{b_0}{L_0}\,( \, \Xi\,(\phi_E\,\Psi_A)\,\Psi_B\,) \, \bigr\rangle
+\frac{1}{2}\,\bigl\langle\,\Psi_C\,\Psi_D\,\frac{\xi_0 b_0}{L_0}\,( \, \phi_E\,\Psi_A\,\Psi_B\,) \, \bigr\rangle\,,
\end{align}
and these contributions precisely cancel the extra terms:
\begin{equation}
\mathcal{A}_{FFFFB}^{(1P)}=-\Delta\mathcal{A}_{FFFFB}.
\end{equation}
Thus we conclude that the contributions $\mathcal{A}_{FFFFB}$ calculated from open superstring field theory coincides with the amplitude in the world-sheet theory \eqref{WS-FFFFB}:
\begin{equation}
\mathcal{A}_{FFFFB} = \mathcal{A}_{FFFFB}^{{\rm WS}},
\end{equation}
where
\begin{equation}
\mathcal{A}_{FFFFB}
=\mathcal{A}_{FFFFB}^{(2P)}+\mathcal{A}_{FFFFB}^{(1P)}.
\end{equation}

\subsection{Fermion-boson-fermion-boson-boson amplitudes}
We next present the calculations of five-point amplitudes of two fermions and three bosons
with the fermion-boson-fermion-boson-boson ordering.
Since disks with two Ramond punctures and three NS punctures have
two even moduli and two odd, we need two insertions of picture-changing operators
in the world-sheet theory.
We choose them to act on $\Phi_B$ and $\Phi_D$,
and then the amplitude $\mathcal{A}_{FBFBB}^{{\rm WS}}$ in the world-sheet theory is written as
\begin{align}
\mathcal{A}_{FBFBB}^{{\rm WS}}=&\
-\,\llangle\,\Psi_A\,X_0 \Phi_B\,\frac{b_0}{L_0}\,\Bigl(\,
\Psi_C\,\frac{b_0}{L_0}\,( \, X_0 \Phi_D\,\eta\phi_E\,) \,  \Bigr) \,\rrangle
- \llangle\, X_0 \Phi_B\,\Psi_C\,\frac{b_0}{L_0}\,\Bigl(\,
X_0 \Phi_D\,\frac{b_0}{L_0}\,( \, \eta\phi_E\,\Psi_A\,) \, \Bigr) \,\rrangle
\nonumber\\
&\
- \llangle\,\Psi_C\,X_0 \Phi_D\,\frac{b_0}{L_0}\,\Bigl(\,
\eta\phi_E\,\frac{b_0}{L_0}\,(\, \Psi_A\,X_0 \Phi_B\,)\,\Bigr) \,\rrangle
- \llangle\, X_0 \Phi_D\,\eta\phi_E\,\frac{b_0}{L_0}\,\Bigl( \,
\Psi_A\,\frac{b_0}{L_0}\,(\,
X_0 \Phi_B\,\Psi_C\,)\,\Bigr) \,\rrangle
\nonumber\\
&\
- \llangle\,\eta\phi_E\,\Psi_A\,\frac{b_0}{L_0}\,\Bigl(\,
\,X_0 \Phi_B\,
\frac{b_0}{L_0}\,(\,\Psi_C\,X_0 \Phi_D\,)\,\Bigr) \,\rrangle\,.
\end{align}
This amplitude is translated into the language of the large Hilbert space as
\begin{align}
\mathcal{A}_{FBFBB}^{{\rm WS}}=&\ 
{}-\bigl\langle\,\Psi_A\,Q\phi_B\,\frac{b_0}{L_0}\, \Bigl( \, 
\Psi_C\,\frac{b_0}{L_0}\,(\,Q\phi_D\, \phi_E\,) \, \Bigr) \,\bigr\rangle
+ \bigl\langle\, Q\phi_B\,\Psi_C\,\frac{b_0}{L_0}\,\Bigl( \, 
Q\phi_D\,\frac{b_0}{L_0}\,( \, \phi_E\,\Psi_A\,) \, \Bigr) \,\bigr\rangle
\nonumber\\
&\
+ \bigl\langle\,\Psi_C\,Q\phi_D\,\frac{b_0}{L_0}\,\Bigl( \, 
\phi_E\,\frac{b_0}{L_0}\,( \,  \Psi_A\,Q\phi_B\,) \, \Bigr) \,\bigr\rangle
+ \bigl\langle\, Q\phi_D\,\phi_E\,\frac{b_0}{L_0}\,\Bigl( \, 
\Psi_A\,\frac{b_0}{L_0}\, ( \, Q\phi_B\,\Psi_C\,\Bigr) \,\bigr\rangle
\nonumber\\
&\
{}- \bigl\langle\,\phi_E\,\Psi_A\,\frac{b_0}{L_0}\,\Bigl( \, 
Q\phi_B\,\frac{b_0 }{L_0}\,( \, \Psi_C\,Q\phi_D\,) \, \Bigr) \,\bigr\rangle.
\end{align}
The goal of this subsection is to reproduce this expression from the calculation in open superstring field theory. 

\subsubsection{Contributions from diagrams with two propagators}
We denote the corresponding amplitude in open superstring field theory by $\mathcal{A}_{FBFBB}$.
We begin with calculating contributions from Feynman diagrams
with three cubic vertices and two propagators for~$\mathcal{A}_{FBFBB}$.
Such diagrams are depicted in figure~\ref{FBFBB-1p}.
\begin{figure}[htbp]
  \begin{center}
    \begin{tabular}{c}
    
      \begin{minipage}{0.30\hsize}
        \begin{center}
          \includegraphics[clip, width=3.5cm]{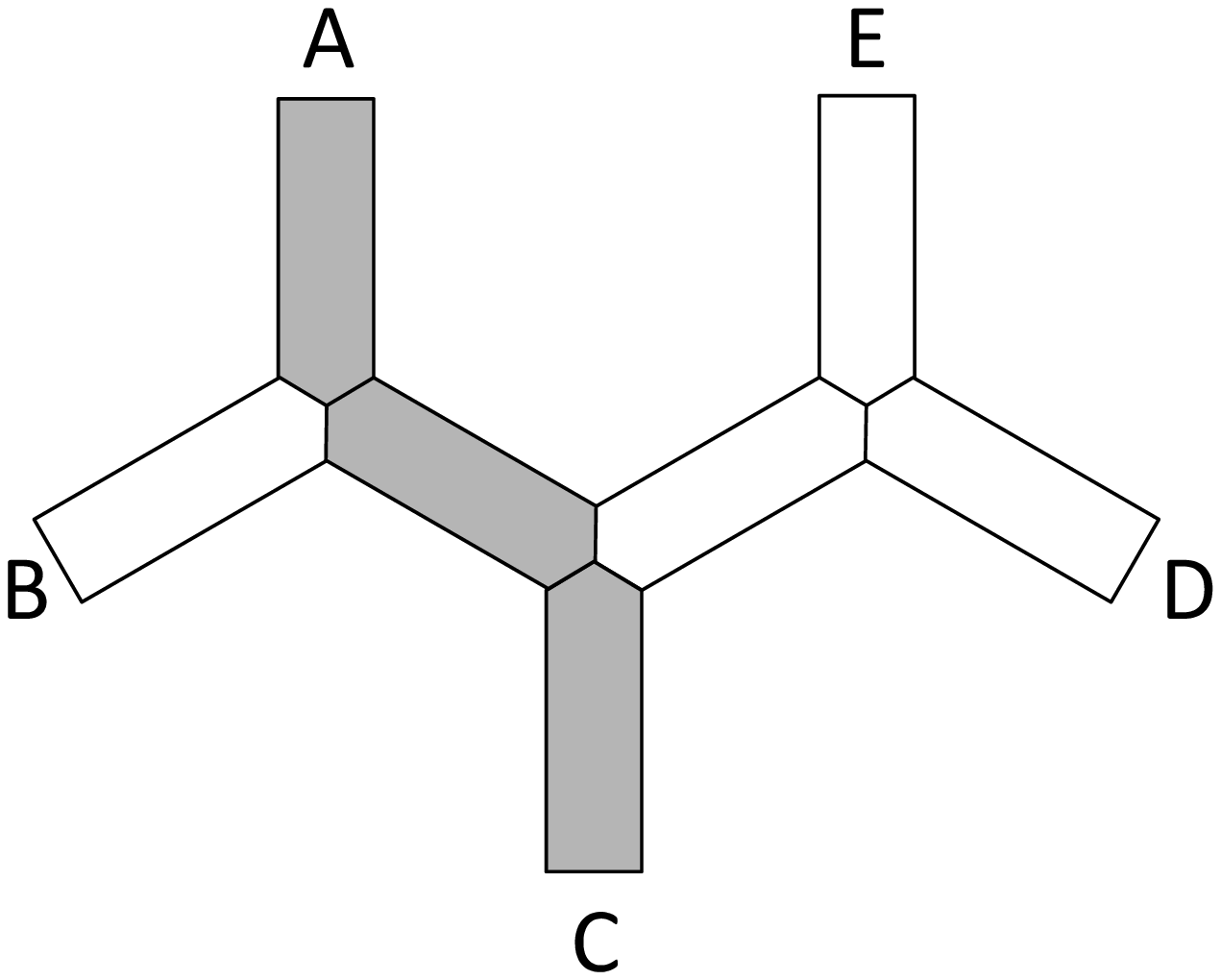}
          \hspace{1.6cm} $(2P)(a)$
                  \end{center}
      \end{minipage}
      \begin{minipage}{0.30\hsize}
        \begin{center}
          \includegraphics[clip, width=3.5cm]{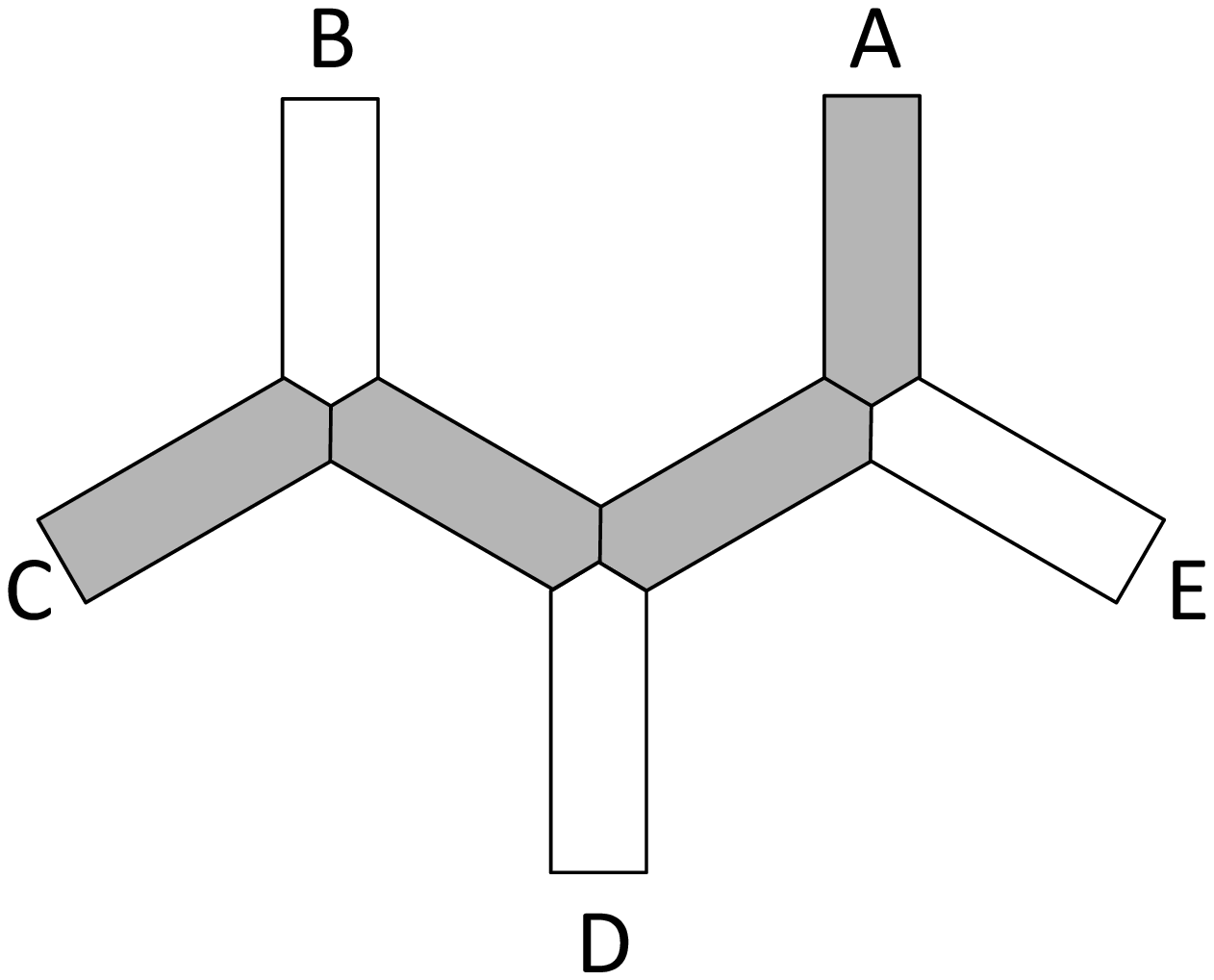}
          \hspace{1.6cm}$(2P)(b)$
        \end{center}
      \end{minipage}
      \begin{minipage}{0.30\hsize}
        \begin{center}
          \includegraphics[clip, width=3.5cm]{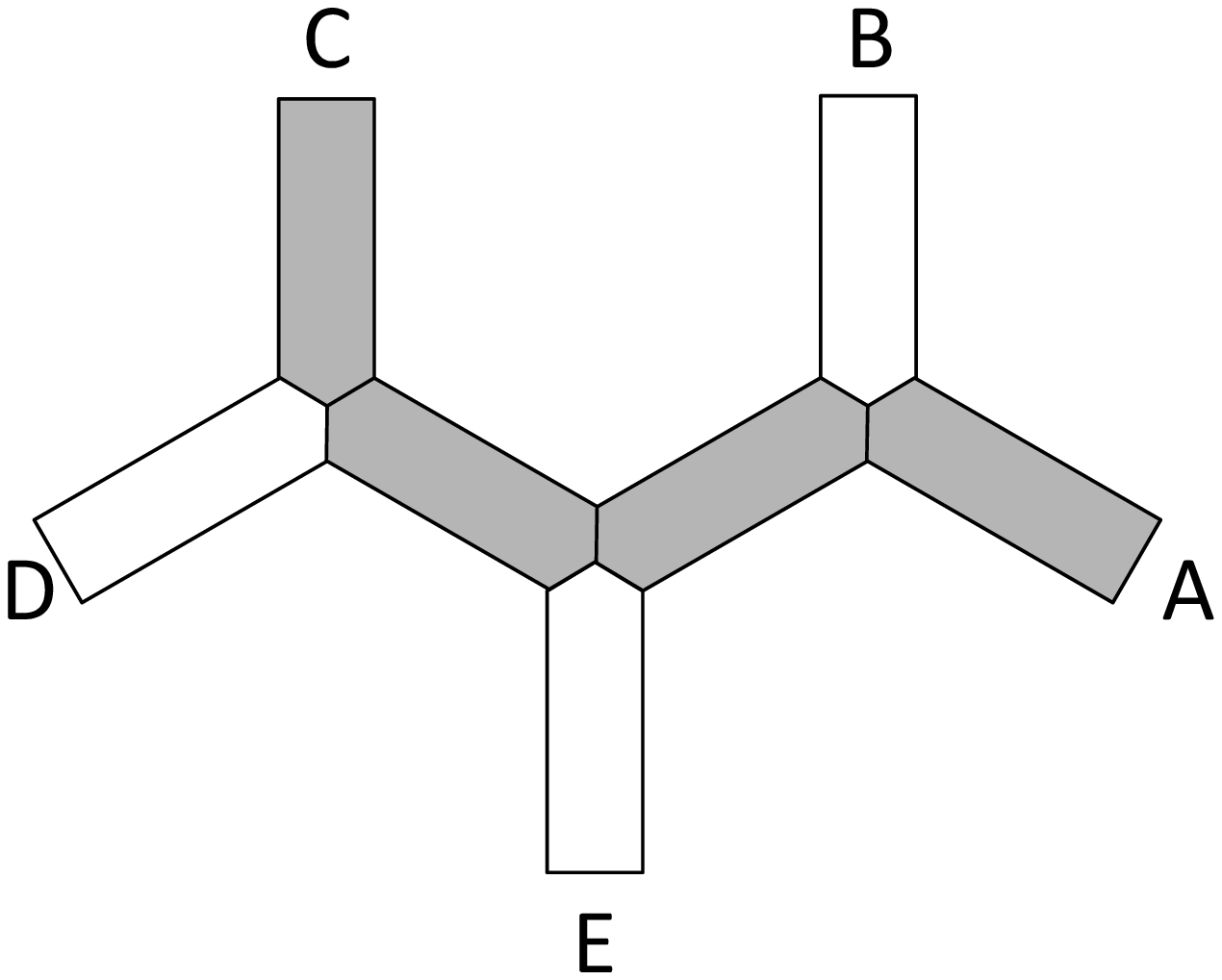}
          \hspace{1.6cm} $(2P)(c)$
        \end{center}
      \end{minipage}
      \\
      \begin{minipage}{0.30\hsize}
        \begin{center}
          \includegraphics[clip, width=3.5cm]{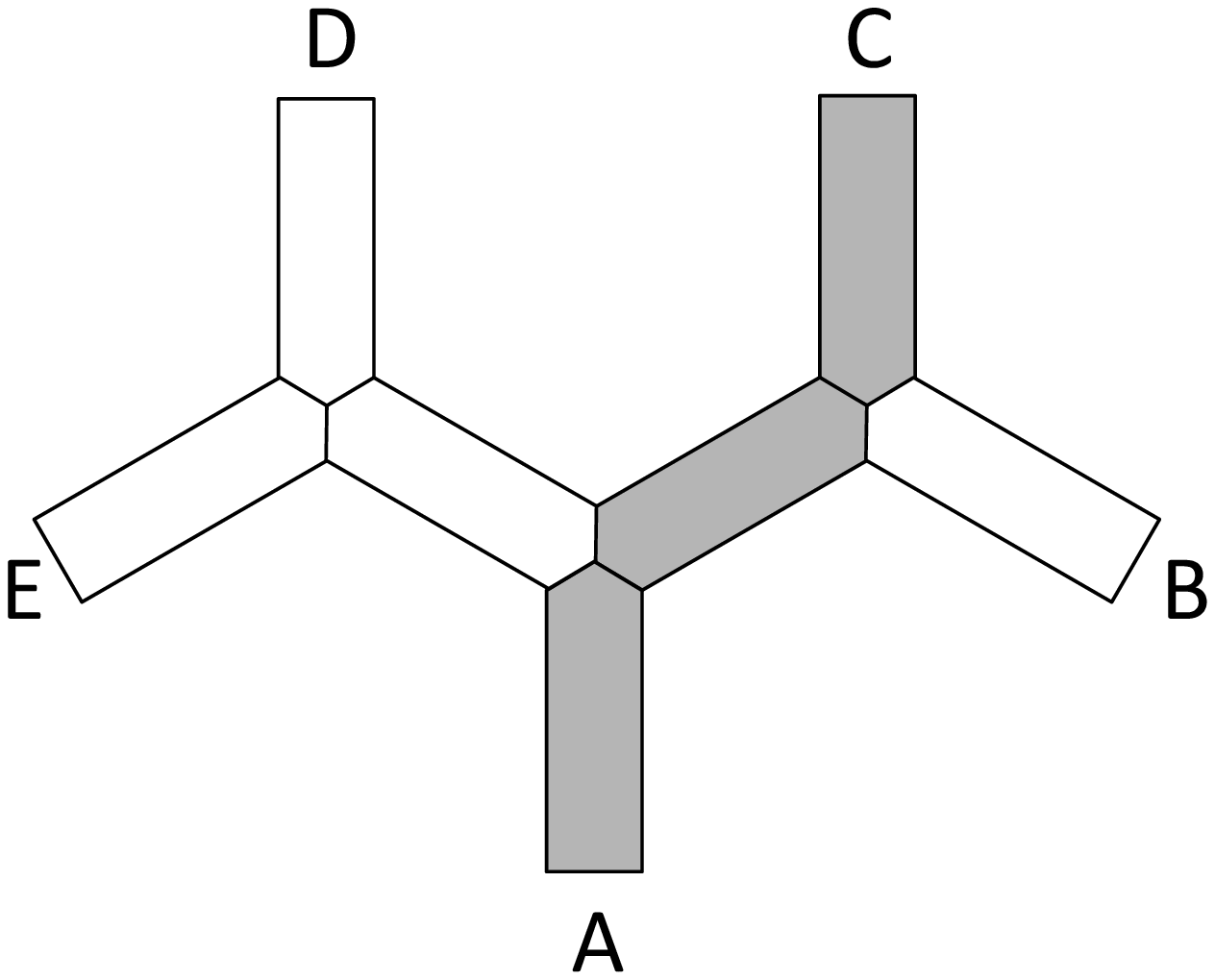}
          \hspace{1.6cm} $(2P)(d)$
        \end{center}
      \end{minipage}
      \begin{minipage}{0.30\hsize}
        \begin{center}
          \includegraphics[clip, width=3.5cm]{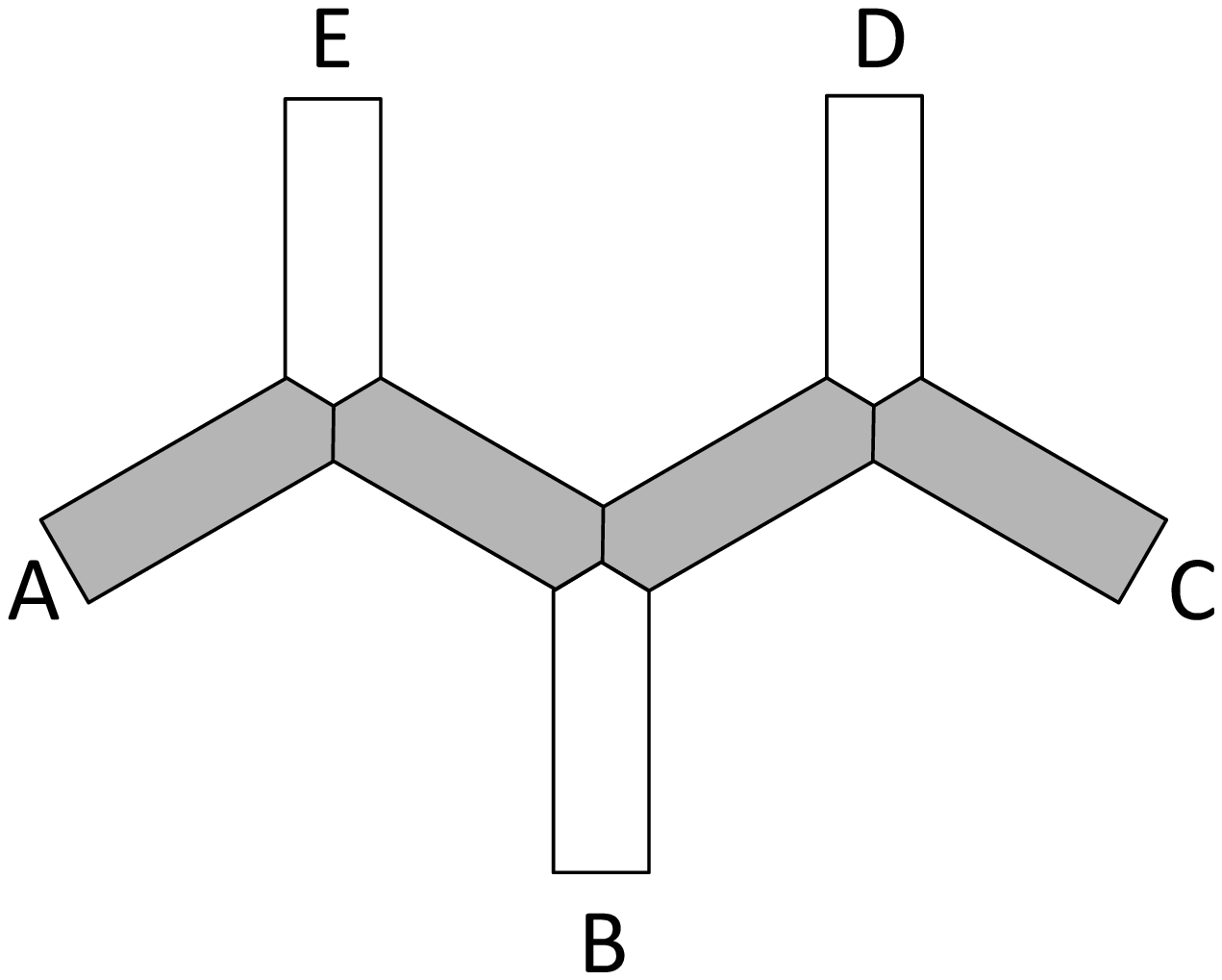}
          \hspace{1.6cm} $(2P)(e)$
        \end{center}
      \end{minipage}
      
    \end{tabular}
    \caption{Feynman diagrams with two cubic vertices and two propagators for $\mathcal{A}_{FBFBB}$.}
    \label{FBFBB-2p}
  \end{center}
\end{figure}

The diagram $(2P)(a)$ in figure \ref{FBFBB-2p} consists of two cubic vertices from $S_{R}^{(1)}$, one cubic vertex from $S_{NS}^{(1)}$, one Ramond propagator, and one NS propagator.
It is calculated as
\begin{align}
 \mathcal{A}_{FBFBB}^{(2P)(a)}\ =&\
\bigl\langle\,\Psi_A\,\phi_B\ \ 
\overbracket[0.5pt]{\!\!\! \Psi\,\bigr\rangle
\bigl(-\,\bigl\langle\,\Psi\!\!\!}\ \ 
\Psi_C\ \
\overbracket[0.5pt]{\!\!\! \phi\,\bigr\rangle \bigr)
\Big(\,-\,\frac{1}{2}\bigl\langle\,\phi\!\!\!}\ \
(Q\phi_D\,\eta\phi_E + \eta\phi_D\,Q\phi_E)\,\bigr\rangle\Big) 
\nonumber\\
=&\
-\frac{1}{2}\,\bigl\langle\,\Psi_A\,\phi_B\,\frac{b_0 X \eta}{L_0}\, \Bigl( \, 
\Psi_C\,\frac{\xi_0 b_0}{L_0}\,
(Q\phi_D\,\eta\phi_E + \eta\phi_D\,Q\phi_E)\,\Bigr) \, \bigr\rangle. \label{FBFBB-2p-a-before}
\end{align}
The corresponding term in the amplitude $\mathcal{A}_{FBFBB}^{{\rm WS}}$ in the world-sheet theory is
\begin{equation}
- \, \bigl\langle \, \Psi_A \, Q\phi_B \, \frac{b_0}{L_0} \, \Bigl( \, \Psi_C \, \frac{b_0}{L_0}\, (\, Q\phi_D \, \phi_E \, )\, \Bigr) \, \bigr\rangle, 
\end{equation}
and we move the BRST operators to reproduce this term
as in the calculation of the diagram~$(2P)(d)$ for $\mathcal{A}_{FBFBB}$.
We find
\begin{align}
\mathcal{A}_{FBFBB}^{(2P)(a)}=&\
-\bigl\langle\,\Psi_A\,Q\phi_B\,\frac{b_0}{L_0}\, \Bigl( \, 
\Psi_C\,\frac{b_0}{L_0}\,( \, Q\phi_D\,\phi_E\,) \, \Bigr) \, \bigr\rangle
\nonumber\\
&\
+ \frac{1}{2}\,\bigl\langle\,\Psi_A\,Q\phi_B\,\frac{b_0}{L_0}\,(\,
\Psi_C\,\phi_D\,\phi_E\,)\,\bigr\rangle
-\,\frac{1}{2}\,\bigl\langle\,\Psi_A\,\phi_B\,\frac{b_0 \Xi}{L_0}\,
(\, \Psi_C\,(Q\phi_D\,\eta\phi_E + \eta\phi_D\,Q\phi_E)\,)\,\bigr\rangle
\nonumber\\
&\
- \frac{1}{2}\,\bigl\langle\,\phi_D\,\phi_E\,\frac{b_0}{L_0}\,
(\, \Psi_A\,Q\phi_B\,\Psi_C\,)\, \bigr\rangle
+\,\frac{1}{2}\,\bigl\langle\,(Q\phi_D\,\eta\phi_E + \eta\phi_D\,Q\phi_E)\,
\frac{b_0}{L_0}\,(\,\Xi\,(\Psi_A\,\phi_B)\,\Psi_C\,)\,\bigr\rangle\,. \label{FBFBB-(2P)(a)}
\end{align}

The diagram $(2P)(b)$ in figure~\ref{FBFBB-2p} consists of three cubic vertices from $S_{R}^{(1)}$ and two Ramond propagators.
It is calculated as
\begin{align}
 \mathcal{A}_{FBFBB}^{(2P)(b)}\ =&\
\bigl(-\,\bigl\langle\,\phi_B\,\Psi_C\ \
\overbracket[0.5pt]{\!\!\! \Psi\,\bigr\rangle \bigr)
\bigl\langle\,\Psi\!\!\!}\ \ 
\phi_D\ \
\overbracket[0.5pt]{\!\!\! \Psi\,\bigr\rangle
\bigl\langle\,\Psi\!\!\!}\ \ 
\phi_E\,\Psi_A\,\bigr\rangle
\nonumber\\
=&\
- \bigl\langle\,\phi_B\,\Psi_C\,\frac{b_0 X \eta}{L_0}\,\Bigl( \, 
\phi_D\,\frac{b_0 X \eta}{L_0}\,(\,\phi_E\,\Psi_A\,)\, \Bigr) \, \bigr\rangle. \label{FBFBB-2p-b-before}
\end{align}
The corresponding term in the amplitude $\mathcal{A}_{FBFBB}^{{\rm WS}}$ in the world-sheet theory is
\begin{equation}
\bigl\langle \, Q\phi_B \, \Psi_C \, \frac{b_0}{L_0}\,\Bigl( \, Q\phi_D \, \frac{b_0}{L_0}\,(\, \phi_E \, \Psi_A \,)\,\Bigr) \, \bigr\rangle,
\end{equation}
and we move the BRST operators from the Ramond propagators to reproduce this term
as in the calculation of the diagram $(2P)(b)$ for $\mathcal{A}_{FFBBB}$.
We find
\begin{align}
\mathcal{A}_{FBFBB}^{(2P)(b)}=&\
\bigl\langle\, Q\phi_B\,\Psi_C\,\frac{b_0}{L_0}\, \Bigl( \, 
Q\phi_D\,\frac{b_0}{L_0}\,( \, \phi_E\,\Psi_A\,) \, \Bigr) \, \bigr\rangle
\nonumber\\
&\
- \bigl\langle\,\phi_B\,\Psi_C\,\frac{b_0 X}{L_0}\, ( \, 
\eta\phi_D\,\Xi\,(\eta\phi_E\,\Psi_A)\,) \, \bigr\rangle
- \bigl\langle\,\phi_B\,\Psi_C\,\frac{b_0 X}{L_0}\, ( \, 
\phi_D\,\eta\phi_E\,\Psi_A\,) \, \bigr\rangle
\nonumber\\
&\
+ \bigl\langle\,\phi_B\,\Psi_C\,\frac{b_0 \Xi}{L_0}\, ( \, 
Q\phi_D\,\eta\phi_E\,\Psi_A\,) \, \bigr\rangle
\nonumber\\
&\
- \bigl\langle\,\eta\phi_E\,\Psi_A\,\frac{b_0 \Xi}{L_0}\, ( \, 
X\,(\eta\phi_B\,\Psi_C)\,\phi_D\,) \, \bigr\rangle
- \bigl\langle\,\eta\phi_E\,\Psi_A\,\frac{b_0}{L_0}\, ( \, 
\Xi\,(\phi_B\,\Psi_C)\,Q\phi_D\,) \, \bigr\rangle\,. \label{FBFBB-(2P)(b)}
\end{align}

The diagram $(2P)(c)$ in figure \ref{FBFBB-2p} consists of three cubic vertices from $S_{R}^{(1)}$ and two Ramond propagators.
It is calculated as
\begin{align}
\mathcal{A}_{FBFBB}^{(2P)(c)}
=&\
\bigl\langle\,\Psi_C\,\phi_D\ \
\overbracket[0.5pt]{\!\!\! \Psi\,\bigr\rangle
\bigl\langle\,\Psi\!\!\!}\ \ 
\phi_E\ \
\overbracket[0.5pt]{\!\!\! \Psi\,\bigr\rangle
\bigl(-\,\bigl\langle\,\Psi\!\!\!}\ \ 
\Psi_A\,\phi_B\,\bigr\rangle \bigr)
\nonumber\\
=&\
-\,\bigl\langle\,\Psi_C\,\phi_D\,\frac{b_0 X \eta}{L_0}\, \Bigl( \, 
\phi_E\,\frac{b_0 X \eta}{L_0}\, ( \, \Psi_A\,\phi_B\,) \, \Bigr) \, \bigr\rangle.
\end{align}
As in the calculation of the diagram $(2P)(b)$ for $\mathcal{A}_{FFBBB}$,
we move the BRST operators from the Ramond propagators
to act on $\phi_B$ and $\phi_D$.
We find
\begin{align}
 \mathcal{A}_{FBFBB}^{(2P)(c)}\ 
=&\
\bigl\langle\,\Psi_C\,Q\phi_D\,\frac{b_0}{L_0}\,\Bigl( \, 
\phi_E\,\frac{b_0}{L_0}\, ( \, \Psi_A\,Q\phi_B\,) \, \Bigr) \, \bigr\rangle
\nonumber\\
&\
-\,\bigl\langle\,\Psi_C\,Q\phi_D\,\frac{b_0}{L_0}\, ( \, 
\eta\phi_E\,\Xi\,(\Psi_A\,\phi_B)\,) \, \bigr\rangle
-\,\bigl\langle\,\Psi_C\,\phi_D\,\frac{b_0 \Xi}{L_0}\, ( \, 
\eta\phi_E\,X\,(\Psi_A\,\eta\phi_B)\,) \, \bigr\rangle
\nonumber\\
&\
+\,\bigl\langle\,\Psi_A\,\phi_B\,\frac{b_0 \Xi}{L_0}\, ( \, 
\Psi_C\,Q\phi_D\,\eta\phi_E\,) \, \bigr\rangle
+\,\bigl\langle\,\Psi_A\,\eta\phi_B\,\frac{b_0 X}{L_0}\, ( \, 
\Xi\,(\Psi_C\,\phi_D)\,\eta\phi_E\,) \, \bigr\rangle\,. \label{FBFBB-(2P)(c)}
\end{align}

The diagram $(2P)(d)$ in figure \ref{FBFBB-2p} consists of two cubic vertices from $S_{R}^{(1)}$, one cubic vertex from $S_{NS}^{(1)}$, one Ramond propagator, and one NS propagator. It is calculated as
\begin{align}
\mathcal{A}_{FBFBB}^{(2P)(d)}\ =&\
 \Big(-\,\frac{1}{2}\,\bigl\langle\,(Q\phi_D\,\eta\phi_E + \eta\phi_D\,Q\phi_E)\ \
\overbracket[0.5pt]{\!\!\! \phi\,\bigr\rangle\Big)
\bigl(-\,\bigl\langle\,\phi\!\!\!}\ \ 
\Psi_A\ \
\overbracket[0.5pt]{\!\!\! \Psi\,\bigr\rangle \bigr)
\bigl\langle\,\Psi\!\!\!}\ \ 
\phi_B\,\Psi_C\,\bigr\rangle
\nonumber\\
=&\
-\,\frac{1}{2}\,\bigl\langle\,(Q\phi_D\,\eta\phi_E + \eta\phi_D\,Q\phi_E)\,
\frac{\xi_0 b_0}{L_0}\,\Bigl( \, 
\Psi_A\,\frac{b_0 X \eta}{L_0}\,(\,
\phi_B\,\Psi_C\,) \, \Bigr) \, \bigr\rangle.
\end{align}
As in the calculation of the diagram $(2P)(d)$ for $\mathcal{A}_{FFBBB}$,
we move the BRST operators to act on $\phi_B$ and $\phi_D$. We find
\begin{align}
\mathcal{A}_{FBFBB}^{(2P)(d)}\ 
=&\
+\bigl\langle\, Q\phi_D\,\phi_E\,
\frac{b_0}{L_0}\,\Bigl( \, \Psi_A\,\frac{b_0}{L_0}\,(\,
Q\phi_B\,\Psi_C\,)\, \Bigr) \, \bigr\rangle
\nonumber\\
&\
+\,\frac{1}{2}\,\bigl\langle\, \phi_D\,\phi_E\,
\frac{b_0}{L_0}\,( \, \Psi_A\,
Q\phi_B\,\Psi_C\,)\,\bigr\rangle
+\,\frac{1}{2}\,\bigl\langle\,(Q\phi_D\,\eta\phi_E + \eta\phi_D\,Q\phi_E)\,
\frac{b_0}{L_0}\,(\,\Psi_A\,\Xi\,(\phi_B\,\Psi_C)\,)\,\bigr\rangle
\nonumber\\
&\
-\,\frac{1}{2}\,\bigl\langle\, 
Q\phi_B\,\Psi_C\frac{b_0}{L_0}\,(\,\phi_D\,\phi_E\,\Psi_A\,)\,\bigr\rangle
-\,\frac{1}{2}\,\bigl\langle\,\phi_B\,\Psi_C\,\frac{b_0 \Xi}{L_0}\,(\,
(Q\phi_D\,\eta\phi_E + \eta\phi_D\,Q\phi_E)\,\Psi_A\,)\,\bigr\rangle\,. \label{FBFBB-(2P)(d)}
\end{align}

The diagram $(2P)(e)$ in figure \ref{FBFBB-2p} consists of three cubic vertices from $S_{R}^{(1)}$ and two Ramond propagators.
It is calculated as
\begin{align}
 \mathcal{A}_{FBFBB}^{(2P)(e)}\ =&\
\bigl(-\,\bigl\langle\,\phi_E\,\Psi_A\ \ 
\overbracket[0.5pt]{\!\!\! \Psi\,\bigr\rangle  \bigr)
\bigl\langle\,\Psi\!\!\!}\ \ 
\phi_B\ \
\overbracket[0.5pt]{\!\!\! \Psi\,\bigr\rangle
\bigl(-\,\bigl\langle\,\Psi\!\!\!}\ \ 
\Psi_C\,\phi_D\,\bigr\rangle \bigr)
\nonumber\\
=&\
\bigl\langle\,\phi_E\,\Psi_A\,\frac{b_0 X \eta}{L_0}\, \Bigl( \, 
\phi_B\,
\frac{b_0 X \eta}{L_0}\,( \, \Psi_C\,\phi_D\,) \, \Bigr) \, \bigr\rangle.
\end{align}
The calculation is almost the same as that of the diagram $(2P)(d)$ for $\mathcal{A}_{FFBBB}$.
We find
\begin{align}
 \mathcal{A}_{FBFBB}^{(2P)(e)}\
=&\
-\bigl\langle\,\phi_E\,\Psi_A\,\frac{b_0}{L_0}\,\Bigl( \, Q\phi_B\, 
\frac{b_0}{L_0}\,( \, \Psi_C\,Q\phi_D\,) \, \Bigr) \, \bigr\rangle
\nonumber\\
&\
-\,\bigl\langle\,\eta\phi_E\,\Psi_A\,\frac{b_0}{L_0}\,( \, Q\phi_B\,
\Xi\,(\Psi_C\,\phi_D)\,) \, \bigr\rangle
+\,\bigl\langle\,\eta\phi_E\,\Psi_A\,\frac{b_0 \Xi}{L_0}\,( \, \phi_B\,
X\,(\Psi_C\,\eta\phi_D)\,) \, \bigr\rangle
\nonumber\\
&\
+\,\bigl\langle\,\Psi_C\,\phi_D\,\frac{b_0 \Xi}{L_0}\, ( \, 
\eta\phi_E\,\Psi_A\,Q\phi_B\,) \, \bigr\rangle
-\bigl\langle\,\Psi_C\,\eta\phi_D\,\frac{b_0 X}{L_0}\, ( \, 
\Xi\,(\eta\phi_E\,\Psi_A)\,\phi_B\,) \, \bigr\rangle\,. \label{FBFBB-(2P)(e)}
\end{align}

\subsubsection{Contributions from diagrams with one propagator}

\begin{figure}[htb]
  \begin{center}
    \begin{tabular}{c}
    
      \begin{minipage}{0.30\hsize}
        \begin{center}
          \includegraphics[clip, width=3.5cm]{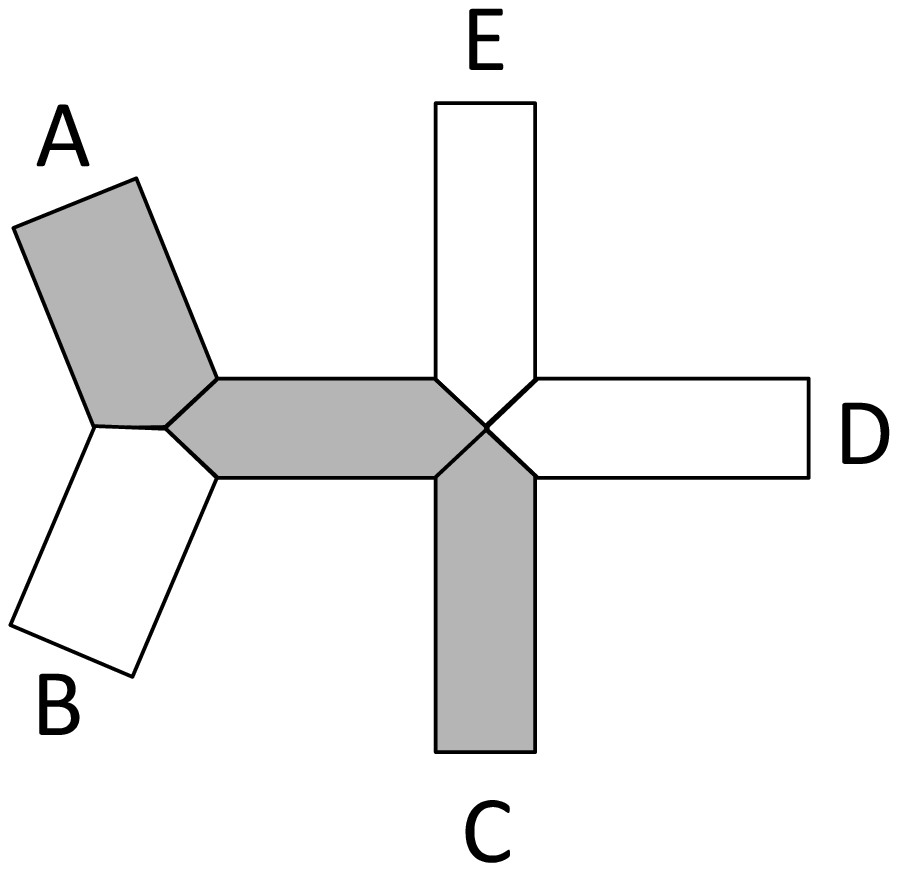}
          \hspace{1.6cm} $(1P)(a)$
                  \end{center}
      \end{minipage}
      \begin{minipage}{0.30\hsize}
        \begin{center}
          \includegraphics[clip, width=3.5cm]{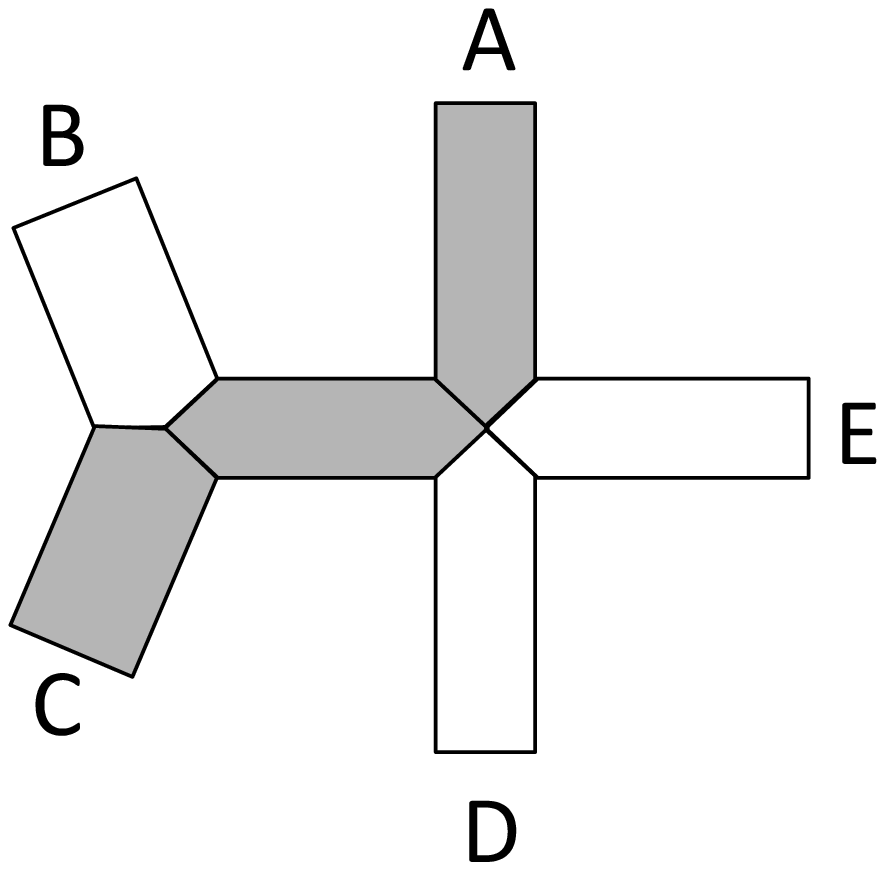}
          \hspace{1.6cm}$(1P)(b)$
        \end{center}
      \end{minipage}
      \begin{minipage}{0.30\hsize}
        \begin{center}
          \includegraphics[clip, width=3.5cm]{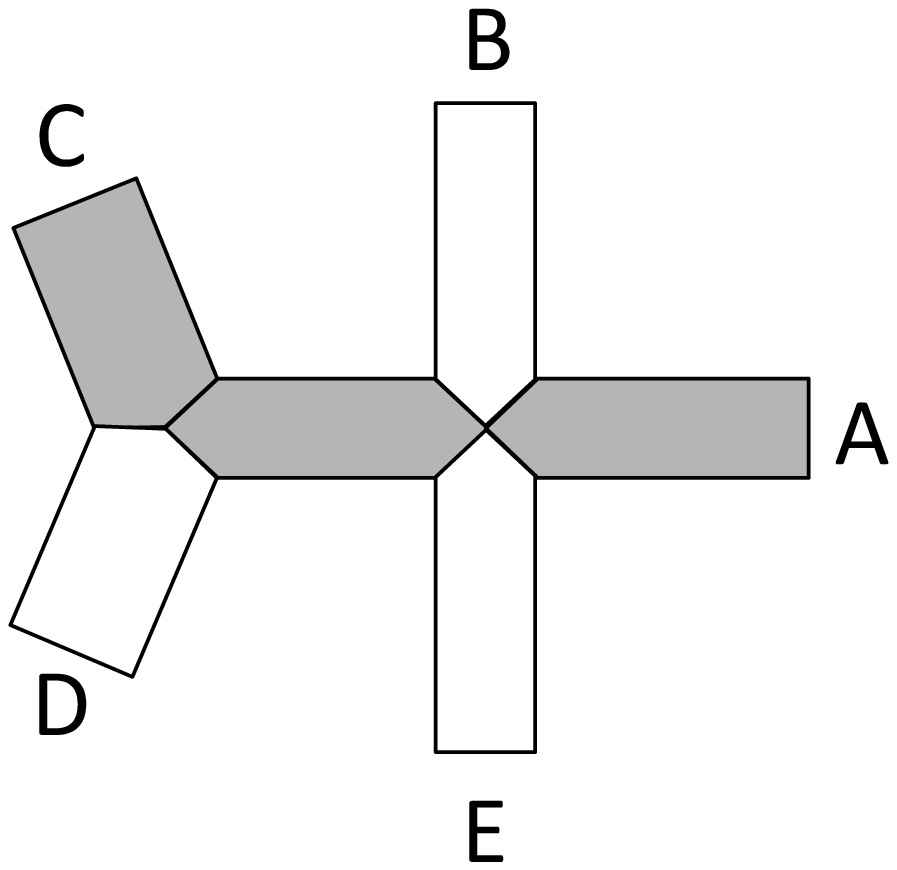}
          \hspace{1.6cm} $(1P)(c)$
        \end{center}
      \end{minipage}
      \\
      \begin{minipage}{0.30\hsize}
        \begin{center}
          \includegraphics[clip, width=3.5cm]{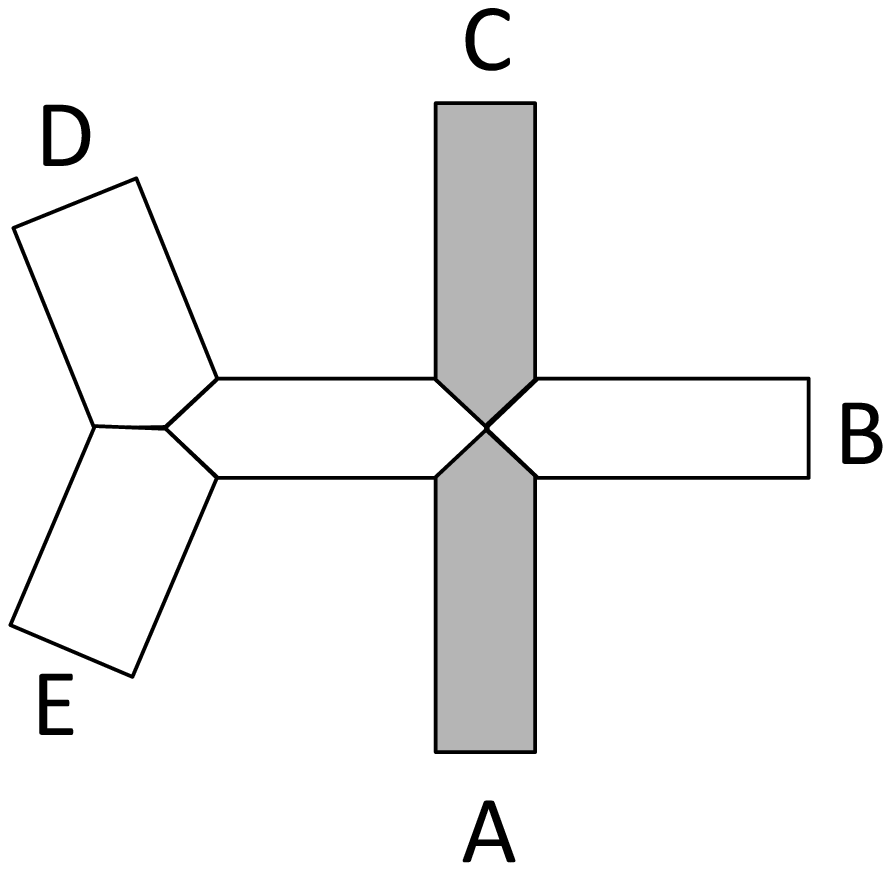}
          \hspace{1.6cm} $(1P)(d)$
        \end{center}
      \end{minipage}
      \begin{minipage}{0.30\hsize}
        \begin{center}
          \includegraphics[clip, width=3.5cm]{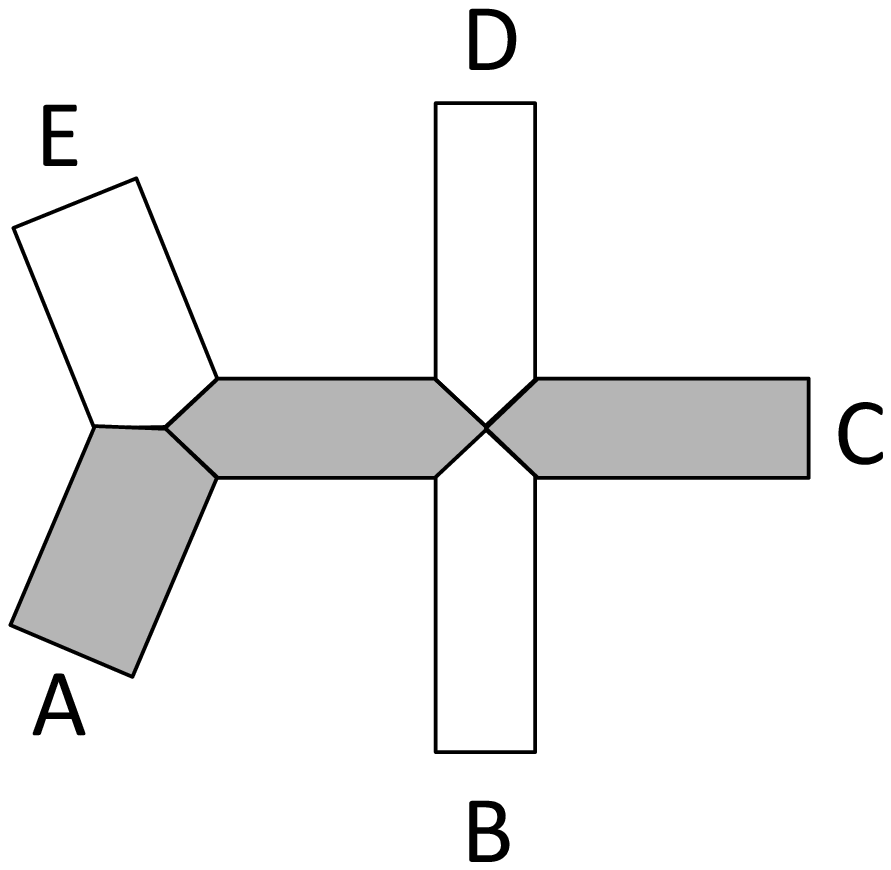}
          \hspace{1.6cm} $(1P)(e)$
        \end{center}
      \end{minipage}
      
    \end{tabular}
    \caption{Feynman diagrams with one cubic vertex, one quartic vertex, and one propagator for $\mathcal{A}_{FBFBB}$.}
    \label{FBFBB-1p}
  \end{center}
\end{figure}

Let us next calculate contributions from Feynman diagrams
with one propagator, one cubic vertex, and one quartic vertex for~$\mathcal{A}_{FBFBB}$.
Such diagrams are depicted in figure~\ref{FBFBB-1p}.

The diagram $(1P)(a)$ in figure \ref{FBFBB-1p} consists of one cubic vertex from $S_{R}^{(1)}$ and the quartic vertex with the fermion-fermion-boson-boson ordering from $S_{R}^{(2)}$.
It is calculated as
\begin{align}
 \mathcal{A}_{FBFBB}^{(1P)(a)}\ =&\
\bigl\langle\,\Psi_A\,\phi_B\ \
\overbracket[0.5pt]{\!\!\! \Psi\,\bigr\rangle
\Big(-\,\frac{1}{2}\,\bigl\langle\,\Psi\!\!\!}\ \ 
(\Xi\,(\Psi_C\,\eta\phi_D)\,\phi_E
-\Xi\,(\Psi_C\,\phi_D)\,\eta\phi_E)\,\bigr\rangle\Big)
\nonumber\\
=&\
\frac{1}{2}\,\bigl\langle\,\Psi_A\,\phi_B\,\frac{b_0 X \eta}{L_0}\,
(\Xi\,(\Psi_C\,\eta\phi_D)\,\phi_E
-\Xi\,(\Psi_C\,\phi_D)\,\eta\phi_E)\,\bigr\rangle
\nonumber\\
=&\
\frac{1}{2}\,\bigl\langle\,\Psi_A\,\phi_B\,\frac{b_0 X}{L_0}\, ( \, 
\Psi_C\,(\eta\phi_D\,\phi_E - \phi_D\eta\phi_E)\,) \, \bigr\rangle
-\,\bigl\langle\,\Psi_A\,\phi_B\,\frac{b_0 X}{L_0}\, ( \, 
\Xi\,(\Psi_C\,\eta\phi_D)\,\eta\phi_E\,) \, \bigr\rangle\,. \label{FBFBB-(1P)(a)}
\end{align}

The diagram $(1P)(b)$ in figure \ref{FBFBB-1p} consists of one cubic vertex from $S_{R}^{(1)}$, the quartic vertex with the fermion-fermion-boson-boson ordering from $S_{R}^{(2)}$, and one Ramond propagator.
It is calculated as
\begin{align}
 \mathcal{A}_{FBFBB}^{(1P)(b)}\ =&\
(-\,\bigl\langle\,\phi_B\,\Psi_C\ \
\overbracket[0.5pt]{\!\!\! \Psi\,\bigr\rangle)
\Big(\frac{1}{2}\,\bigl\langle\,\Psi\!\!\!}\ \ 
(
\phi_D\,\Xi\,(\eta\phi_E\,\Psi_A)
+\,\eta\phi_D\,\Xi\,(\phi_E\,\Psi_A))\,\bigr\rangle\Big)
\nonumber\\
=&\
\frac{1}{2}\,\bigl\langle\,\phi_B\,\Psi_C\,\frac{b_0 X \eta}{L_0}\,
(
\phi_D\,\Xi\,(\eta\phi_E\,\Psi_A)
+\,\eta\phi_D\,\Xi\,(\phi_E\,\Psi_A)
)\,\bigr\rangle
\nonumber\\
=&\
\frac{1}{2}\,\bigl\langle\,\phi_B\,\Psi_C\,\frac{b_0 X}{L_0}\,( \, 
(\phi_D\,\eta\phi_E - \eta\phi_D\,\phi_E)\,\Psi_A\,)\, \bigr\rangle
+\,\bigl\langle\,\phi_B\,\Psi_C\,\frac{b_0 X}{L_0}\, ( \, 
\eta\phi_D\,\Xi\,(\eta\phi_E\,\Psi_A)\,) \, \bigr\rangle\,. \label{FBFBB-(1P)(b)}
\end{align}

The diagram $(1P)(c)$ in figure \ref{FBFBB-1p} consists of one cubic vertex from $S_{R}^{(1)}$,
one quartic vertex with the fermion-boson-fermion-boson ordering from $S_{R}^{(2)}$,
and one Ramond propagator.
It is calculated as
\begin{align}
 \mathcal{A}_{FBFBB}^{(1P)(c)}\ =&\
\bigl\langle\,\Psi_C\,\phi_D\ \
\overbracket[0.5pt]{\!\!\! \Psi\,\bigr\rangle
\Big(\,\frac{1}{2}\,\bigl\langle\,\Psi\!\!\!}\ \ 
(\phi_E\,\Xi\,(\Psi_A\,\eta\phi_B)
-\,\eta\phi_E\,\Xi\,(\Psi_A\,\phi_B))\,\bigr\rangle\Big)
\nonumber\\
&\
-\,\bigl\langle\,\Psi_C\,\phi_D\ \
\overbracket[0.5pt]{\!\!\! \Psi\,\bigr\rangle
\Big(\,\frac{1}{2}\,\bigl\langle\,\Psi\!\!\!}\ \ 
(
\Xi\,(\eta\phi_E\,\Psi_A)\,\phi_B
+\,\Xi\,(\phi_E\,\Psi_A)\,\eta\phi_B)\,\bigr\rangle\Big)
\nonumber\\
=&\
-\,\frac{1}{2}\,\bigl\langle\,\Psi_C\,\phi_D\,\frac{b_0 X \eta}{L_0}\,
(\phi_E\,\Xi\,(\Psi_A\,\eta\phi_B)
-\,\eta\phi_E\,\Xi\,(\Psi_A\,\phi_B))\,\bigr\rangle
\nonumber\\
&\
+\,\frac{1}{2}\,\bigl\langle\,\Psi_C\,\phi_D\,\frac{b_0 X \eta}{L_0}\,
(
\Xi\,(\eta\phi_E\,\Psi_A)\,\phi_B
+\,\Xi\,(\phi_E\,\Psi_A)\,\eta\phi_B)\,\bigr\rangle
\nonumber\\
=&\
-\,\bigl\langle\,\Psi_C\,\phi_D\,\frac{b_0 X}{L_0}\, ( \, 
\eta\phi_E\,\Xi\,(\Psi_A\,\eta\phi_B)\,) \, \bigr\rangle
-\,\bigl\langle\,\Psi_C\,\phi_D\,\frac{b_0 X}{L_0}\, ( \, 
\Xi\,(\eta\phi_E\,\Psi_A)\,\eta\phi_B\,) \, \bigr\rangle\,. \label{FBFBB-(1P)(c)}
\end{align}

The diagram $(1P)(d)$ in figure \ref{FBFBB-1p} consists of one cubic vertex from $S_{NS}^{(1)}$,
one quartic vertex with the fermion-boson-fermion-boson ordering from $S_{R}^{(2)}$,
and one NS propagator.
It is calculated as
\begin{align}
 \mathcal{A}_{FBFBB}^{(1P)(d)}\ =&\
\Big(\,-\,\frac{1}{2}\,\bigl\langle\,(Q\eta\phi_D\,\eta\phi_E + \eta\phi_D\,Q\phi_E)\ \
\overbracket[0.5pt]{\!\!\! \phi\,\bigr\rangle\Big)
\Big(\,-\,\frac{1}{2}\,\bigl\langle\,\phi\!\!\!}\ \ 
(
\Psi_A\,\Xi\,(\eta\phi_B\,\Psi_C)
+\,\Xi\,(\Psi_A\,\eta\phi_B)\,\Psi_C)\,\bigr\rangle\Big)
\nonumber\\
&\
\Big(\,-\,\frac{1}{2}\,\bigl\langle\,(Q\eta\phi_D\,\eta\phi_E + \eta\phi_D\,Q\phi_E)\ \
\overbracket[0.5pt]{\!\!\! \phi\,\bigr\rangle\Big)
\Big(\,\frac{1}{2}\,\bigl\langle\,\eta\phi\!\!\!}\ \ 
(
\Psi_A\,\Xi\,(\phi_B\,\Psi_C)
+ \Xi\,(\Psi_A\,\phi_B)\,\Psi_C)\,\bigr\rangle\Big)
\nonumber\\
=&\
\frac{1}{2}\,\bigl\langle\,(Q\eta\phi_D\,\eta\phi_E + \eta\phi_D\,Q\phi_E)\,
\frac{\xi_0 b_0}{L_0}\,(\Psi_A\,\Xi\,(\eta\phi_B\,\Psi_C)
+\,\Xi\,(\Psi_A\,\eta\phi_B)\,\Psi_C)\,\bigr\rangle\,. \label{FBFBB-(1P)(d)}
\end{align}

The diagram $(1P)(e)$ in figure \ref{FBFBB-1p} consists of one cubic vertex from $S_{R}^{(1)}$,
one quartic vertex with the fermion-boson-fermion-boson ordering from $S_{R}^{(2)}$,
and one Ramond propagator.
It is calculated as
\begin{align}
 \mathcal{A}_{FBFBB}^{(1P)(e)}\ =&\
(-\,\bigl\langle\,\phi_E\,\Psi_A\ \
\overbracket[0.5pt]{\!\!\! \Psi\,\bigr\rangle)
\Big(\,\frac{1}{2}\,\bigl\langle\,\Psi\!\!\!}\ \ 
(
\phi_B\,\Xi\,(\Psi_C\,\eta\phi_D)
-\,\eta\phi_B\,\Xi\,(\Psi_C\,\phi_D)
)\,\bigr\rangle\Big)
\nonumber\\
&\
+\, 
(-\,\bigl\langle\,\phi_E\,\Psi_A\ \
\overbracket[0.5pt]{\!\!\! \Psi\,\bigr\rangle)
\Big(\,-\,\frac{1}{2}\,\bigl\langle\,\Psi\!\!\!}\ \ 
(
\Xi\,(\eta\phi_B\,\Psi_C)\,\phi_D
+\,\Xi\,(\phi_B\,\Psi_C)\,\eta\phi_D
)\,\bigr\rangle\Big)
\nonumber\\
=&\
\bigl\langle\,\phi_E\,\Psi_A\,\frac{b_0 X}{L_0}\, ( \, 
\eta\phi_B\,\Xi\,(\Psi_C\,\eta\phi_D)\,) \, \bigr\rangle
+\,\bigl\langle\,\phi_E\,\Psi_A\,\frac{b_0 X}{L_0}\, ( \, 
\Xi\,(\eta\phi_B\,\Psi_C)\,\eta\phi_D\,) \, \bigr\rangle\,. \label{FBFBB-(1P)(e)}
\end{align}

\subsubsection{Contributions from diagrams without propagators}
\begin{figure}[H]
  \begin{center}
    \begin{tabular}{c}
    
      \begin{minipage}{0.33\hsize}
        \begin{center}
          \includegraphics[clip, width=3.5cm]{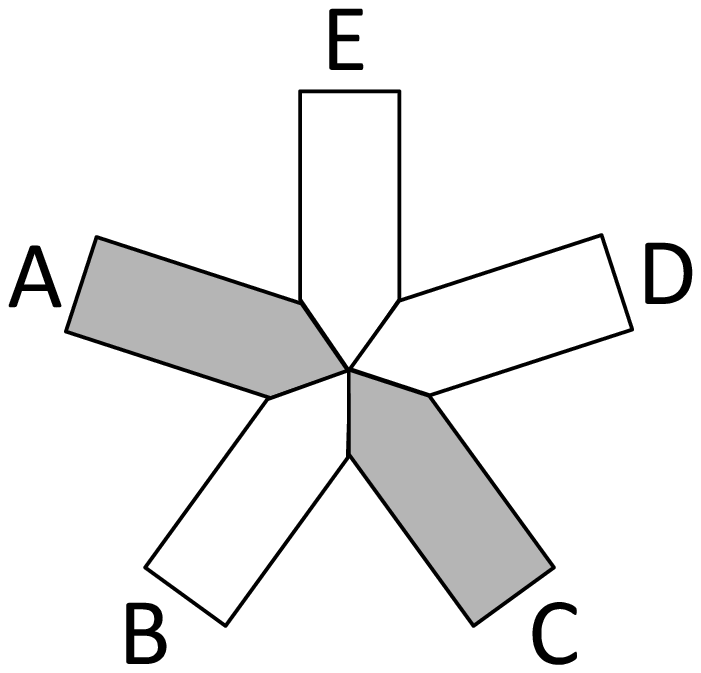}
          \hspace{1.6cm} $(NP)$
                  \end{center}
      \end{minipage}
          \end{tabular}
    \caption{Feynman diagram with a quintic vertex for $\mathcal{A}_{FBFBB}$.}
    \label{FBFBB-np}
  \end{center}
\end{figure}

The last Feynman diagram for~$\mathcal{A}_{FBFBB}$
consists of a quintic a quintic vertex from $S_{R}^{(3)}$.
It is depicted in figure~\ref{FBFBB-np}. From the quintic interaction $S_{FBFBB}$ in~\eqref{quintic-FBFBB},
\begin{align}
S_{FBFBB}
=&-\frac{1}{3}\, \bigl\langle\, \phi\,\Psi\,\Xi(\, \eta\phi\,\Xi(\,\eta\phi\,\Psi\,)\,)\, \bigr\rangle
-\frac{1}{3}\, \bigl\langle\,\phi\,\Psi\,\Xi(\,\eta\phi\,\Xi(\,\Psi\,\eta\phi\,)\,) \,\bigr\rangle
-\frac{1}{3}\, \bigl\langle\,\phi\,\Psi\,\Xi(\,\Xi(\eta\phi\,\Psi\,)\,\eta\phi\,) \,\bigr\rangle \nonumber\\
&-\frac{1}{3}\, \bigl\langle\,\phi\,\Xi(\,\eta\phi\,(\Xi(\,\Psi\,\eta\phi\,)\,)\,\Psi \,\bigr\rangle
-\frac{1}{3}\, \bigl\langle\, \phi\,\Xi(\,\Xi(\,\eta\phi\,\Psi\,)\,\eta\phi\,)\,\Psi\,\bigr\rangle
-\frac{1}{3}\, \bigl\langle\,\phi\,\Xi(\, \Xi(\,\Psi\,\eta\phi\,)\,\eta\phi\,) \,\Psi\,\bigr\rangle \nonumber \\
&-\frac{1}{3}\, \bigl\langle\,\phi\,\Xi(\,\eta\phi\,\Psi\,)\,\Xi(\eta\phi\Psi\,)\,\,\bigr\rangle
-\frac{1}{3}\, \bigl\langle\, \phi\,\Xi(\,\Psi\,\eta\phi\,)\,\Xi(\,\eta\phi\,\Psi\,) \,\bigr\rangle
-\frac{1}{3}\, \bigl\langle\, \phi\,\Xi(\,\Psi\,\eta\phi\,)\,\Xi(\,\Psi\,\eta\phi\,)\,\bigr\rangle \nonumber \\
&+\frac{1}{6}\, \bigl\langle\,\phi\,\Psi\,\Xi(\,\eta\phi\,\phi\,\Psi\,) \,\bigr\rangle
-\frac{1}{6}\, \bigl\langle\,\phi\,\Psi\,\Xi(\,\phi\,\eta\phi\,\Psi\,) \,\bigr\rangle \nonumber \\
&+\frac{1}{6}\, \bigl\langle\,\phi\,\Xi(\,\Psi\,\eta\phi\,\phi\,)\,\Psi \,\bigr\rangle 
-\frac{1}{6}\, \bigl\langle\,\phi\,\Xi(\,\Psi\,\phi\,\eta\phi\,)\,\Psi \,\bigr\rangle,
\end{align}
it is calculated as
\begin{align}
\mathcal{A}_{FBFBB}^{(NP)}
=&-\frac{1}{3}\, \bigl\langle\, \phi_B\,\Psi_C\,\Xi(\, \eta\phi_D\,\Xi(\,\eta\phi_E\,\Psi_A\,)\,)\, \bigr\rangle
-\frac{1}{3}\, \bigl\langle\,\phi_E\,\Psi_A\,\Xi(\,\eta\phi_B\,\Xi(\,\Psi_C\,\eta\phi_D\,)\,) \,\bigr\rangle \nonumber \\
&-\frac{1}{3}\, \bigl\langle\,\phi_E\,\Psi_A\,\Xi(\,\Xi(\eta\phi_B\,\Psi_C\,)\,\eta\phi_D\,) \,\bigr\rangle 
-\frac{1}{3}\, \bigl\langle\,\phi_D\,\Xi(\,\eta\phi_E\,(\Xi(\,\Psi_A\,\eta\phi_B\,)\,)\,\Psi_B \,\bigr\rangle \nonumber \\
&-\frac{1}{3}\, \bigl\langle\, \phi_D\,\Xi(\,\Xi(\,\eta\phi_E\,\Psi_A\,)\,\eta\phi_B\,)\,\Psi_C\,\bigr\rangle
-\frac{1}{3}\, \bigl\langle\,\phi_B\,\Xi(\, \Xi(\,\Psi_C\,\eta\phi_D\,)\,\eta\phi_E\,) \,\Psi_A\,\bigr\rangle \nonumber \\
&-\frac{1}{3}\, \bigl\langle\,\phi_D\,\Xi(\,\eta\phi_E\,\Psi_A\,)\,\Xi(\eta\phi_B\,\Psi_C\,)\,\bigr\rangle
-\frac{1}{3}\, \bigl\langle\, \phi_B\,\Xi(\,\Psi_C\,\eta\phi_D\,)\,\Xi(\,\eta\phi_E\,\Psi_A\,) \,\bigr\rangle \nonumber \\
&-\frac{1}{3}\, \bigl\langle\, \phi_E\,\Xi(\,\Psi_A\,\eta\phi_B\,)\,\Xi(\,\Psi_C\,\eta\phi_D\,)\,\bigr\rangle \nonumber \\
&+\frac{1}{6}\, \bigl\langle\,\phi_B\,\Psi_C\,\Xi(\,\eta\phi_D\,\phi_E\,\Psi_A\,) \,\bigr\rangle
-\frac{1}{6}\, \bigl\langle\,\phi_B\,\Psi_C\,\Xi(\,\phi_D\,\eta\phi_E\,\Psi_A\,) \,\bigr\rangle \nonumber \\
&+\frac{1}{6}\, \bigl\langle\,\phi_B\,\Xi(\,\Psi_C\,\eta\phi_D\,\phi_E\,)\,\Psi_A \,\bigr\rangle 
-\frac{1}{6}\, \bigl\langle\,\phi_B\,\Xi(\,\Psi_C\,\phi_D\,\eta\phi_E\,)\,\Psi_A \,\bigr\rangle\,.
\end{align}
We transform each term such that the first string field is $\Psi_A$
and move $\eta$ acting on $\phi_B$ to find
\begin{align}
\mathcal{A}_{FBFBB}^{(NP)}
=&
-\frac{1}{2}\,\bigl\langle\,\Psi_A\,\Xi(\,\phi_B\,\Psi_C\,)\,\phi_D\,\eta\phi_E\,\bigr\rangle
-\frac{1}{2}\,\bigl\langle\,\Psi_A\,\phi_B\,\Xi(\,\Psi_C\,\eta\phi_D\,\phi_E\,)\,\bigr\rangle \nonumber \\
&+\frac{1}{2}\,\bigl\langle\,\Psi_A\,\phi_B\,\Xi(\,\Psi_C\,\phi_D\,\eta\phi_E\,)\,\bigr\rangle
+\frac{1}{2}\,\bigl\langle\,\Psi_A\,\Xi(\,\phi_B\,\Psi_C\,)\,\eta\phi_D\,\phi_E\,\bigr\rangle \nonumber \\
&+\bigl\langle\,\Psi_A\,\Xi(\,\Xi\,(\,\phi_B\,\Psi_C\,)\,\eta\phi_D\,)\,\eta\phi_E\,\bigr\rangle
-\bigl\langle\,\Psi_A\,\Xi(\,\phi_B\,\Xi(\,\Psi_C\,\eta\phi_D\,)\,)\,\eta\phi_E\,\bigr\rangle \nonumber \\
&+\bigl\langle\,\Psi_A\,\phi_B\,\Xi(\,\Xi(\,\Psi_C\,\eta\phi_D\,)\,\eta\phi_E\,)\,\bigr\rangle \,.
\label{FBFBB NP}
\end{align}

\subsubsection{Contributions from all diagrams}
In the calculations of the diagrams with two propagators,
we found that all the terms of~$\mathcal{A}_{FBFBB}^{\rm WS}$ are reproduced,
but we also found extra terms with one propagator are generated.
As we did in subsection~\ref{cancellation}, we combine these extra terms
and the contributions from the diagrams with one propagator and then decompose them according to the location of the propagator as follows:
\begin{equation}
\begin{split}
& ( \mathcal{A}_{FBFBB}^{(2P)} -\mathcal{A}_{FBFBB}^{\rm WS} )
+\mathcal{A}_{FBFBB}^{(1P)}
 \\
& = \mathcal{A}_{FBFBB}^{(AB|CDE)}
+\mathcal{A}_{FBFBB}^{(BC|DEA)}
+\mathcal{A}_{FBFBB}^{(CD|EAB)}
+\mathcal{A}_{FBFBB}^{(DE|ABC)}
+\mathcal{A}_{FBFBB}^{(EA|BCD)} \,,
\end{split}
\end{equation}
where
\begin{equation}
\begin{split}
\mathcal{A}_{FBFBB}^{(2P)}
& = \mathcal{A}_{FBFBB}^{(2P)(a)}
+\mathcal{A}_{FBFBB}^{(2P)(b)}
+\mathcal{A}_{FBFBB}^{(2P)(c)}
+\mathcal{A}_{FBFBB}^{(2P)(d)}
+\mathcal{A}_{FBFBB}^{(2P)(e)} \,, \\
\mathcal{A}_{FBFBB}^{(1P)}
& = \mathcal{A}_{FBFBB}^{(1P)(a)}
+\mathcal{A}_{FBFBB}^{(1P)(b)}
+\mathcal{A}_{FBFBB}^{(1P)(c)}
+\mathcal{A}_{FBFBB}^{(1P)(d)}
+\mathcal{A}_{FBFBB}^{(1P)(e)} \,.
\end{split}
\end{equation}

Let us first consider the contribution $\mathcal{A}_{FBFBB}^{(AB|CDE)}$. The sources of this contribution are the diagrams $(2P)(a)$, $(2P)(c)$, and~$(1P)(a)$.
We collect the second and third terms of $\mathcal{A}_{FBFBB}^{(2P)(a)}$ in~\eqref{FBFBB-(2P)(a)}, the fourth and fifth terms of~$\mathcal{A}_{FBFBB}^{(2P)(c)}$ in~\eqref{FBFBB-(2P)(c)}, and $\mathcal{A}_{FBFBB}^{(1P)(a)}$ in~\eqref{FBFBB-(1P)(a)}, and we find that the sum of these terms reduces to a form without propagators:
\begin{align}
\mathcal{A}_{FBFBB}^{(AB|CDE)}=
\frac{1}{2}\,\bigl\langle\,\Psi_A\,\phi_B\,\Xi\,(
\Psi_C\,\eta\,(\phi_D\,\phi_E)\big)\,\bigr\rangle\,. \label{np-A}
\end{align}

The sources of the contribution~$\mathcal{A}_{FBFBB}^{(BC|DEA)}$ are the diagrams $(2P)(b)$, $(2P)(d)$, and~$(1P)(b)$.
We collect
the third, fourth, and fifth terms of~$\mathcal{A}_{FBFBB}^{(2P)(b)}$ in~\eqref{FBFBB-(2P)(b)}, the fourth and fifth terms of~$\mathcal{A}_{FBFBB}^{(2P)(d)}$ in~\eqref{FBFBB-(2P)(d)}, and $\mathcal{A}_{FBFBB}^{(1P)(b)}$ in~\eqref{FBFBB-(1P)(b)}, and we find that the sum of these terms reduces to a form without propagators:
\begin{align}
\mathcal{A}_{FBFBB}^{(BC|DEA)}
=-\,\frac{1}{2}\,\bigl\langle\,\Psi_A\,\Xi\,(\phi_B\,\Psi_C)\,
\eta\,(\phi_D\,\phi_E)\,\bigr\rangle\,. \label{np-B}
\end{align}

The sources of the contribution~$\mathcal{A}_{FBFBB}^{(CD|EAB)}$ are the diagrams $(2P)(c)$, $(2P)(e)$, and~$(1P)(c)$.
We collect relevant terms in~\eqref{FBFBB-(2P)(c)}, \eqref{FBFBB-(2P)(e)}, and~\eqref{FBFBB-(1P)(c)},
and we find that the sum of these terms reduces to a form without propagators:
\begin{align}
\mathcal{A}_{FBFBB}^{(CD|EAB)}
=-\,\bigl\langle\,\Psi_A\,\phi_B\,\Xi\,(\Psi_C\,\phi_D)\,\eta\phi_E\,\bigr\rangle
-\,\bigl\langle\,\Psi_A\,\eta\phi_B\,\Xi\,(\Xi\,(\Psi_C\,\phi_D)\,
\eta\phi_E\big)\,\bigr\rangle\,. \label{np-C}
\end{align}

The sources of the contribution~$\mathcal{A}_{FBFBB}^{(DE|ABC)}$ are the diagrams $(2P)(a)$, $(2P)(d)$, and $(1P)(d)$.
We collect relevant terms in~\eqref{FBFBB-(2P)(a)}, \eqref{FBFBB-(2P)(d)}, and~\eqref{FBFBB-(1P)(d)}, and this time the sum of these terms turns out to vanish:
\begin{equation}
\mathcal{A}_{FBFBB}^{(CD|EAB)}=0 \,.
\end{equation}

The sources of the contribution~$\mathcal{A}_{FBFBB}^{(EA|BCD)}$ are the diagrams $(2P)(b)$, $(2P)(e)$, and~$(1P)(e)$.
We collect the relevant terms in~\eqref{FBFBB-(2P)(b)}, \eqref{FBFBB-(2P)(e)}, and~\eqref{FBFBB-(1P)(e)},
and we find that the sum of these terms reduces to a form without the propagators:
\begin{align}
\mathcal{A}_{FBFBB}^{(EA|BCD)}=&
\bigl\langle\,\Psi_A\,\phi_B\,\Xi\,(\Psi_C\,\phi_D)\,\eta\phi_E\,\bigr\rangle
-\,\bigl\langle\,\Psi_A\,\Xi\,(
\Xi\,(\eta\phi_B\,\Psi_C)\,\phi_D\big)\,\eta\phi_E\,\bigr\rangle \nonumber \\
&-\,\bigl\langle\,\Psi_A\,\Xi\,(
\eta\phi_B\,\Xi\,(\Psi_C\,\phi_D)\big)\,\eta\phi_E\,\bigr\rangle\,. \label{np-E}
\end{align}

To summarize, the sum of all the contributions is given by
\begin{align}
&\ \mathcal{A}_{FBFBB}^{(AB|CDE)}
+\mathcal{A}_{FBFBB}^{(BC|DEA)}
+\mathcal{A}_{FBFBB}^{(CD|EAB)}
+\mathcal{A}_{FBFBB}^{(DE|ABC)}
+\mathcal{A}_{FBFBB}^{(EA|BCD)} \nonumber \\
&=\
 \frac{1}{2}\,\bigl\langle\,\Psi_A\,\phi_B\,\Xi\,(
\Psi_C\,\eta\,(\phi_D\,\phi_E)\big)\,\bigr\rangle
-\,\frac{1}{2}\,\bigl\langle\,\Psi_A\,\Xi\,(\phi_B\,\Psi_C)\,
\eta\,(\phi_D\,\phi_E)\,\bigr\rangle
\nonumber\\
&\quad \ 
- \bigl\langle\,\Psi_A\,\phi_B\,\Xi\,(\Psi_C\,\phi_D)\,\eta\phi_E\,\bigr\rangle
-\,\bigl\langle\,\Psi_A\,\eta\phi_B\,\Xi\,(\Xi\,(\Psi_C\,\phi_D)\,
\eta\phi_E\big))\,\bigr\rangle
\nonumber\\
&\quad \
+ \bigl\langle\,\Psi_A\,\phi_B\,\Xi\,(\Psi_C\,\phi_D)\,\eta\phi_E\,\bigr\rangle
-\,\bigl\langle\,\Psi_A\,\Xi\,(
\Xi\,(\eta\phi_B\,\Psi_C)\,\phi_D\big)\,\eta\phi_E\,\bigr\rangle
\nonumber\\
&\quad \
-\,\bigl\langle\,\Psi_A\,\Xi\,(
\eta\phi_B\,\Xi\,(\Psi_C\,\phi_D)\big)\,\eta\phi_E\,\bigr\rangle \,.
\end{align}
We move $\eta$ acting on $\phi_B$ to avoid the combination $\eta \phi_B$ and find
\begin{align}
&\mathcal{A}_{FBFBB}^{(AB|CDE)}
+\mathcal{A}_{FBFBB}^{(BC|DEA)}
+\mathcal{A}_{FBFBB}^{(CD|EAB)}
+\mathcal{A}_{FBFBB}^{(DE|ABC)}
+\mathcal{A}_{FBFBB}^{(EA|BCD)} \nonumber \\
&=\frac{1}{2}\,\bigl\langle\,\Psi_A\,\Xi(\,\phi_B\,\Psi_C\,)\,\phi_D\,\eta\phi_E\,\bigr\rangle
+\frac{1}{2}\,\bigl\langle\,\Psi_A\,\phi_B\,\Xi(\,\Psi_C\,\eta\phi_D\,\phi_E\,)\,\bigr\rangle \nonumber \\
&\quad -\frac{1}{2}\,\bigl\langle\,\Psi_A\,\phi_B\,\Xi(\,\Psi_C\,\phi_D\,\eta\phi_E\,)\,\bigr\rangle
-\frac{1}{2}\,\bigl\langle\,\Psi_A\,\Xi(\,\phi_B\,\Psi_C\,)\,\eta\phi_D\,\phi_E\,\bigr\rangle \nonumber \\
&\quad-\bigl\langle\,\Psi_A\,\Xi(\,\Xi\,(\,\phi_B\,\Psi_C\,)\,\eta\phi_D\,)\,\eta\phi_E\,\bigr\rangle
+\bigl\langle\,\Psi_A\,\Xi(\,\phi_B\,\Xi(\,\Psi_C\,\eta\phi_D\,)\,)\,\eta\phi_E\,\bigr\rangle \nonumber \\
&\quad-\bigl\langle\,\Psi_A\,\phi_B\,\Xi(\,\Xi(\,\Psi_C\,\eta\phi_D\,)\,\eta\phi_E\,)\,\bigr\rangle \,.
\end{align}
The sum of these terms is precisely canceled by the contribution~$\mathcal{A}_{FBFBB}^{(NP)}$
in the form~\eqref{FBFBB NP}:
\begin{equation}
\mathcal{A}_{FBFBB}^{(AB|CDE)}
+\mathcal{A}_{FBFBB}^{(BC|DEA)}
+\mathcal{A}_{FBFBB}^{(CD|EAB)}
+\mathcal{A}_{FBFBB}^{(DE|ABC)}
+\mathcal{A}_{FBFBB}^{(EA|BCD)}
= -\mathcal{A}_{FBFBB}^{(NP)}.
\end{equation}
We thus conclude that the amplitude $\mathcal{A}_{FBFBB}^{{\rm WS}}$ in the world-sheet theory is correctly reproduced by the amplitude $\mathcal{A}_{FBFBB}$ calculated in open superstring field theory:
\begin{equation}
\mathcal{A}_{FBFBB}=\mathcal{A}_{FBFBB}^{{\rm WS}},
\end{equation}
where
\begin{equation}
\mathcal{A}_{FBFBB}=\mathcal{A}_{FBFBB}^{(2P)}+\mathcal{A}_{FBFBB}^{(1P)}+\mathcal{A}_{FBFBB}^{(NP)}.
\end{equation}

\end{document}